\documentclass[twocolumn]{aastex631}

\usepackage{amsmath}

\newcommand\tess{TESS}
\newcommand\gaia{\textit{Gaia}}
\newcommand\kms{$\textrm{km~s}^{-1}$}
\newcommand\ms{$\textrm{m~s}^{-1}$}
\newcommand\cms{$\textrm{cm~s}^{-1}$}
\newcommand\gcmcubed{$\textrm{g~cm}^{-3}$}

\newcommand\teff{T$_{\rm{eff}}$}
\newcommand\vsini{$v$~sin~$i$}

\newcommand{\unit}[1]{\ensuremath{\, \mathrm{#1}}} %Remove if necessary
\newcommand\earthmass{$M_{\oplus}$}
\newcommand\earthradius{$R_{\oplus}$}
\newcommand\jupitermass{$M_{J}$}

\newcommand\solmass{$M_{\odot}$}
\newcommand\solradius{$R_{\odot}$}

\received{September 22, 2022}
\accepted{December 15, 2022}
\shortauthors{Kanodia et al. 2023}
\shorttitle{TOI-5205: A mid-M dwarf hosting a Jovian Planet}

\submitjournal{AJ}

\begin{document}

\title{TOI-5205 b: A Short-period Jovian Planet Transiting a Mid-M Dwarf}

\author[0000-0001-8401-4300]{Shubham Kanodia}
\affil{Earth and Planets Laboratory, Carnegie Institution for Science, 5241 Broad Branch Road, NW, Washington, DC 20015, USA}
\affil{Department of Astronomy \& Astrophysics, 525 Davey Laboratory, The Pennsylvania State University, University Park, PA 16802, USA}
\affil{Center for Exoplanets and Habitable Worlds, 525 Davey Laboratory, The Pennsylvania State University, University Park, PA 16802, USA}

\author[0000-0001-9596-7983]{Suvrath Mahadevan}
\affil{Department of Astronomy \& Astrophysics, 525 Davey Laboratory, The Pennsylvania State University, University Park, PA 16802, USA}
\affil{Center for Exoplanets and Habitable Worlds, 525 Davey Laboratory, The Pennsylvania State University, University Park, PA 16802, USA}
\affil{ETH Zurich, Institute for Particle Physics \& Astrophysics, Switzerland}

\author[0000-0002-2990-7613]{Jessica Libby-Roberts}
\affil{Department of Astronomy \& Astrophysics, 525 Davey Laboratory, The Pennsylvania State University, University Park, PA 16802, USA}
\affil{Center for Exoplanets and Habitable Worlds, 525 Davey Laboratory, The Pennsylvania State University, University Park, PA 16802, USA}

\author[0000-0001-7409-5688]{Gudmundur Stefansson}
\affil{Henry Norris Russell Fellow}
\affil{Department of Astrophysical Sciences, Princeton University, 4 Ivy Lane, Princeton, NJ 08540, USA}

\author[0000-0003-4835-0619]{Caleb I. Ca\~nas}
\altaffiliation{NASA Postdoctoral Program Fellow}
\affil{NASA Goddard Space Flight Center, 8800 Greenbelt Road, Greenbelt, MD 20771, USA}
\affil{Department of Astronomy \& Astrophysics, 525 Davey Laboratory, The Pennsylvania State University, University Park, PA 16802, USA}
\affil{Center for Exoplanets and Habitable Worlds, 525 Davey Laboratory, The Pennsylvania State University, University Park, PA 16802, USA}

\author[0000-0002-4487-5533]{Anjali A. A. Piette}
\affil{Earth and Planets Laboratory, Carnegie Institution for Science, 5241 Broad Branch Road, NW, Washington, DC 20015, USA}

\author[0000-0001-7119-1105]{Alan Boss}
\affil{Earth and Planets Laboratory, Carnegie Institution for Science, 5241 Broad Branch Road, NW, Washington, DC 20015, USA}

\author{Johanna Teske}
\affil{Earth and Planets Laboratory, Carnegie Institution for Science, 5241 Broad Branch Road, NW, Washington, DC 20015, USA}

\author[0000-0001-9046-2265]{John Chambers}
\affil{Earth and Planets Laboratory, Carnegie Institution for Science, 5241 Broad Branch Road, NW, Washington, DC 20015, USA}

\author[0000-0003-2307-0629]{Greg Zeimann}
\affil{Hobby Eberly Telescope, University of Texas, Austin, Austin, TX, 78712, USA}

\author[0000-0002-0048-2586]{Andrew Monson}
\affil{Steward Observatory, The University of Arizona, 933 N.\ Cherry Avenue, Tucson, AZ 85721, USA}

\author[0000-0003-0149-9678]{Paul Robertson}
\affil{Department of Physics \& Astronomy, University of California Irvine, Irvine, CA 92697, USA}

\author[0000-0001-8720-5612]{Joe P.\ Ninan}
\affil{Department of Astronomy and Astrophysics, Tata Institute of Fundamental Research, Homi Bhabha Road, Colaba, Mumbai 400005, India}

\author[0000-0002-9082-6337]{Andrea S.J.\ Lin}
\affil{Department of Astronomy \& Astrophysics, 525 Davey Laboratory, The Pennsylvania State University, University Park, PA 16802, USA}
\affil{Center for Exoplanets and Habitable Worlds, 525 Davey Laboratory, The Pennsylvania State University, University Park, PA 16802, USA} 

\author[0000-0003-4384-7220]{Chad F. Bender}
\affil{Steward Observatory, The University of Arizona, 933 N.\ Cherry Avenue, Tucson, AZ 85721, USA}

\author[0000-0001-9662-3496]{William D. Cochran}
\affil{McDonald Observatory and Department of Astronomy, The University of Texas at Austin, USA}
\affil{Center for Planetary Systems Habitability, The University of Texas at Austin, USA}

\author[0000-0002-2144-0764]{Scott A. Diddams}
\affil{Electrical, Computer \& Energy Engineering, University of Colorado, 425 UCB, Boulder, CO 80309, USA}
\affil{Department of Physics, University of Colorado, 2000 Colorado Avenue, Boulder, CO 80309, USA}
\affil{Time and Frequency Division, National Institute of Standards and Technology, 325 Broadway, Boulder, CO 80305, USA}

\author[0000-0002-5463-9980]{Arvind F.\ Gupta}
\affil{Department of Astronomy \& Astrophysics, 525 Davey Laboratory, The Pennsylvania State University, University Park, PA 16802, USA}
\affil{Center for Exoplanets and Habitable Worlds, 525 Davey Laboratory, The Pennsylvania State University, University Park, PA 16802, USA}

\author[0000-0003-1312-9391]{Samuel Halverson}
\affil{Jet Propulsion Laboratory, 4800 Oak Grove Drive, Pasadena, CA 91109, USA}

\author[0000-0002-6629-4182]{Suzanne Hawley}
\affil{Department of Astronomy, Box 351580, University of Washington, Seattle, WA 98195 USA}

\author[0000-0002-4475-4176]{Henry A. Kobulnicky}
\affil{Department of Physics \& Astronomy, University of Wyoming, Laramie, WY 82070, USA}

\author[0000-0001-5000-1018]{Andrew J. Metcalf}
\affiliation{Space Vehicles Directorate, Air Force Research Laboratory, 3550 Aberdeen Ave. SE, Kirtland AFB, NM 87117, USA}

\author[0000-0001-9307-8170]{Brock A. Parker}
\affil{Department of Physics \& Astronomy, University of Wyoming, Laramie, WY 82070, USA}

\author[0000-0002-5300-5353]{Luke Powers}
\affil{Department of Astronomy \& Astrophysics, 525 Davey Laboratory, The Pennsylvania State University, University Park, PA 16802, USA}
\affil{Center for Exoplanets and Habitable Worlds, 525 Davey Laboratory, The Pennsylvania State University, University Park, PA 16802, USA}

\author[0000-0002-4289-7958]{Lawrence W. Ramsey}
\affil{Department of Astronomy \& Astrophysics, 525 Davey Laboratory, The Pennsylvania State University, University Park, PA 16802, USA}
\affil{Center for Exoplanets and Habitable Worlds, 525 Davey Laboratory, The Pennsylvania State University, University Park, PA 16802, USA}

\author[0000-0001-8127-5775]{Arpita Roy}
\affil{Space Telescope Science Institute, 3700 San Martin Dr, Baltimore, MD 21218, USA}
\affil{Department of Physics and Astronomy, Johns Hopkins University, 3400 N Charles Street, Baltimore, MD 21218, USA}

\author[0000-0002-4046-987X]{Christian Schwab}
\affil{School of Mathematical and Physical Sciences, Macquarie University, Balaclava Road, North Ryde, NSW 2109, Australia}

\author[0000-0002-5817-202X]{Tera N. Swaby}
\affil{Department of Physics \& Astronomy, University of Wyoming, Laramie, WY 82070, USA}

\author[0000-0002-4788-8858]{Ryan C. Terrien}
\affil{Carleton College, One North College St., Northfield, MN 55057, USA}

\author[0000-0001-9209-1808]{John Wisniewski}
\affil{Department of Physics \& Astronomy, George Mason University, 4400 University Drive, MS 3F3, Fairfax, VA 22030, USA}

\correspondingauthor{Shubham Kanodia}
\email{shbhuk@gmail.com}

\begin{abstract}
We present the discovery of TOI-5205~b, a transiting Jovian planet orbiting a solar metallicity M4V star, which was discovered using Transiting Exoplanet Survey Satellite photometry and then confirmed using a combination of precise radial velocities, ground-based photometry, spectra, and speckle imaging.  TOI-5205~b has one of the highest mass ratios for M dwarf planets with a mass ratio of almost 0.3$\%$, as it orbits a host star that is just $0.392 \pm 0.015$ \solmass{}. Its planetary radius is $1.03 \pm 0.03~R_J$, while the mass is $1.08 \pm 0.06~M_J$.  Additionally, the large size of the planet orbiting a small star results in a transit depth of $\sim 7\%$, making it one of the deepest transits of a confirmed exoplanet orbiting a main-sequence star. The large transit depth makes TOI-5205~b a compelling target to probe its atmospheric properties, as a means of tracing the potential formation pathways. While there have been radial-velocity-only discoveries of giant planets around mid-M dwarfs, this is the first transiting Jupiter with a mass measurement discovered around such a low-mass host star. The high mass of TOI-5205~b stretches conventional theories of planet formation and disk scaling relations that cannot easily recreate the conditions required to form such planets. 

% Also it's solar metallicity!
\end{abstract}

%% Keywords should appear after the \end{abstract} command. 
%% See the online documentation for the full list of available subject
%% keywords and the rules for their use.

% Add Stefan-Boltzmann citation

\section{Introduction} \label{sec:intro}
M dwarfs are the most common type of stars in the Galaxy \citep{henry_solar_2006, reyle_10_2021}, and host a higher number of planets on average compared to FGK stars \citep{mulders_increase_2015}. Yet due to their lower stellar (and disk) masses---and associated slower formation time scales---gas giants are expected to be infrequent around M dwarfs \citep{laughlin_core_2004, ida_toward_2005}. Recently, \cite{burn_new_2021} generated a synthetic planet population across a range of stellar masses and metallicities, to find that nominal scaling relations for disk properties and migration rates cannot reproduce the existence of gas giants for stellar masses $< 0.5$ \solmass{}.

New discoveries from the Transiting Exoplanet Survey Satellite \citep[TESS;][]{ricker_transiting_2014}, have helped find numerous gas giants around M dwarfs despite their rarity \citep[e.g.,][]{canas_warm_2020, jordan_hats-74ab_2022, canas_toi-3714_2022, kanodia_toi-3757_2022}, by observing millions of M dwarfs that are also bright enough for radial velocity (RV) mass measurements of transiting planet candidates \citep{stassun_tess_2018}. Despite the enhanced detection signatures, the sample of confirmed transiting gas giants with precise mass measurements around M dwarfs consists of only $< 10$ planets. All of these transiting gas giants around M dwarfs orbit early-M host stars, most of which are also metal-rich stars \citep{gan_toi-530b_2022, kanodia_toi-3757_2022}. These trends agree with the mass budget argument, which necessitates massive stars (and disks) with high dust content to form the 10 \earthmass~ cores \citep{pollack_formation_1996} in a timely manner before the disk dissipates. The alternative formation mechanism invokes disk instabilities for massive protoplanetary disks to form these gas giants more quickly \citep{boss_rapid_2006}. 

However, as we move from early-M dwarfs toward the mid-M dwarfs, the internal structures of these stars change \citep{limber_dwarf_1960}. Around 0.35 \solmass{}, the partially convective M dwarfs (convective core + radiative envelope + convective outer envelope) transition to fully convective stars. This transition is associated with slow oscillations in stellar properties (radius, luminosity, etc.), which can potentially impact the orbital evolution of planets around these stars \citep{vanderplas_understanding_2018, feiden_gaia_2021}. In this manuscript, we present the discovery of the first transiting Jovian exoplanet, which also has a mass measurement, orbiting a mid-M dwarf -- TOI-5205.

To characterise the host star and confirm the planetary nature of TOI-5205~b, we use a combination of TESS and ground-based photometry (RBO, TMMT and APO/ARCTIC),  high-contrast speckle imaging (WIYN/NESSI), precision RVs from the Habitable-zone Planet Finder spectrograph (HPF) and low-resolution optical spectra from the Low Resolution Spectrograph 2 (LRS2). In Section \ref{sec:observations} we detail these observations, while in Section \ref{sec:stellar} we discuss the stellar parameters. Subsequently, in Section \ref{sec:joint} we detail the data analysis, including the joint fitting of the photometry and RVs. In Section \ref{sec:discussion}
we discuss the mass budget for protoplanetary disks that would be required to form such a massive planet, and place it in context of other planets around M dwarfs. Finally, we summarise our findings in Section \ref{sec:conclusion}.

\section{Observations}\label{sec:observations}
\subsection{TESS}\label{sec:TESS}

\begin{figure}[!t] 
\centering
\includegraphics[width=0.5\textwidth]{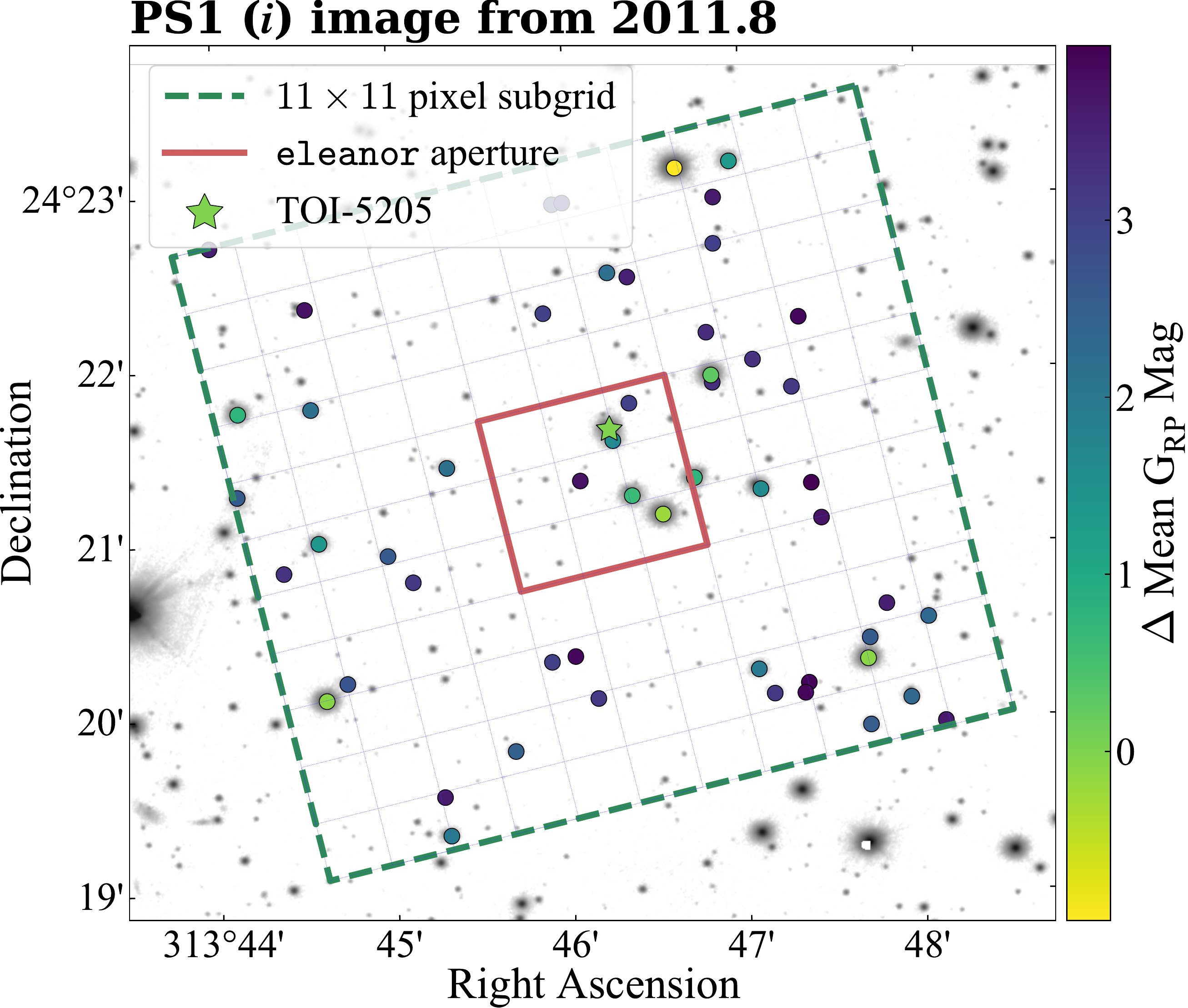}
\caption{We overlay an 11 x 11 pixel footprint from \tess{} Sector 15 (blue grid) on a Pan-STARRS1 image from $\sim 2011$ \citep{chambers_pan-starrs1_2016}. The \tess{} aperture is outlined in red and we highlight TOI-5205 with a star. Each \tess{} pixel is $\sim 21\arcsec$ in size. The TESS observations of TOI-5205 are contaminated by the presence of background stars, thereby necessitating ground-based transits to constrain the true transit depth.} \label{fig:tess_map}
\end{figure}

\begin{figure}[!t] 
\centering
\includegraphics[width=0.5\textwidth]{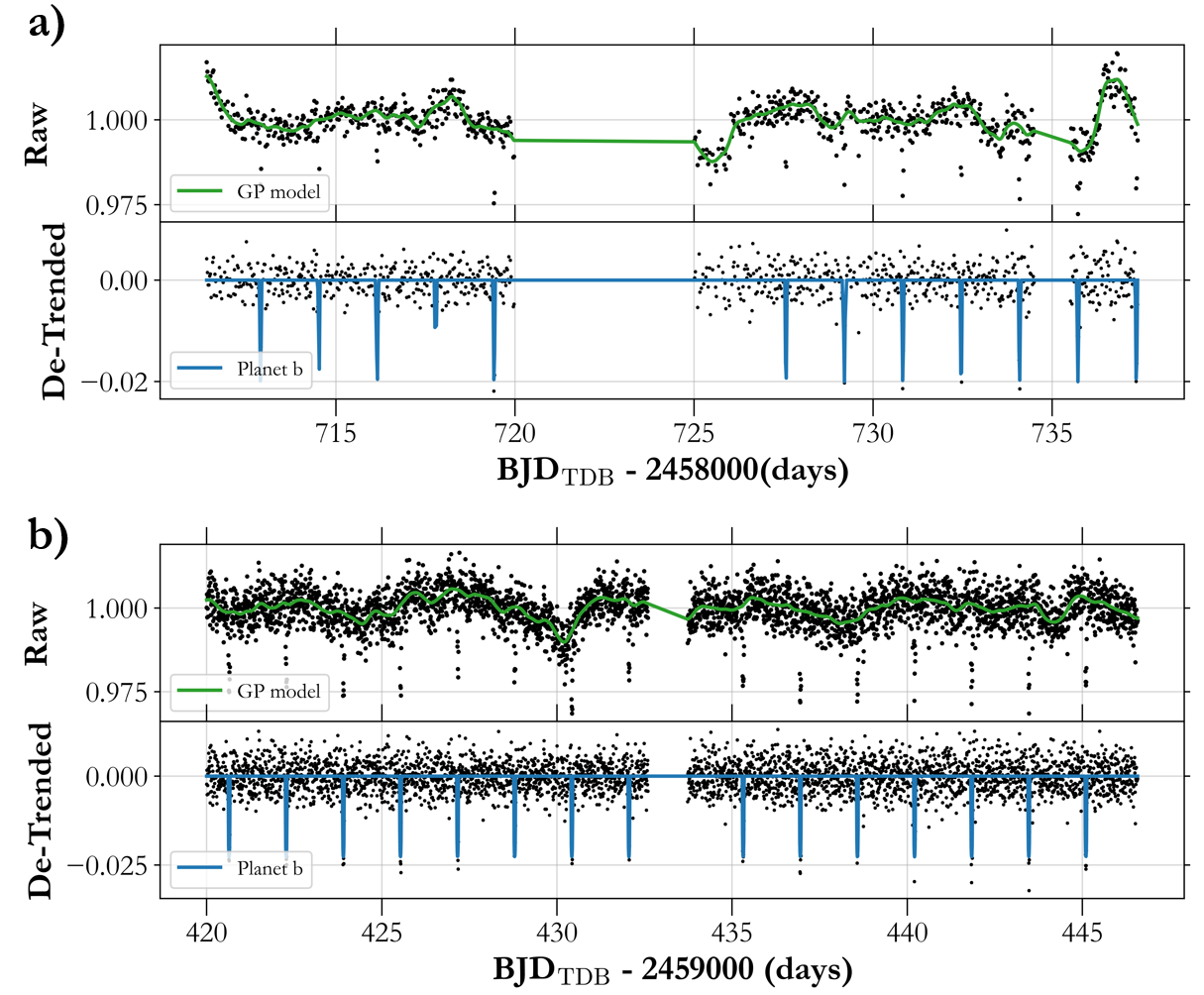}
\caption{Time series plot for \tess{} based on long cadence \texttt{eleanor} photometry from Sector 15 (Panel a with 1800 s exposure time) and Sector 41 (Panel b with 600 s exposure time), along with a stellar rotation GP kernel (\texttt{RotationTerm} from \texttt{celerite2}) in green. The detrended (GP subtracted) photometry is shown in the bottom panel, with the TOI-5205~b transits overlaid in blue.} \label{fig:tess_lc}
\end{figure}

TOI-5205 (TIC-419411415, \gaia{} DR3 1842656663520849024) is a mid-M dwarf observed by TESS in Sector 15 in Camera 1 (\autoref{fig:tess_map}) from 2019 August 15 to 2019 September 11 at $\sim 30$ minute cadence (\autoref{fig:tess_lc}a), and Sector 41 in Camera 1 from 2021 July 23 to 2021 August 20 at $\sim 10$ minute cadence (\autoref{fig:tess_lc}b). The planet candidate was identified using the Quick Look Pipeline (QLP) algorithm developed by \cite{huang_photometry_2020}, under the `faint-star search' \citep{kunimoto_tess_2022} with a period of $\sim 1.63$ d.

We extract the light curve from the TESS full-frame images (FFIs) using using \texttt{eleanor} \citep{feinstein_eleanor_2019}, which uses the TESScut\footnote{\url{https://mast.stsci.edu/tesscut/}} service to obtain a cut-out of \(31\times31\) pixels from the calibrated FFIs centered on TOI-5205. The  light curve is derived from the \texttt{CORR\_FLUX} values, in which \texttt{eleanor} uses linear regression with pixel position, measured background, and time to remove signals correlated with these parameters. The default aperture is a  $2\times1$ pixel rectangle, which does not include the target star. Instead, we set the \texttt{aperturemode} to `large' in \texttt{eleanor} which uses a $3\times3$ pixel square aperture that includes the target star and obtains a  combined differential photometric precision (CDPP) of $\sim 3850$ and $\sim 4730$ ppm for the two sectors, respectively (\autoref{fig:tess_lc}). The CDPP is formally the RMS of the photometric noise on transit timescales, and was originally defined for \textit{Kepler} \citep{jenkins_overview_2010}. We also try a custom aperture in \texttt{eleanor} of size 2x1 pixels, which includes only the two top-right pixels from the large aperture shown below. This gives us comparable posteriors to the photometry fit, while having a slightly degraded CDPP. For subsequent analysis, we use the `large' aperture shown in \autoref{fig:tess_map}.

TOI-5205 is present in a crowded field with 10 stars located $< 30 \arcsec$ away, with the closest star (TIC 1951446034) located about 4.2\arcsec~ away and $\sim 1.7$ mag fainter in the TESS bandpass (\autoref{fig:tess_map}). Based on \gaia{} DR3 astrometry, TIC 1951446034  is not co-moving and is instead $\sim 30\times$ more distant than TOI-5205 \citep[$\sim$ 2300 pc;][]{vallenari_gaia_2022}. The \texttt{eleanor} aperture includes many of these field stars, which present a significant source of dilution to the TESS light curve and necessitate ground-based follow-up that can resolve these background stars. We discuss this dilution further in Section \ref{sec:joint} where we include a dilution term while fitting the TESS photometry.

\begin{figure*}[!t] 
\centering
\includegraphics[width=1.0\textwidth]{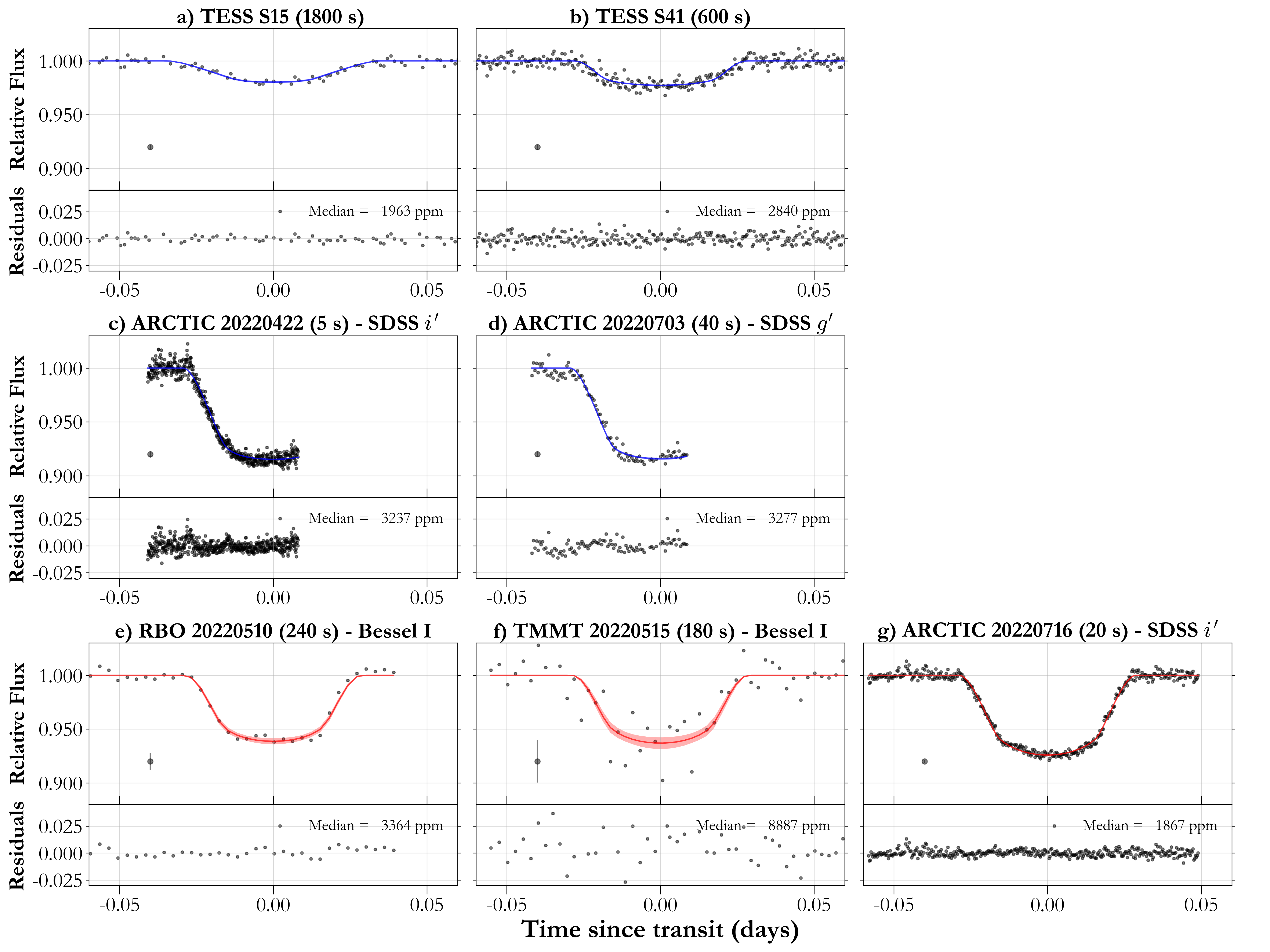}
\caption{Photometric observations for TOI-5205~b; in all the plots, the grey points show the detrended data, while the model is shown in colour, along with the 1-$\sigma$ confidence intervals as translucent bands. We also include the representative median statistical uncertainty at x = -0.04, but the errorbar is smaller than the point for certain instruments. \textbf{a-b)} The TESS light curve phase-folded to the best fit orbital period for sectors 15 and 41, respectively. \textbf{c-d)} Ground based observations from ARCTIC for TOI-5205~b that are used to estimate transit depth, shape and ephemeris (The data behind the ARCTIC transits is included along with the manuscript). \textbf{e-g)} The RBO transit from 2022 May 10, TMMT transit from 2022 May 15, and the ARCTIC transit for 2022 July 16 are included to improve the ephemeris estimate, but not to estimate the transit depth (model shown in red) because of dilution from the background companion (Section \ref{sec:photometry})}. \label{fig:transits}
\end{figure*}

\subsection{Ground-based transit Follow-up}\label{sec:photometry}

\begin{deluxetable}{ccccc}
{\tabletypesize{\scriptsize}
\tablecaption{Summary of ground based photometric follow up \label{tab:photometric}}
\centering
\tablehead{\colhead{Obs Date} & \colhead{Filter} & \colhead{Exposure} & \colhead{PSF}  & \colhead{Field of View} \\
\colhead{(YYYY-MM-DD)} & \colhead{} & \colhead{Time (s)} & \colhead{FWHM (")}  & \colhead{(')} } 
\startdata
\multicolumn{5}{c}{\hspace{-0.2cm}RBO (0.6 m)}  \\
2022-05-10 & Bessell I & 240 & 2.6 -- 6.0 & 8.94 $\times$ 8.94 \\
\hline
\multicolumn{5}{c}{\hspace{-0.2cm}TMMT (0.3 m)}  \\
2022-05-15 & Bessell I & 180 & 3.8 -- 4.5 & 40.75 $\times$ 40.75 \\
\hline
\multicolumn{5}{c}{\hspace{-0.2cm}APO (3.5 m)}  \\
2022-04-22 & SDSS \textit{i'} & 5 & 2.0 -- 3.2 &  7.9 $\times$ 7.9 \\
2022-07-03 & SDSS \textit{g'} & 40 & 1.4 -- 2.2 &  7.9 $\times$ 7.9 \\
2022-07-16 & SDSS \textit{i'} & 20 & 3.4 -- 8.3 & 7.9 $\times$ 7.9 \\
\enddata
}
\end{deluxetable}

\subsubsection{3.5 m ARC telescope}\label{sec:arctic}
We observed three transits of TOI-5205~b using the  Astrophysical Research Consortium (ARC) Telescope Imaging Camera \citep[ARCTIC;][]{huehnerhoff_astrophysical_2016} at the ARC 3.5 m Telescope at Apache Point Observatory (APO) on the nights of 2022 April 22, 2022 July 3, and 2022 July 16. All these observations were conducted using quad-amplifier and fast readout mode using $4 \times 4$ on-chip binning mode to achieve a gain of 2 $\mathrm{e^{-}/ADU}$, a plate scale of $0.456 \mathrm{\arcsec/pixel}$, and a readout time of 2.7~s. The relevant observation parameters are included in \autoref{tab:photometric}.

\textbf{2022 April 22}: We observed an ingress of TOI-5205~b (\autoref{fig:transits}c) in SDSS \textit{i'} while the target was rising from an airmass of 1.41 to 1.16. To spatially resolve and separate out the background star ($\sim 4.2\arcsec$ away), we moderately defocus the star instead of using the engineered diffuser available on ARCTIC \citep{stefansson_toward_2017}. These observations were conducted towards the end of the night, with the transit being interrupted by morning twilight. To prevent saturating the detector with the bright sky we used a short exposure time of 5 seconds. We processed the photometry using \texttt{AstroImageJ} \citep{collins_astroimagej_2017} and the final reduction used a photometric aperture radius of 6 pixels ($2.74\arcsec$), an inner sky radius of 15 pixels ($6.8\arcsec$) and outer sky radius of 25 pixels ($11.4\arcsec$). This small innermost annulus separates TOI-5205 and the closest background star ($\sim 4.2 \arcsec$). Furthermore, to verify the transit depth we also perform PSF photometry (instead of aperture photometry; following the routine described in Section \ref{sec:4star}), and obtain a comparable transit depth as that from the procedure followed above using aperture photometry.

\textbf{2022 July 3}: To check for chromaticity (Section \ref{sec:comoving}), we also observed TOI-5205~b on 2022 July 3 (\autoref{fig:transits}d) in SDSS \textit{g'} while it was rising from an airmass of 2.35 to 1.45. Similar to the previous observation, we do not use a diffuser, and moderately defocus the star. The data was reduced using aperture photometry in  \texttt{AstroImageJ} using the same annuli as above. We detrend this photometry with the Full Width Half Maximum (FWHM) of the target star across the night. The observation was interrupted due to increasing humidity and cloudy conditions, which forced us to stop observing shortly after transit midpoint.

\textbf{2022 July 16}: We obtained a full transit of TOI-5205~b on 2022 July 16 (\autoref{fig:transits}g) while it was rising from an airmass of 2.69 to 1.01 in SDSS \textit{i'}. During these observations the telescope secondary mirror had hardware issues that prevented us from using the focusser. This led to the stellar PSF changing by $\sim 2$x during the night, which caused significant systematics in the photometry that had to be detrended out by the airmass and FWHM during the night. Due to the lack of focusser control, our PSF FWHM is much larger than on previous nights, necessitating larger aperture radii of 12, 18, and 25 pixels or 5.5, 8.2, and $11.4\arcsec$ respectively. The large science aperture includes varying levels of contamination from the closest background star across the night. We therefore do not use this dataset to refine our transit depth, but only the ephemeris. 

% \iffalse
\subsubsection{0.6 m RBO}
We observed a transit of TOI-5205~b on 2022 May 10 (\autoref{fig:transits}e) using the $0.6 m$ telescope at the Red Buttes Observatory (RBO) in Wyoming \citep{kasper_remote_2016}. The RBO telescope is a f/8.43 Ritchey-Chrétien Cassegrain constructed by DFM Engineering, Inc. 

The target rose from an airmass of 2.1 to 1.1. The observations were performed using the Bessell I filter with 2x2 pixel on-chip binning and exposure times of 240 s. The binned plate scale for RBO is 0.73$\unit{\arcsec/pixel}$.

\subsubsection{0.3 m TMMT}
We observed a transit on 2022 May 15 (\autoref{fig:transits}f) using the using the Three-hundred MilliMeter ($300 \unit{mm}$) Telescope \citep[TMMT;][]{monson_standard_2017} at  Las Campanas Observatory in Chile. TMMT is a f/7.8 FRC300 from Takahashi on a German equatorial AP1600 GTO mount with an Apogee Alta U42-D09 CCD Camera,  FLI ATLAS focuser, and Centerline filter wheel.

The target rose from an airmass of 4.88 at the start of observations to a minimum airmass of 1.67, and then set to an airmass of 1.69 at the end of observations. The observations were performed using Bessell I filter with $1 \times 1$ on-chip binning and exposure times of 180 $\unit{s}$. In the $1 \times 1$ binning mode, TMMT has a gain of $1.35 \unit{e/ADU}$, a plate scale of $1.194 \unit{\arcsec/pixel}$, and a readout time of $6 \unit{s}$. 

Considering the dilution from the neighbouring companion, we use the RBO, TMMT, and third ARCTIC trnasits only to refine the ephemeris, and not to estimate the transit depth (shown in red in \autoref{fig:transits}).

\subsection{Estimating JHK magnitudes using FourStar}\label{sec:4star}
%Describe observation, frames, plate scale
We acquired near-infrared imaging using the FourStar Infrared Camera on the 6.5 m Magellan Baade telescope \citep{persson_fourstar_2013} during the night of 2022 July 13.  The plate scale for FourStar is 0.16\arcsec~ per pixel, while the seeing during observations was $\sim 0.9 \arcsec$, which was useful to clearly separate the nearby background sources in a short 2.911 second exposure in the J, H and Ks filters.  Each filtered observation used a 5-point dice-5 dither pattern and processed using a custom FourStar reduction package (FSRED).  We used the \texttt{daophot} suite of programs to perform point-spread function (PSF) fitting photometry \citep{stetson_daophot_1987, stetson_ccd_1988}.  The PSF photometry was compared to un-blended 2MASS stars in the field to determine the photometric zeropoints in each filter. The final JHK magnitudes are listed in Table \ref{tab:stellarparam}.

\subsection{Speckle Imaging with NESSI at WIYN}\label{sec:nessi}

\begin{figure}[] 
\centering
\includegraphics[width=0.5\textwidth]{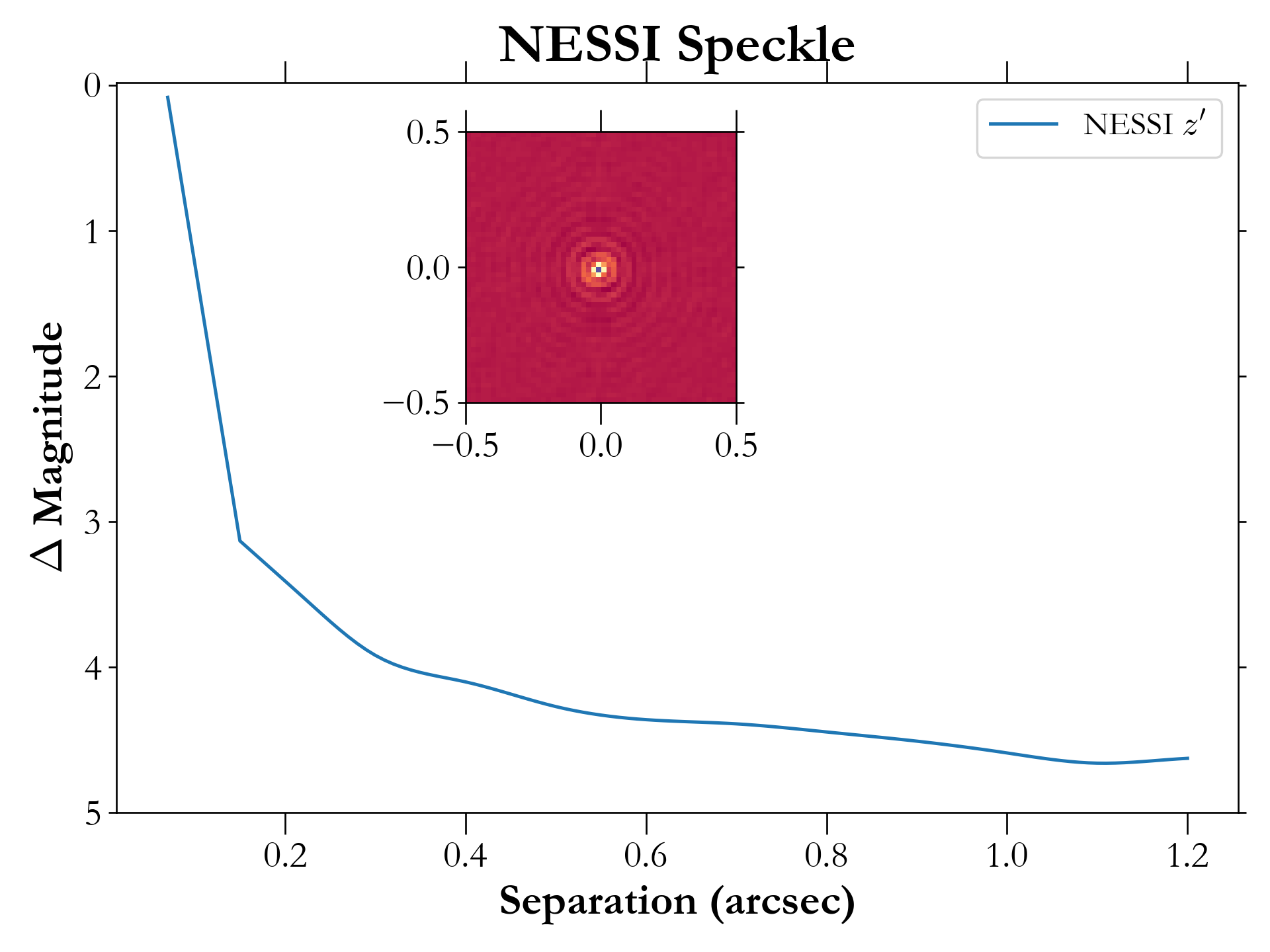}
\caption{5$\sigma$ contrast curve for TOI-5205 observed from NESSI in the Sloan \(z^\prime\)filter showing no bright companions within \(1.2''\) from the host star. The \(z^\prime\) image is shown as an inset 1$\arcsec$ across.} \label{fig:NESSI}
\end{figure}

To search for faint stellar companions or background sources that might have contributed to or diluted the detected transit signal, we acquired observations of TOI-5205 with the NN-EXPLORE Exoplanet Stellar Speckle Imager \citep[NESSI;][]{scott_nn-explore_2018} on the WIYN 3.5m telescope at Kitt Peak National Observatory on 5 May 2021. A sequence of 40 ms diffraction-limited images was taken in the Sloan \(z^\prime\) filter during the 9-minute observation, and these were then reconstructed following the procedures described by \cite{howell_speckle_2011}. We detect no nearby sources with magnitudes brighter than $\Delta z^{\prime}$ = 4.0 for separations $>0.3''$. The contrast curve and reconstructed speckle image are shown in \autoref{fig:NESSI}.

\subsection{LRS2}\label{sec:lrs2}
To confirm the spectral type and stellar parameters for TOI-5205, we also observe the target using the Low Resolution Spectrograph 2 \citep[LRS2; ][]{lee_lrs2_2010, chonis_lrs2_2016} on the Hobby-Eberly Telescope \citep[HET;][]{ramsey_early_1998} at McDonald Observatory, in West Texas. LRS2 is a low-resolution (R$\sim$1900) optical integral-field unit (IFU) spectrograph composed of two arms that simultaneously observe two 6\arcsec $\times$12\arcsec~fields of view separated by 100\arcsec. The blue arm (LRS2-B) consists of a pair of channels with spectral ranges of $\sim$3640--4670 $\rm\AA$ and $\sim$4540--7000 $\rm\AA$, while the red arm (LRS2-R) is composed of two channels covering $\sim$ 6430--8450 $\rm\AA$ and $\sim$ 8230--10560 $\rm\AA$. The LRS2-R data were obtained with a 1800 second exposure on 2022 June 11 (1.4\arcsec~ seeing), while the LRS2-B data were taken on 2022 August 3 (1.6\arcsec~ seeing) with the same exposure time.

\begin{figure*}[] 
\centering
\includegraphics[width=\textwidth]{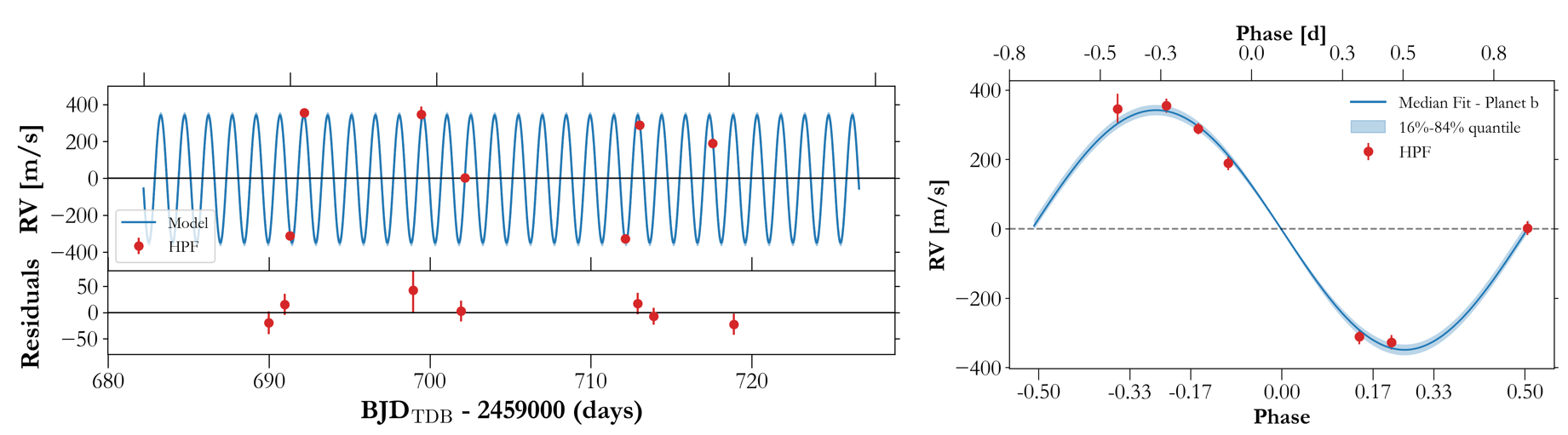}
\caption{\textbf{Left:} Time series of RV observations of TOI-5205 with HPF (red). The best-fitting model derived from the joint fit to the photometry and RVs is plotted in blue, including the 16-84$\%$ confidence interval in lighter blue. The bottom panel shows the residuals after subtracting the model. \textbf{Right:} HPF RV observations phase folded on the best fit orbital period from the joint fit from Section \ref{sec:joint}. While we let the eccentricity float in this fit, the results are consistent with a circular orbit (\autoref{tab:planetprop}). \label{fig:rv}} 
\end{figure*}

The raw data were processed with \texttt{Panacea}\footnote{\url{https://github.com/grzeimann/Panacea}}, an automated reduction pipeline for LRS2 written by G. Zeimann  (Zeimann et al., in preparation).  The initial processing includes bias-correction, wavelength calibration from arc lamps taken within 7 nights of the observation, fiber trace calculation from flat field exposures over $\pm$7 nights, fiber normalization from twilight exposures over $\pm$7 nights, fiber extraction, and an initial flux calibration from default response curves and measures of the mirror illumination as well as the the exposure throughput from guider images.  After the initial reduction, we used
\texttt{LRS2Multi}\footnote{\url{https://github.com/grzeimann/LRS2Multi}}, a \texttt{python} interface to perform advanced reduction steps and calibrations for \texttt{Panacea} products.  Using \texttt{LRS2Multi}, we identified the target star, defined a 3.5\arcsec\ aperture and used fibers beyond that aperture to build our sky model for each exposure.  We subtracted the initial sky, and then constructed a principle component basis of 25 components with the residuals to further subtract sky residuals that occur from variable spectral point spread functions for each fiber.  This is especially important for the LRS2-R channels.  We extracted the target spectrum from the sky-subtracted frames and normalised the LRS2-B to the LRS2-R spectrum using a 100 \AA~ window in the overlap between the two spectrographs.  Noting that the default response may not be accurate enough for spectrophotometry, we reduced and calibrated standard stars from June-2021 through Aug-2022 and measured the average flux calibration correction.  The response correction was smoothed by a median filter with a 250 pixel kernel and was applied to our extracted spectrum.  The correction was relatively small and smoothly declining with a $\sim$10\%\ positive correction in the blue and a $\sim$10\%\ negative correction in the red.  Finally, the telluric correction was chosen from three empirical models constructed from a dozen HR telluric standard stars. We note that the  relative chromatic flux calibration should be good to $\sim$5\% for $\sim$3700--10200 $\rm\AA$ based on the standard star analysis above, with the exception of regions with strong telluric absorption and where individual channels overlap.  The final LRS2 spectra was used to estimate the spectral type of the star (\autoref{fig:pyHammer}, Section \ref{sec:stellarchar}).

\subsection{Radial velocity follow-up with HPF}\label{sec:hpfrvs}

We started RV observations of TOI-5205 with HPF \citep{mahadevan_habitable-zone_2012, mahadevan_habitable-zone_2014} on 2022 April 20. HPF is a high resolution near-infrared (\(8080-12780\)\ \AA), fiber-fed \citep{kanodia_overview_2018} precision RV spectrograph with a stabilized environment \citep{stefansson_versatile_2016}. HPF is located at HET, which is a fixed-altitude telescope with a roving pupil design, and is fully queue-scheduled, where all the observations are executed by the HET resident astronomers \citep{shetrone_ten_2007}. We correct for bias, non-linearity, cosmic rays, and calculate the slope/flux and variance images  from the raw HPF data, using the algorithms described in the package \texttt{HxRGproc} \citep{ninan_habitable-zone_2018}. We do not utilise simultaneous calibration using the near-infrared (NIR) Laser Frequency Comb for HPF \citep{metcalf_stellar_2019} due to concerns about the impact of scattered calibration light given the faintness of our target. Instead, we obtain a wavelength solution for the target exposures by interpolating the wavelength solution from other LFC exposures on the night of the observations.  This has been shown to enable precise wavelength calibration and drift correction with a precision of $\sim30$ \cms{} per observation \citep{stefansson_sub-neptune-sized_2020}, a value much smaller than our expected per observation RV uncertainty (instrumental + photon noise) for this object of 22 \ms{} (in 969 s exposures, and 15 \ms{} in binned 30 minute exposures). 

\begin{deluxetable}{ccc}
\tablecaption{RVs (binned in $\sim$ 30 minute exposures) of TOI-5205. \label{tab:rvs}}
\tablehead{\colhead{$\unit{BJD_{TDB}}$}  &  \colhead{RV}   & \colhead{$\sigma$}  \\
           \colhead{(d)}   &  \colhead{\ms{}} & \colhead{\ms{}}}
\startdata
2459689.97105 & -339.74 & 21.68 \\ 
2459690.96744 & 326.19 & 20.27 \\ 
2459698.96070 & 317.19 & 43.67 \\ 
2459701.93977 & -26.88 & 20.05 \\ 
2459712.90820 & -355.59 & 20.44 \\ 
2459713.90256 & 259.72 & 16.08 \\ 
2459718.89386 & 160.39 & 19.95 \\ 
\enddata
\end{deluxetable}

To derive the RVs from the extracted spectra, we use the template-matching method \citep[e.g.,][]{anglada-escude_harps-terra_2012}. This has been implemented under the \texttt{SpEctrum Radial Velocity AnaLyser} pipeline \citep[\texttt{SERVAL};][]{zechmeister_spectrum_2018}, which has since been modified for HPF \citep{stefansson_sub-neptune-sized_2020}. Under this method, we first create a master template from the target star observations, and then determine the Doppler shift for each individual observation by moving it in velocity space, comparing it with the template, and minimizing the \(\chi^2\) statistic. The master template is created using all of the HPF observations for TOI-5205, after masking out the telluric and sky-emission lines. The telluric regions are identified by a synthetic telluric-line mask generated from \texttt{telfit} \citep{gullikson_correcting_2014}, a Python wrapper to the Line-by-Line Radiative Transfer Model package \citep{clough_atmospheric_2005}. We use \texttt{barycorrpy} \citep{kanodia_python_2018} to perform the barycentric correction on the individual spectra, which is the Python implementation  of the algorithms from \cite{wright_barycentric_2014}. 

We obtained a total of 7 visits on this target between 2022 April 20 and 2022 May 19 (\autoref{fig:rv}). Each visit was divided into 2 exposures of 969 s each, where the median S/N of each HPF exposure was 40 per pixel at 1070 nm. The individual exposures were then combined by weighted averaging, with the final binned RVs being listed in  \autoref{tab:rvs}.

\section{Stellar Parameters}\label{sec:stellar}

The stellar properties for TOI-5205 are crucial for understanding the system. Because it sits near this transition zone between fully and partially convective M dwarfs, the typical M dwarf scaling relations have additional scatter and often diverge. We have undertaken a thorough, multi-faceted approach to constraining the stellar properties and testing their robustness, the details of which are included in the Appendix. We summarise the main results here. From \gaia{} magnitudes and LRS2 spectra, we estimate a spectral subtype of M4 $\pm$ 1 for TOI-5205. From photometric relations we obtain an effective temperature of 3430 $\pm$ 54 K, solar metallicity, and a stellar radius of 0.394 $\pm$ 0.011 \solradius{}. We then use a mass-radius relationship for M dwarfs to obtain a mass of 0.392 $\pm$ 0.015 \solmass{}. We use H-$\alpha$ equivalent width measurements from LRS2 spectra and the lack of a detectable rotation period in the photometry to conclude that TOI-5205 is not an active star. Additionally, we rule out a number of false positive scenarios (such as background and hierarchical eclipsing systems) using a combination of archival images, HPF spectra, NESSI high contrast imaging, chromatic estimates of the transit depth. The procedure followed to perform this analysis and characterise the host star is explained in Appendix \ref{sec:stellarchar}.

\begin{deluxetable*}{lccc}
{\tabletypesize{\small}
\tablecaption{Summary of stellar parameters for TOI-5205 \label{tab:stellarparam}}
\tablehead{\colhead{~~~Parameter}&  \colhead{Description}&
\colhead{Value}&
\colhead{Reference}}
\startdata
\multicolumn{4}{l}{\hspace{-0.2cm} Main identifiers:}  \\
~~~TOI & \tess{} Object of Interest & 5205 & \tess{} mission \\
~~~TIC & \tess{} Input Catalogue  & 419411415 & Stassun \\
% ~~~2MASS & \(\cdots\) & 2MASS J20550491+2421387 & 2MASS  \\
~~~Gaia DR3 & \(\cdots\) & 1842656663520849024 & Gaia DR3\\
\multicolumn{4}{l}{\hspace{-0.2cm} Equatorial Coordinates and Proper Motion:} \\
~~~$\alpha_{\mathrm{J2016}}$ &  Right Ascension (RA) &  20:55:04.96 & Gaia DR3\\
~~~$\delta_{\mathrm{J2016}}$ &  Declination (Dec) & +24:21:39.54 & Gaia DR3\\
~~~$\mu_{\alpha}$ &  Proper motion (RA, \unit{mas/yr}) &  $41.68 \pm 0.02$ & Gaia DR3\\
~~~$\mu_{\delta}$ &  Proper motion (Dec, \unit{mas/yr}) & $52.07 \pm 0.02$ & Gaia DR3 \\
~~~$\varpi$ &  Parallax (mas)  & $11.464 \pm 0.026$ & Gaia DR3 \\
~~~$d$ &  Distance in pc  & $86.865 \pm 0.05$ & Anders \\
\multicolumn{4}{l}{\hspace{-0.2cm} Broadband photometry:}  \\
~~~$G$ & $G$ mag & $14.903\pm 0.003$ & Gaia DR3 \\
~~~$g$ & PS1 g mag & $16.877 \pm 0.008$ & PS1\\
~~~$r$ & PS1 r mag & $15.694 \pm 0.008$ & PS1\\
~~~$i$ &  PS1 i mag  & $14.21 \pm 0.01$ & PS1\\
~~~$z$ &  PS1 z mag  & $13.55 \pm 0.02$ & PS1 \\
~~~$y$ &  PS1 y mag  & $13.207 \pm 0.005$ & PS1 \\
~~~$J$ & $J$ mag & $ 11.90\pm 0.02$ & This work\\
~~~$H$ & $H$ mag & $ 11.28\pm 0.02$ & This work\\
~~~$K_s$ & $K_s$ mag & $ 11.04\pm0.02$ & This work\\
\multicolumn{4}{l}{\hspace{-0.2cm} Derived  photometry:} \\
~~~$A_G$ & Extinction in mag & $0.12\pm0.02$ & Anders\\
~~~$M_G$ & Absolute $G$ mag & $10.09\pm0.02$ & Anders\\
\multicolumn{4}{l}{\hspace{-0.2cm} Stellar Parameters:}\\
~~~$T_{\mathrm{eff}}^a$ &  Effective temperature in \unit{K} & $3430 \pm 54$ & This work\\
~~~$\mathrm{[Fe/H]}$ &  Metallicity  & Solar & This work\\
~~~$\log(g)^a$ & Surface gravity in cgs units & $4.84 \pm 0.03$ & This work\\
~~~Sp Type$^b$ & Spectral Type & M4.0 $\pm$ 1.0 & This work\\
~~~$R_*$$^c$ &  Radius in $R_{\odot}$ & $0.394\pm0.011$ & This work\\
~~~$M_*$$^d$ &  Mass in $M_{\odot}$ & $0.392\pm0.015$ & This work\\
~~~$L_*$ &  Luminosity in $L_{\odot}$ & $0.0194\pm0.0016$ & This work\\
~~~$\rho_*$ &  Density in $\unit{g/cm^{3}}$ & $9.0\pm0.5$ & This work\\
\multicolumn{4}{l}{\hspace{-0.2cm} Other Stellar Parameters:}           \\
~~~$v \sin i_*$ &  Rotational velocity in \unit{km/s}  & $< 2$ & This work\\
~~~$\Delta$RV &  Absolute radial velocity in \unit{km/s} & $-65.9\pm0.3$ & This work\\
~~~$U, V, W$ &  Galactic velocities in \unit{km/s} &  $-48.29\pm0.12, -50.50\pm0.27, 14.90\pm0.07$ & This work\\
~~~$U, V, W^e$ &  Galactic velocities (LSR) in \kms{} & $-37.19\pm0.86, -38.26\pm0.74, 22.15\pm0.61$ & This work\\
\enddata
\tablenotetext{}{References are: Stassun \citep{stassun_tess_2018}, Gaia DR3 \citep{vallenari_gaia_2022}, PS1 \citep{chambers_pan-starrs1_2016}, Anders \citep{anders_photo-astrometric_2022}}
\tablenotetext{a}{Using the \teff{} - $M_G$ relation from \cite{rabus_discontinuity_2019}.}
\tablenotetext{b}{Spectral typing using relations based on Gaia colour \citep{kiman_exploring_2019}}
\tablenotetext{c}{Using $R_*$ - $M_K$ relation from \cite{mann_how_2015, mann_erratum_2016}.}
\tablenotetext{d}{Using $M_*$ - $R_*$ relation from \cite{schweitzer_carmenes_2019}.}

\tablenotetext{e}{The barycentric UVW velocities are converted into local standard of rest (LSR) velocities using the constants from \cite{schonrich_local_2010}.}
}
\end{deluxetable*}

\subsection{Transition between partially and fully convective stars}\label{sec:transition}

M dwarfs with masses $\sim$ 0.35 \solmass{} have internal structures that transition from being partially convective (for the more massive stars) to fully convective  \citep[for the less massive ones; ][]{limber_structure_1958, baraffe_closer_2018}. On the more massive end, the partially convective stars have convective cores and envelopes separated by a radiative zone. As these stars fuse $^3$He in the convective core, the $^3$He abundance rises with temperature when in nonequilibrium \citep[Figure 2;][]{baraffe_closer_2018}, and causes the convective core to increase in radius, and ultimately merge with the outer convective envelope that has a lower $^3$He abundance \citep{macdonald_explanation_2018, feiden_gaia_2021}. This merger is accompanied by a sudden drop in the $^3$He abundance in the core, which reduces the reaction rate, causing the core to contract and separate from the envelope \citep[Figure 5 from][]{feiden_gaia_2021}. When the core contracts, the temperature begins to rise again, producing an increase in the abundance of $^3$He, and an episodic cycling over Gyr timescales. Due to these repeated mergers and contractions, the abundance of the convective envelope increases until the core-envelope merger is not accompanied by a sudden decrease in abundance (and associated nuclear reaction rate). At this point, the star attains a fully convective steady state. The timescale to attain this fully convective state for stars in this transition zone depends on the mass and metallicity of the star \citep{kroupa_theoretical_1997, feiden_gaia_2021}. Unsurprisingly, these oscillations are accompanied by slow and small variations in the radius and luminosity of the star \citep{van_saders_3he-driven_2012, macdonald_explanation_2018}. This transition zone is also accompanied by an inflection in the mass-luminosity relation\footnote{Empirically this was first noticed as an increase in the stellar luminosity function for the local neighbourhood  $M_V \sim 11.5$, which was then attributed to the combination of a smooth initial mass function, and an inflection in the mass-luminosity relationship due to this transition.} for M dwarfs as was noted by \cite{kroupa_low-luminosity_1990} and \cite{ delfosse_accurate_2000}. As an aside, this feature in the mass-luminosity relation causes a local maxima in the slope, which can reproduce the additional scatter in the \teff{} - $R_*$ relation for mid-M dwarfs in \cite{mann_how_2015}.

Based on \gaia{} DR2, \cite{jao_gap_2018} presented the discovery of the now eponymous gap near $M_G \sim 10.2$ in the \gaia{} colour-magnitude diagram (CMD; $M_G$ vs. $G_{\rm{BP}} - G_{\rm{RP}}$). This is a narrow diagonal region with an under-density of stars, the width of which is a function of $G_{\rm{BP}} - G_{\rm{RP}}$ colour \citep{jao_fine_2020}.  While the gap is associated with a 10--20\% decrement in the number of stars, it is hardly seen redwards of $G_{\rm{BP}} - G_{\rm{RP}} \sim 2.7$. Theoretical models have been used to approximately reproduce the  properties of the gap in the CMD relying on the $^3$He instability, and attribute this under-density to the transition between partial and fully-convective M dwarfs \citep{feiden_gaia_2021}. 

While TOI-5205 does not lie in this gap based on \gaia{} photometry, it is one of the few known planet hosting stars in its vicinity, i.e. near this transition zone between fully and partially convective M dwarfs \citep{silverstein_lhs_2022}. TOI-5205 has a $G_{\rm{BP}} - G_{\rm{RP}}$ of $\sim 2.8$, and $M_G$ of 10.09$^{+0.01}_{-0.03}$ from \gaia{} DR3, which would place it redwards of this diagonal gap ($M_G$ vs. $G_{\rm{BP}} - G_{\rm{RP}}$ space). The background companion to TOI-5205 at $\sim 4\arcsec$ could contaminate the prism spectra used to obtain the colour estimates. \cite{creevey_gaia_2022} mention that a CCD window of 3.5\arcsec $\times$ 2.1\arcsec~ is used while extracting the spectra, the orientation for which is quasi-random on the sky over different epochs. However the background companion is much hotter \citep[\teff{} $\sim 5450$ K;][]{stassun_revised_2019} than TOI-5205 (\teff{} $\sim 3400$ K; \autoref{tab:stellarparam}), and therefore bluer.

\section{Joint Fitting of Photometry and RVs}\label{sec:joint}
We perform a joint fit of the photometry and RVs using the \texttt{python} package \texttt{exoplanet} \citep{foreman-mackey_exoplanet-devexoplanet_2021} which relies on \texttt{PyMC3}, the Hamiltonian Monte Carlo (HMC) package \citep{salvatier_probabilistic_2016}. The HMC method has shown to be computationally efficient in spanning multi-dimensional parameter spaces to estimate parameter posteriors. The \texttt{exoplanet} package uses \texttt{starry} \citep{luger_starry_2019, agol_analytic_2020} to model the transits, and relies on the analytical models from \cite{mandel_analytic_2002}, and separate quadratic limb-darkening terms for each instrument. The limb-darkening priors use the reparameterization suggested by \cite{kipping_efficient_2013} for uninformative sampling. We perform a joint fit with all the photometry and RVs, where we fit each phased transit (\autoref{fig:transits}) with separate limb-darkening coefficients. We also include a simple-white noise model in the form a jitter term for each photometry dataset.  Our likelihood function for the \tess{} photometry includes a Gaussian Process (GP) kernel to model the quasi-periodic signal (\autoref{fig:tess_lc}). This signal is discussed further in Appendix \ref{sec:rotation}.

% \startlongtable
\begin{deluxetable*}{llc}
\tablecaption{Derived Parameters for the TOI-5205 System.   \label{tab:planetprop}}
\tablehead{\colhead{~~~Parameter} &
\colhead{Units} &
\colhead{Value$^a$} 
}
\startdata
\sidehead{Orbital Parameters:}
~~~Orbital Period\dotfill & $P$ (days) \dotfill & 1.630757$\pm0.000001$\\
~~~Eccentricity\dotfill & $e$ \dotfill & 0.020$^{+0.020}_{-0.014}$ \\
~~~Argument of Periastron\dotfill & $\omega$ (radians) \dotfill & -0.74$^{+3.25}_{-1.74}$ \\
~~~Semi-amplitude Velocity\dotfill & $K$ (\ms{})\dotfill &
346$\pm14$\\
~~~Systemic Velocity$^b$\dotfill & $\gamma_{\mathrm{HPF}}$ (\ms{})\dotfill & -28$\pm11$\\
~~~RV trend\dotfill & $dv/dt$ (\ms{} yr$^{-1}$)   & 0.05$^{+4.92}_{-5.08}$   \\ 
~~~RV jitter\dotfill & $\sigma_{\mathrm{HPF}}$ (\ms{})\dotfill & 14.7$^{+16.6}_{-10.1}$\\
\sidehead{Transit Parameters:}
~~~Transit Midpoint \dotfill & $T_C$ (BJD\textsubscript{TDB})\dotfill & 2459443.47179$\pm0.00019$\\
~~~Scaled Radius\dotfill & $R_{p}/R_{*}$ \dotfill & 
0.2720$^{+0.0039}_{-0.0043}$\\
~~~Scaled Semi-major Axis\dotfill & $a/R_{*}$ \dotfill & 10.94$^{+0.22}_{-0.21}$\\
~~~Orbital Inclination\dotfill & $i$ (degrees)\dotfill & 88.21$^{+0.24}_{-0.22}$\\
~~~Transit Duration\dotfill & $T_{14}$ (days)\dotfill & $0.0583 \pm0.0011$\\
~~~Photometric Jitter$^c$ \dotfill & $\sigma_{\mathrm{TESS~S15}}$ (ppm)\dotfill & $2985_{-85}^{+89}$\\ % jitter_TESS_S15,0.00298575$^{+0.00008843}_{-0.00008445}$
~~~ & $\sigma_{\mathrm{TESS~S41}}$ (ppm)\dotfill & $4241\pm50$\\ % jitter_TESS_S41,0.00424078$^{+0.00005205}_{-0.00005003}$
~~~ & $\sigma_{\mathrm{ARCTIC~20220422}}$ (ppm)\dotfill & $5291\pm160$\\ % jitter_ARCTIC_20220422,0.00529065$^{+0.00016632}_{-0.00015541}$
~~~ & $\sigma_{\mathrm{RBO~20220510}}$ (ppm)\dotfill & $865^{+947}_{-562}$\\ % jitter_RBO_20220510,0.00086535$^{+0.00094721}_{-0.00056266}$
~~~ & $\sigma_{\mathrm{TMMT~20220515}}$ (ppm)\dotfill & $15759^{+1364}_{-1201}$\\ % jitter_TMMT_20220515,0.01575923$^{+0.00136429}_{-0.00120089}$
~~~ & $\sigma_{\mathrm{ARCTIC~20220703}}$ (ppm)\dotfill & $3948^{+472}_{-443}$\\ % jitter_ARCTIC_20220704_SDSSg,0.00394826$^{+0.00047225}_{-0.00044305}$
~~~ & $\sigma_{\mathrm{ARCTIC~20220717}}$ (ppm)\dotfill & $2716\pm120$\\ % jitter_ARCTIC_20220717_SDSSi,0.00271597$^{+0.00012504}_{-0.00011853}$
% ~~~Limb Darkening$^d$ $\dotfill$ & $u_{1}$, $u_{2}$ $\dotfill$ & $0.43\pm0.10$, $0.37^{+0.20}_{-0.28}$  \\
~~~Dilution$^{de}$ \dotfill & $D_{\mathrm{TESS~S15}}$ \dotfill & $0.234\pm0.012$\\
~~~ & $D_{\mathrm{TESS~S41}}$ \dotfill & $0.259\pm0.008$\\ 
\sidehead{Planetary Parameters:}
~~~Mass\dotfill & $M_{p}$ (M$_\oplus$)\dotfill &  $343^{+18}_{-17}$\\
~~~ & $M_{p}$ ($M_J$)\dotfill &  $1.08\pm0.06$\\
~~~Radius\dotfill & $R_{p}$  (R$_\oplus$) \dotfill& $11.6\pm0.3$\\
~~~ & $R_{p}$  ($R_J$) \dotfill& 1.03$\pm0.03$\\
~~~Density\dotfill & $\rho_{p}$ (\gcmcubed{})\dotfill & 1.21$\pm0.11$\\
~~~Semi-major Axis\dotfill & $a$ (AU) \dotfill & $0.0199\pm0.0002$\\
~~~Average Incident Flux$^f$\dotfill & $\langle F \rangle$ (\unit{10^5\ W/m^2})\dotfill &  0.67$\pm$0.06\\
~~~Planetary Insolation & $S$ (S$_\oplus$)\dotfill &  $49\pm4$\\
~~~Equilibrium Temperature$^g$ \dotfill & $T_{\mathrm{eq}}$ (K)\dotfill & 737$\pm15$\\
\enddata
\tablenotetext{a}{The reported values refer to the 16-50-84\% percentile of the posteriors.}
\tablenotetext{b}{In addition to the "Absolute RV" from \autoref{tab:stellarparam}.}
\tablenotetext{c}{Jitter (per observation) added in quadrature to photometric instrument error.}
% \tablenotetext{d}{Where $u_1 + u_2 < 1$, and $u_1 > 0$ according to \cite{kipping_efficient_2013}.}
\tablenotetext{d}{Dilution due to presence of background stars in \tess{} aperture, not accounted for in the \texttt{eleanor} flux.}
\tablenotetext{e}{We treat the dilution terms for RBO 20220510, TMMT 20220515, and ARCTIC 20220717 as nuisance parameters, since those datasets are used only to refine the ephemeris.}
\tablenotetext{f}{We use a Solar flux constant = 1360.8 W/m$^2$, to convert insolation to incident flux.}
\tablenotetext{g}{We assume the planet to be a black body with zero albedo and perfect energy redistribution to estimate the equilibrium temperature. }
\normalsize
\end{deluxetable*}

We include a dilution term (\rm{Dil}) in the photometric model to account for the presence of blended (or spatially unresolved) background stars in the TESS photometry. We assume that the higher spatial resolution ground-based photometry from the first two ARCTIC transits  has no contamination from the background stars (i.e. Dil = 1), and therefore can be used to correct the TESS photometry. This dilution term is fit separately for individual TESS sectors, due to the different placement of the target and background stars on the camera pixels. We fit the dilution using a uniform prior from 0.1 to 1.5 to correct for potential over-compensation of the dilution term. While this is not a problem for the \texttt{eleanor} reduction, occasionally the SPOC data can over-correct for dilution as shown for TOI-824 \citep{burt_toi-824_2020}, especially in crowded fields. The dilution term (\rm{Dil}) is used to inflate the planetary radius (R$_p$) estimate as shown below:

\begin{equation}
    R_{p, \rm{true}} = \frac{ R_{p, \rm{TESS}}}{\sqrt{\rm{Dil}}}
\end{equation}

The first two ARCTIC transits (ingress in $i'$, $g'$) are used to estimate the true transit depth. The ARCTIC dataset from 2022 July 16 suffers from instrument systematics due to wildly varying PSF FWHM from a malfunctioning focusser. This manifests as varying levels of contamination from the nearby star. We use this ARCTIC dataset along with the RBO and TMMT photometry to improve our ephemeris estimate.

Separate from the joint fit, we also use the ARCTIC dataset from 2022 July 16 to estimate the eccentricity using the photo-eccentric effect \citep{dawson_photoeccentric_2012}, which relies on the transit duration and estimates and eccentricity of 0.11$^{+0.32}_{-0.08}$. This is consistent with the eccentricity obtained from the RV orbit (albeit a weaker limit), and suggests a circular orbit, which is unsurprising for a giant planet at such a short orbital period, which would have a circularization time scale of $\sim$ Myr. The precise photometry and duration estimate is then used to calculate a host star density assuming a circular orbit, to confirm the stellar parameters in Section \ref{sec:photrelations}.

We model the RVs using a standard Keplerian model, allowing the eccentricity to float. We also include an RV offset and jitter term of HPF, along with a linear RV trend to account for long term drifts (both instrumental and astrophysical). We use \texttt{scipy.optimize}, to find the initial \textit{maximum a posteriori} (MAP) parameter estimates, which uses the default BFGS algorithm \citep[Broyden–Fletcher–Goldfarb–Shanno algorithm;][]{broyden_convergence_1970, fletcher_new_1970, goldfarb_family_1970, shanno_conditioning_1970}.  These parameter estimates are then used as the initial conditions for parameter estimation using  ``No U-Turn Sampling" \citep[NUTS,][]{hoffman_no-u-turn_2014}, implemented for the HMC sampler \texttt{PyMC3}, where we check for convergence using the Gelman-Rubin statistic \citep[$\hat{\text{R}} \le 1.1$;][]{ford_improving_2006}.

The final derived planet parameters from the joint fit are included in \autoref{tab:planetprop}, with the phased RVs shown in \autoref{fig:rv}.

\section{Discussion}\label{sec:discussion}

\begin{figure*}[!t]
\fig{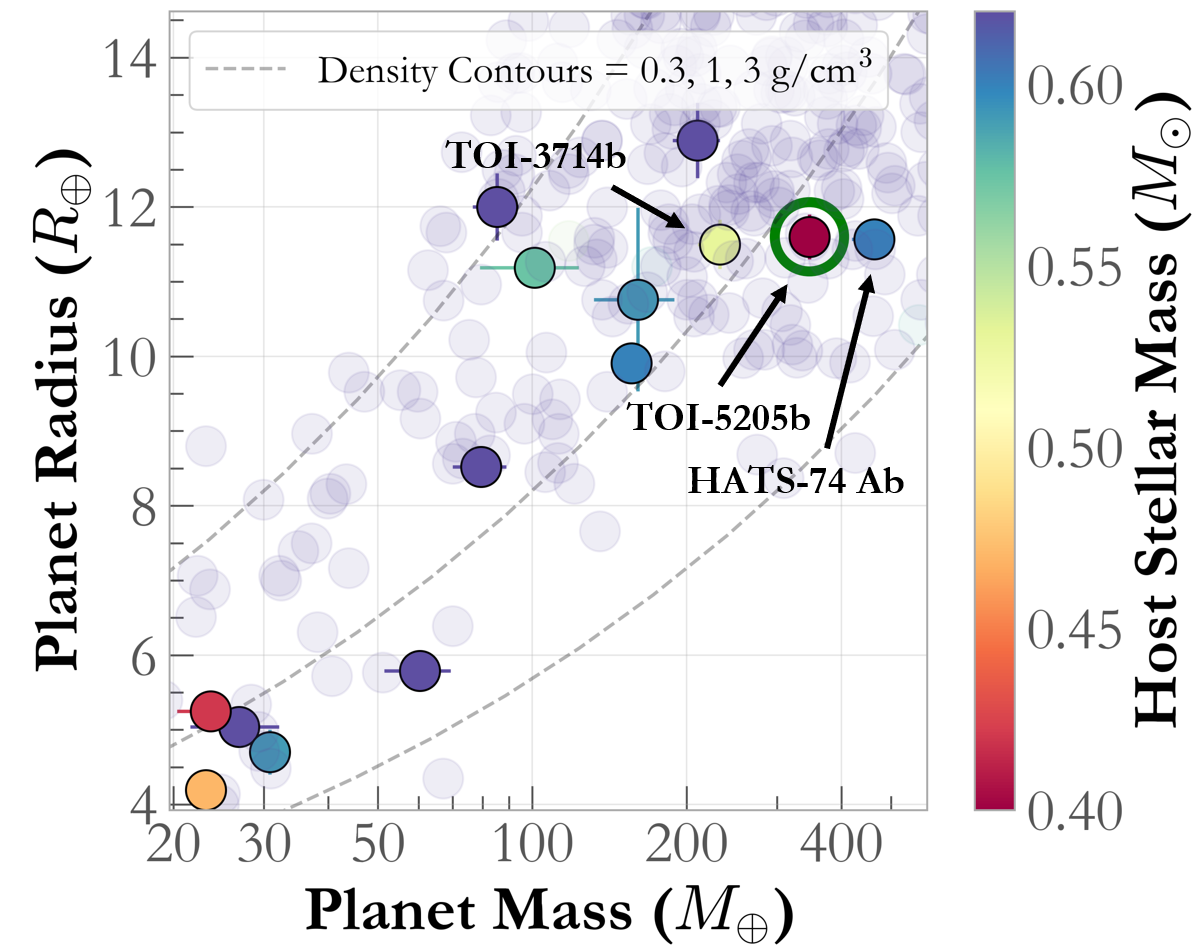}{0.45\textwidth}
{\small a) Planet radius as a function of mass}    \label{fig:RadiusMass}
\fig{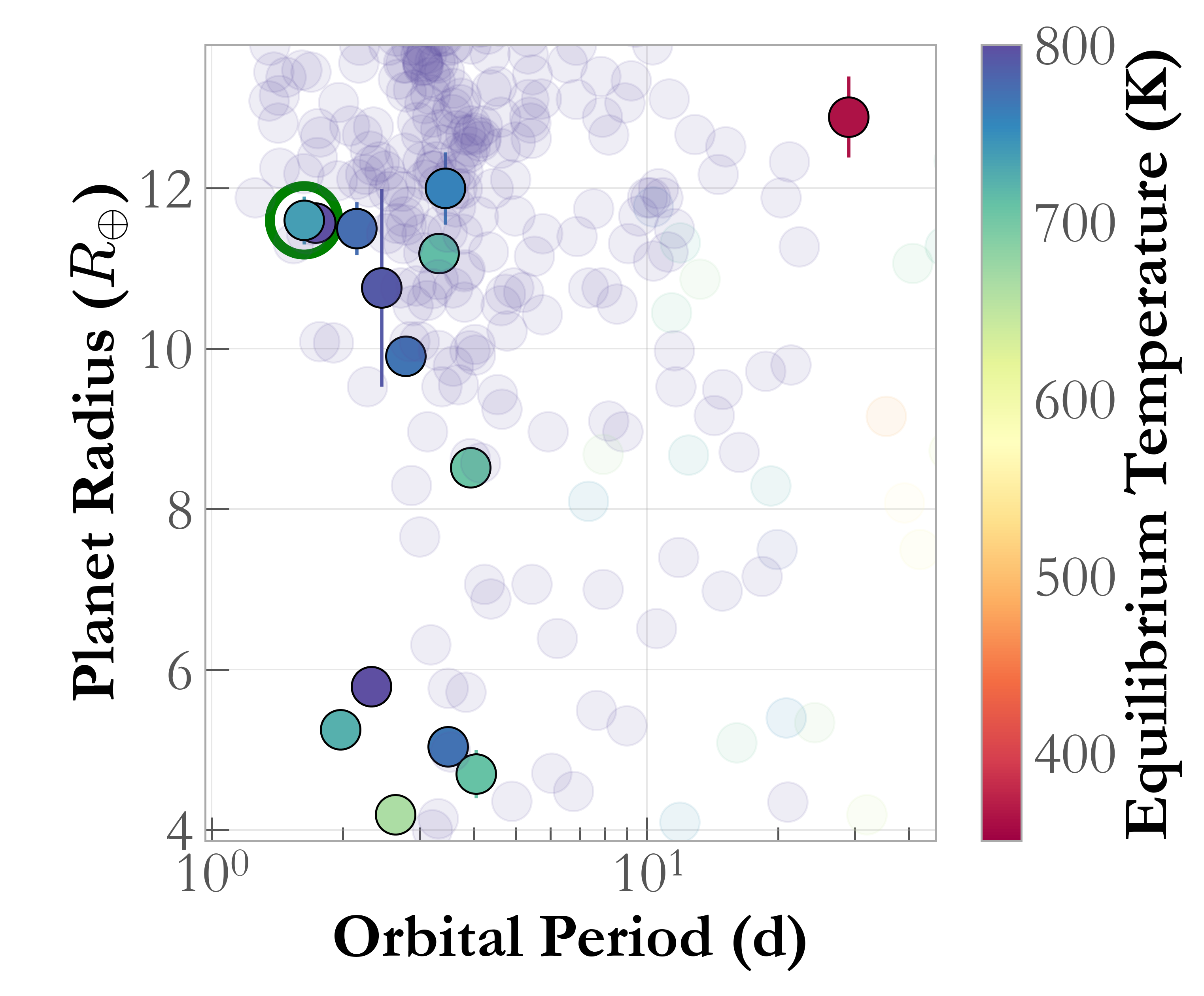}{0.45\textwidth}
{\small b) Planet radius as a function of orbital period} \label{fig:RadiusPeriod} \vspace{0.2 cm}
\fig{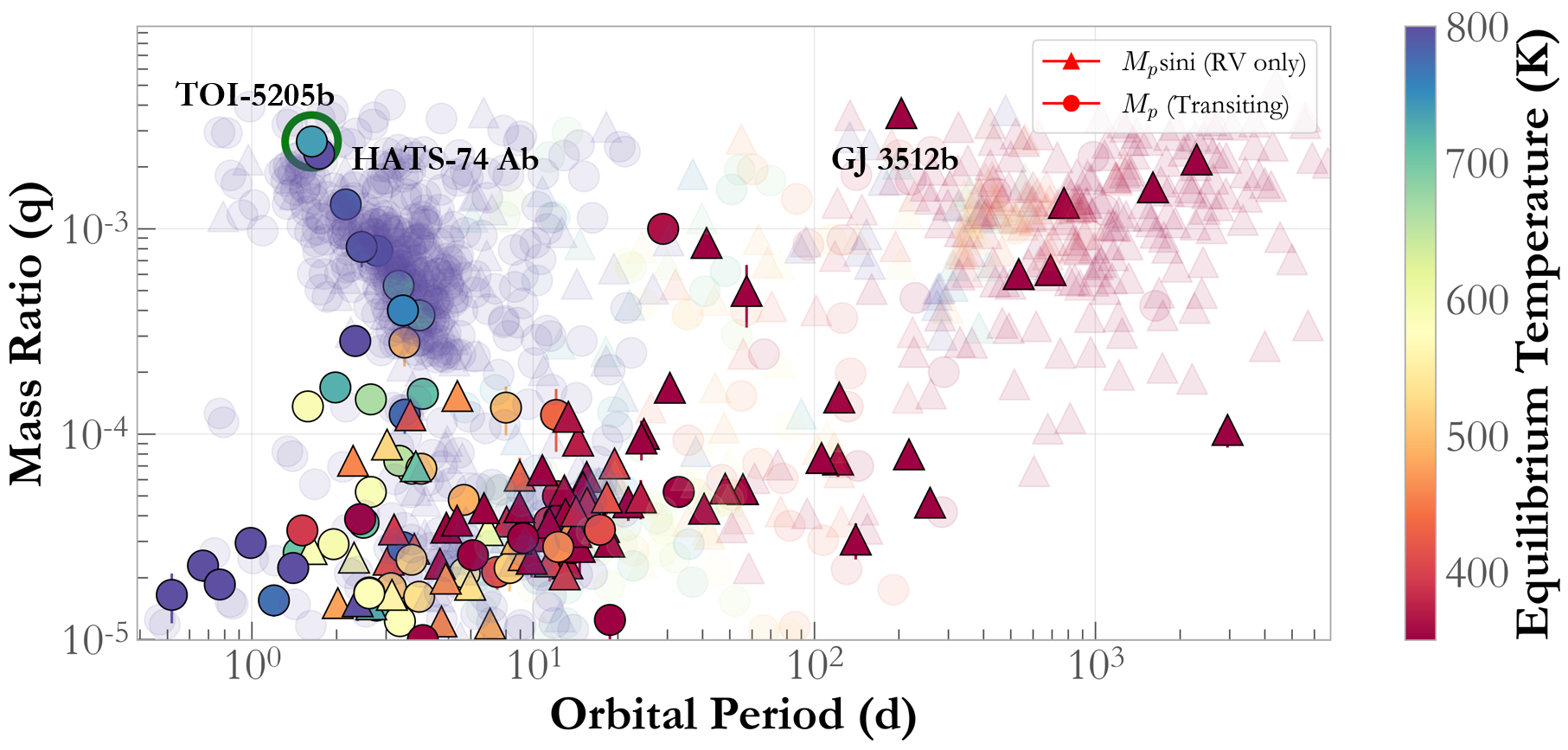}{0.90\textwidth \hspace{0.3 cm}}
{ \small c) Mass ratio as a function of orbital period} \label{fig:MassRatio}
\caption{\small \textbf{a)} We show TOI-5205~b (circled in green) in a mass-radius plane alongside other M dwarf planets (coloured by the stellar mass). We also include planets around FGK stars in the background, along with density contours for 0.3, 1, 3 \gcmcubed{} \citep{PSCompPars}. \textbf{b)} The radius-period plane is shown for the same sample of planets, but coloured by the equilibrium temperature. \textbf{c)} Planet-to-star mass ratio vs orbital period for planets with true mass (transiting; circle) and minimum mass (RV only; triangle) measurements. The planets are colour coded by the equilibrium temperature, the M dwarf planets are solid, whereas those orbiting FGK stars are shown in the background. TOI-5205~b (circled in green) has the highest mass ratio for transiting M dwarf planets. The highest mass ratio M dwarf planet is GJ 3512 b at $\sim 200$ d \citep{morales_giant_2019}.}\label{fig:MRP}
\end{figure*}

While gas giants are predicted to be rare and hard to form under the core-accretion framework \citep{laughlin_core_2004, ida_toward_2005}, they do exist around M dwarfs, as has been evinced by recent discoveries from transiting surveys, especially TESS \citep{johnson_characterizing_2012, hartman_hats-6b_2015, bayliss_ngts-1b_2018, canas_warm_2020, jordan_hats-74ab_2022, canas_toi-3714_2022, kanodia_toi-3757_2022}. In addition to transit discoveries, there have been RV-only detections of gas giants around M dwarfs, e.g., \cite{johnson_california_2010, wittenmyer_gj_2014, astudillo-defru_harps_2017, trifonov_carmenes_2018, feng_search_2020, morales_giant_2019, quirrenbach_carmenes_2022}. Some of these RV detected planets are around mid and late M dwarfs, but typically at longer orbital periods than the transiting planets \citep{schlecker_rv-detected_2022}. Due to the heterogeneous nature of this transiting sample, it is not straightforward to estimate the occurrence rate of such planets and compare them to population synthesis models or protoplanetary disk surveys.

So far all the discoveries of these transiting giant planets have been around early M dwarfs (M0 -- M2), which are consistent with the simulations from \cite{burn_new_2021} that find that gas giants do not form for host stars $< 0.5$ \solmass{}. We also note the recent discovery of the interesting TOI-1227 system, which hosts an inflated Jupiter-sized planet orbiting a very young (11 Myr) late M dwarf (0.17 \solmass{}). However, this planet just has a mass upper limit of 0.5 $M_J$, is still contracting, and will likely eventually shrink down to a super-Neptune \citep{mann_tess_2022}. Additionally, \cite{parviainen_toi-519_2021} validated a substellar object orbiting a mid-M dwarf (TOI-519), and place a 95$\%$ upper mass limit of 14 $M_J$ based on Doppler boosting, ellipsoidal variations, etc. TOI-5205~b defies this trend, as it orbits a mid-M dwarf host and has one of the largest mass ratio\footnote{GJ 3512b has a larger mass ratio at 0.37\%, but it does not transit and hence only a lower limit of its mass is available \citep{morales_giant_2019}.} for M dwarf planets  at 0.27\%. It is a Jovian sized planet (\autoref{fig:MRP}a) with an orbital period of $\sim 1.6$ days (\autoref{fig:MRP}b), and joins the current sample of $\sim 10$ known transiting gas giants around M dwarfs. TOI-5205~b is the first gas giant known to transit a mid-M dwarf, which also results in a transit depth $\delta$ of $\sim 7\%$. While we do not have precise constraints on the metallicity of the host star, photometric relations estimates suggest a host star metallicity close to solar ([Fe/H] = 0; Section \ref{sec:stellarchar}).

\subsection{Planet Formation}

In this section, we present a simple mass budget argument\footnote{\cite{schlecker_rv-detected_2022} discuss some of the other challenges in the formation of gas giants around low mass M dwarfs under the core accretion paradigm, beyond the mass budget discussed here.} to estimate the minimum mass of the primordial protoplanetary disk in which this giant planet formed under the core-accretion paradigm, where models suggest that runaway gaseous accretion should initiate once a protoplanet has reached a solid core mass of $\sim 10$ \earthmass{} \citep{pollack_formation_1996}. We calculate the heavy-element mass for TOI-5205~b using the relations from \cite{thorngren_mass-metallicity_2016} to be $\sim 60$ \earthmass{} (or roughly 10x more metal-enriched than the host star), but also note that there is considerable scatter in their sample that can perhaps be attributed to the vagaries in planet formation and evolution. There are additional uncertainties due to the unknown heavy element composition, and uncertainties in the equation of state used for their model. As it stands, these models predict $\sim 10$ \earthmass{} of heavy elements locked up in the central core, with the rest (60 - 10 $\sim$ 50 \earthmass) diffused in the H/He envelope.

\begin{figure*}[] 
\centering
\includegraphics[width=\textwidth]{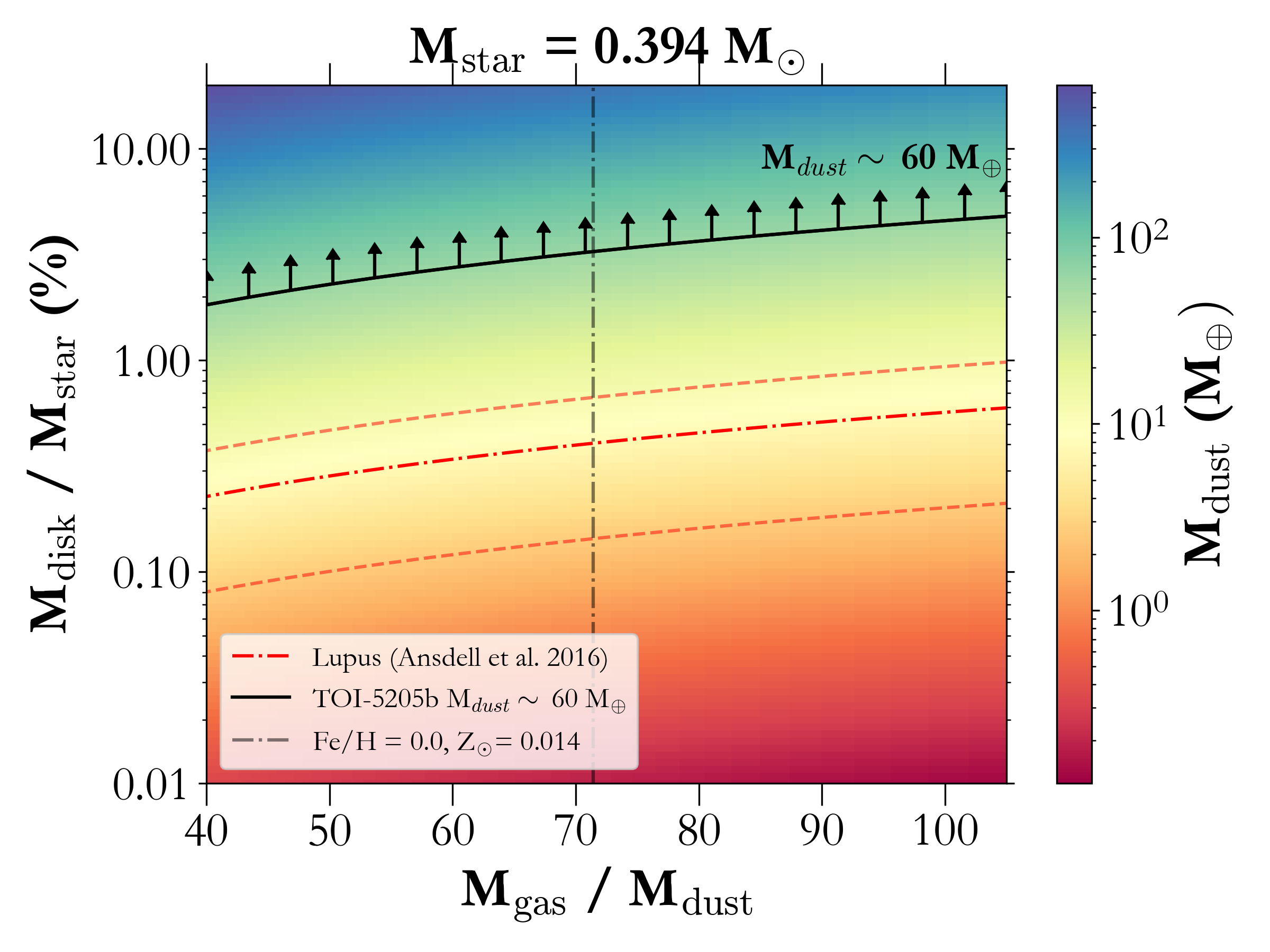}
\caption{The mass budget of the dust present in the disk as a function of disk mass ratio (disk to star; y axis), and the gas-to-dust ratio (x axis).  The black line is the contour corresponding to the estimated heavy-element mass for TOI-5205~b of $\sim 60$ \earthmass{} based on relations from \cite{thorngren_mass-metallicity_2016}. This indicates the disk properties required to form the planet even at 100\% formation efficiency, i.e. if all the dust present in the disk could accumulate in TOI-5205~b. A lower formation efficiency would imply an even larger disk dust mass. The red line shows the dust mass for a disk orbiting a mid-M dwarf as massive as TOI-5205 using the scaling relations from \cite{ansdell_alma_2016} in the young (1 -- 3 Myr) Lupus complex, while the region next to it shows the $1-\sigma$ uncertainty. We also include a vertical line to show solar metallicity of Z$_{\odot} = 0.014$, or a gas-to-dust ratio of $\sim 71$. The heavy-element core for TOI-5205~b (black line) is MUCH more massive than expected from scaling relations based on the Lupus complex.} \label{fig:MassBudget}
\end{figure*}

The dust mass of the disk is typically estimated for mm sized dust particles in Class II disks using flux continuum measurements at $\sim 850~ \mu$m, which is then used to calculate the mass assuming a blackbody with typical temperatures of 20 K. We decompose the total dust mass in the disk as a product of the disk mass ratio and gas-to-dust ratio (\autoref{fig:MassBudget}).  The canonical disk mass scaling (ratio of disk to stellar mass) assumed is $\sim 0.3\%$ based on a study of the Taurus region by \cite{andrews_mass_2013}, along with the gas-to-dust ratio of 70 -- 100 ranging from solar to the interstellar medium \citep[ISM;][]{bohlin_survey_1978}. Following these scaling relations suggests a total of 4--5 \earthmass~ of dust available for planet formation for TOI-5205, which would be insufficient to form a 10 \earthmass~ core to start runaway gaseous accretion even with 100$\%$ planet formation efficiency. In this section we refer to planet formation efficiency as the fraction of the total dust mass of the disk that is used to form TOI-5205~b.  Therefore in subsequent sections we discuss more realistic scaling values based on recent studies.

\subsubsection{Disk mass scaling}

\cite{pascucci_steeper_2016} suggest that the M$_{\rm{dust}}$ / M$_*$ relation becomes steeper with age, and more so for low mass stars. If so, these traditional relations would likely underestimate the initial mass of M dwarf disks. Results from \cite{ansdell_alma_2017} agree with this, where they find that (for a given stellar mass) the mass of dust present in a disk tends to decrease with age. They show this using a comparative analysis of disks in five young star forming regions spanning ages from 1--2 to 5--10 Myr and fitting separate scaling relations to each and then comparing the slopes, thereby corroborating the results from \cite{pascucci_steeper_2016}. They also note a large dispersion in these scaling relations that are not attributed to measurement systematics, but rather intrinsic astrophysical variation (or diversity) in disk properties within populations.

Observations and simulations based on the Orion Nebula Cluster show that for massive optically-thick disks with fluxes $>$ 10 mJy \citep[Figure 13 from][]{eisner_protoplanetary_2018}, the typical continuum flux-disk mass relations tends to under-predict the disk mass by up to an order-of-magnitude. However invoking disk stability arguments, the underestimate is probably less than that because depending on the surface density profile, disks can be $\sim 10\%$ of the stellar mass before they are unstable. These massive optically thick disks (Flux $\sim$ 10 mJy, M$_{\rm{dust}} \sim$ 10 -- 100 \earthmass{}) are seen around M dwarfs as well \citep[Figure 10;][]{eisner_protoplanetary_2018}; which is consistent with the typical scatter of $\sim 1$ dex seen in these (M$_{\rm{disk}}$/M$_*$) relations.

Studies suggest that planet formation is already underway for Class II disks \citep{greaves_have_2010, najita_mass_2014}, and indeed that the primordial disk mass available for giant planet formation early in the disk lifetime (0.1 -- 1 Myr) is likely much larger than than the masses measured for Class II disks, also evinced by measurements of the more massive Class I disks \citep{andrews_circumstellar_2005, vorobyov_embedded_2011}. Additionally, a lot of the solid mass for Class II disks can be locked up in planetesimals and planets, which the mm flux measurements would be insensitive to.

Given the significant scatter that exists in these scaling relations, and the various factors that can be responsible for underestimating the primordial dust mass in disks as mentioned above, it is not entirely unreasonable to postulate a more massive disk around TOI-5205 than that predicted by the standard 0.3 $\%$ M$_{\rm{disk}}$/M$_*$ scaling ratios for Class II disks. In observations pre-ALMA \citep[in Taurus;][]{andrews_mass_2013}, and then confirmed with ALMA --- Lupus,  \citep[][]{ansdell_alma_2016}, Chameleon I \citep[][]{pascucci_steeper_2016}, Upper Sco \citep[][]{barenfeld_alma_2016}, and $\sigma$ Orionis \citep[][]{ansdell_alma_2017} among others. Indeed \cite{andrews_mass_2013} do discuss the presence of outliers in their sample of disks in Taurus, which are anomalously massive at $\sim 10\%$ total disk-to-stellar mass.

\subsubsection{Gas-to-dust Ratio}

While the correlation between Jovian planet occurrence and metallicity of the host star has been well established \citep{gonzalez_stellar_1997, santos_metal-rich_2001, fischer_planet-metallicity_2005, ghezzi_stellar_2010, sousa_spectroscopic_2011}, there is still considerable uncertainty in the gas-to-dust (inverse of metallicity) assumed in planet formation models. This is typically estimated by measuring the mass of the gas in the disk using CO lines, which is then combined with dust mass measurements from mm continuum to obtain the gas-to-dust mass ratio. 

The typical ISM estimate for the gas-to-dust ratio is $\sim 100$ \citep{bohlin_survey_1978}, but a small sample of Taurus disks revealed a mean value closer to $\sim 16$ \citep{williams_parametric_2014}. In fact, \cite{ansdell_alma_2016} find that for disks in Lupus, the ratio might even be closer to 10, which was then corroborated by \cite{miotello_lupus_2017}. While these CO measurements could indicate a low gas-to-dust ratio, they could also be due to the selective loss of CO gas in the disk due to CO condensation, which would not apply to H$_2$. The latter was supported by \cite{rosotti_constraining_2017}, who showed that the accretion rate versus disk mass relationship is consistent when the mass of the disk is estimated using a gas-to-dust ratio of $\sim 100$. Based on this they suggest that this ratio cannot be lower than by a factor of 2 from the canonical ISM value of 100. Most recently, \cite{anderson_new_2022} find that gas mass measurements of CO isotopologues extrapolated to H$_2$ can have significant uncertainties, often by many orders of magnitude, thereby severely underestimating the gas-to-dust ratio.  All of this is to suggest that while the intrinsic gas-to-dust ratio for protoplanetary disks is hard to constrain, it should be within a factor of few of 100. 

For a solar metallicity $Z_{\odot} = 0.014$ disk (gas-to-dust ratio of $\sim 70$), we would require a disk that is about $3\%$ total disk-to-stellar mass,  to have the $\sim 60$ \earthmass{} of heavy-elements estimated for TOI-5205~b. While a detailed planet formation simulation is beyond the scope of this paper, \cite{lin_balanced_2018} suggest a maximum efficiency for giant planet formation under pebble accretion of $\sim 10\%$, which would require a disk that is $\sim 30\%$ in host star mass. Conversely, if the actual heavy-element mass for TOI-5205~b is lower than predicted by \cite{thorngren_mass-metallicity_2016} model, the required disk mass would scale down by the same factor.

\subsubsection{Disk lifetimes (increasing efficiency of planet formation)}
Apart from the low disk masses, the other issue with giant planet formation around M dwarfs is the longer orbital timescales (at a given separation) due to the lower host star mass. This results in a much slower growth rate for planetesimal formation ($\sim 1$ Myr), which must succeed in forming a massive enough core to initiate runaway accretion before the disk disperses.

The typical disk lifetime inferred by studying the incidence of disks in cluster of different ages is $\sim 3$ Myr, with an upper bound of $\sim 10$ Myr \citep{ribas_disk_2014}. It has also been established that this lifetime scales with stellar mass, and while the disks around M dwarfs typically last longer \citep{carpenter_evidence_2006}, they still disperse within $\sim 20$ Myr \citep{pecaut_star_2016}. Recently the discovery of very long-lived ($\gtrsim 20$ Myr), so called `Peter Pan' disks has been reported around M dwarfs \citep{lee_2mass_2020, silverberg_peter_2020, gaidos_planetesimals_2022}. Models suggest that the existence of these disks requires relatively high disk masses and very low external photoevaporation, similar to those found at the periphery of star-forming regions \citep{coleman_peter_2020}. These longer-lived massive disks would offer more time for the formation of solid cores massive enough to initiate runaway gas accretion under the slower core-accretion paradigm 

\subsubsection{Disk instability scenario}

Previous studies use the positive correlation for giant plant occurrence with stellar mass and metallicity \citep{gonzalez_stellar_1997, santos_metal-rich_2001, fischer_planet-metallicity_2005, ghezzi_stellar_2010, sousa_spectroscopic_2011} as evidence of core-accretion \citep{ida_toward_2005, thorngren_mass-metallicity_2016, ghezzi_retired_2018}. However, this correlation with metallicity is only seen for $M_p \lesssim 4$ \jupitermass{}; while stars hosting more massive planets are on average closer to solar metallicity, or even metal-poor \citep{santos_observational_2017, schlaufman_evidence_2018, maldonado_connecting_2019}. This suggests a dichotomy in the formation mechanism centered at $\sim$ 4 \jupitermass{}, with less massive objects classified as planets formed through core accretion, while more massive planets form through disk instability, similar to brown dwarfs and low-mass stars \citep{schlaufman_evidence_2018}. 

Even though the mass of TOI-5205~b is $< 4$ \jupitermass{}, due to the large mass ratio for TOI-5205~b we consider the disk instability scenario. Interestingly enough, this $\sim 10\%$ disk mass regime discussed in the previous section is also the typical disk mass required to enable giant planet formation under the disk instability scenario either close-in \citep{boss_rapid_2006} or farther out \citep{boss_formation_2011}. Disk instability has been proposed as a faster ($\sim 10^3$ yr) alternative to the slower ($\sim 1$ Myr) core-accretion formation scenario for M dwarfs where the lower host star mass translates to longer orbital timescales at a given distance from the star \citep{laughlin_core_2004}. Under this mechanism a massive 10-20$\%$ disk would have to be marginally unstable to start breaking up into lumps of gas and dust. These instabilities typically also require cooler temperatures, which warrants the formation of the planet at large orbital separations (ex-situ formation\footnote{See \cite{helled_giant_2014, dawson_origins_2018, helled_planet_2021} for comprehensive reviews on giant planet formation.}), followed by subsequent inward migration through disk migration \citep{kley_planet-disk_2012} or high eccentricity excitation \citep{beauge_multiple-planet_2012}. Given the scope of current models, we cannot rule out disk instability as a potential formation mechanism for TOI-5205~b.

Overall, we see two possible ways to explain the existence of this planet given current theories of planet formation -- i) A $\sim 60$ \earthmass~ solid heavy-element core: which would require a disk that is $\sim$ 3\% -- 30\% the mass of the host star (for 100\% and 10\% formation efficiency respectively), under which case both core-accretion and disk instability scenarios should be possible. ii) The interior models are biased and over-predict the solid core mass. Under the canonical core accretion scenario, this would suggest a core of 10 \earthmass, and would need a disk that is $\sim$ 0.5\% -- 5\% the mass of the host star (for 100\% and 10\% formation efficiency respectively).

\subsection{Atmospheric characterization}\label{sec:atmosphere}

\begin{figure*}
\gridline{\fig{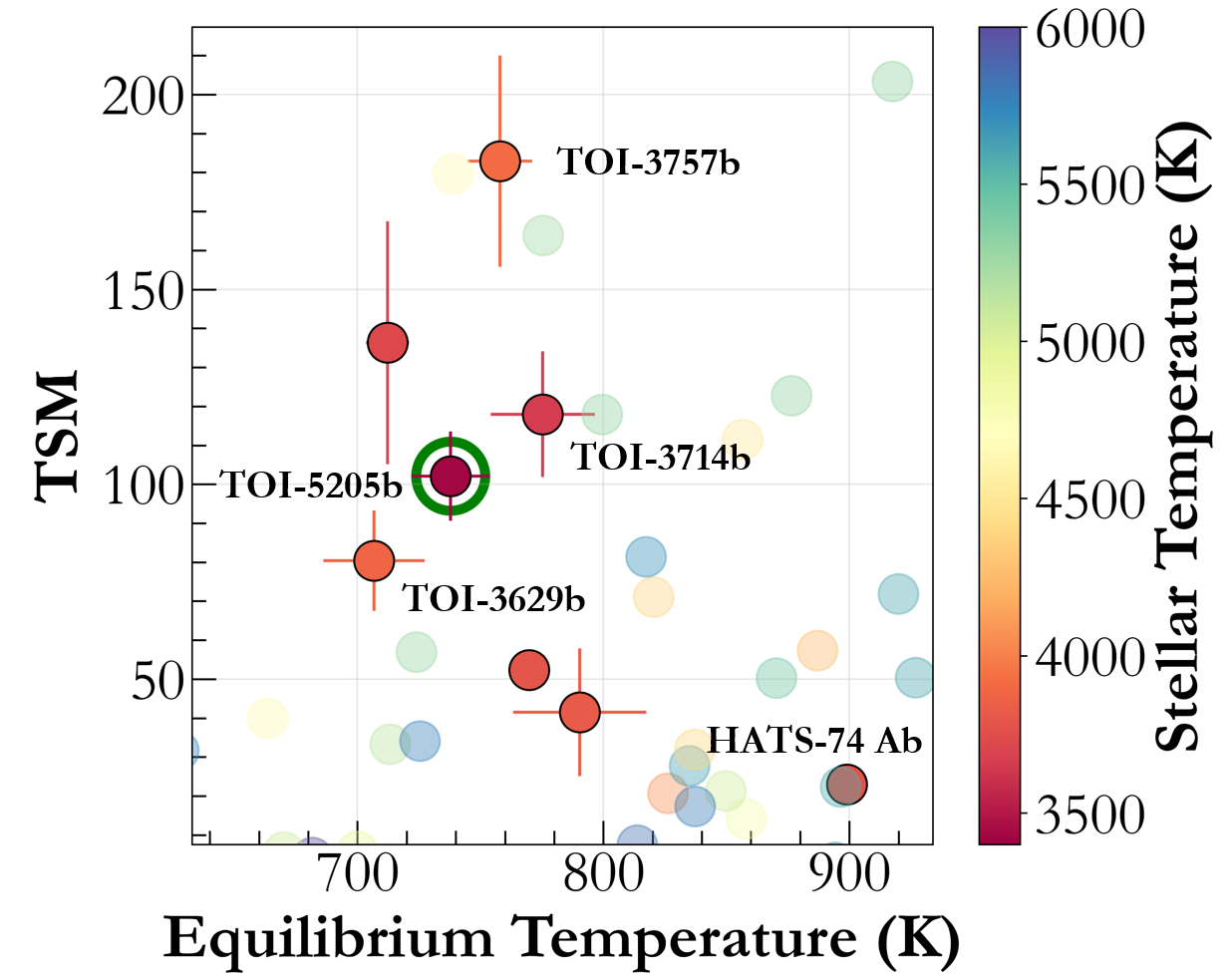}{0.4\textwidth}{{\small a) TSM comparison}}    \label{fig:TSM}
          \fig{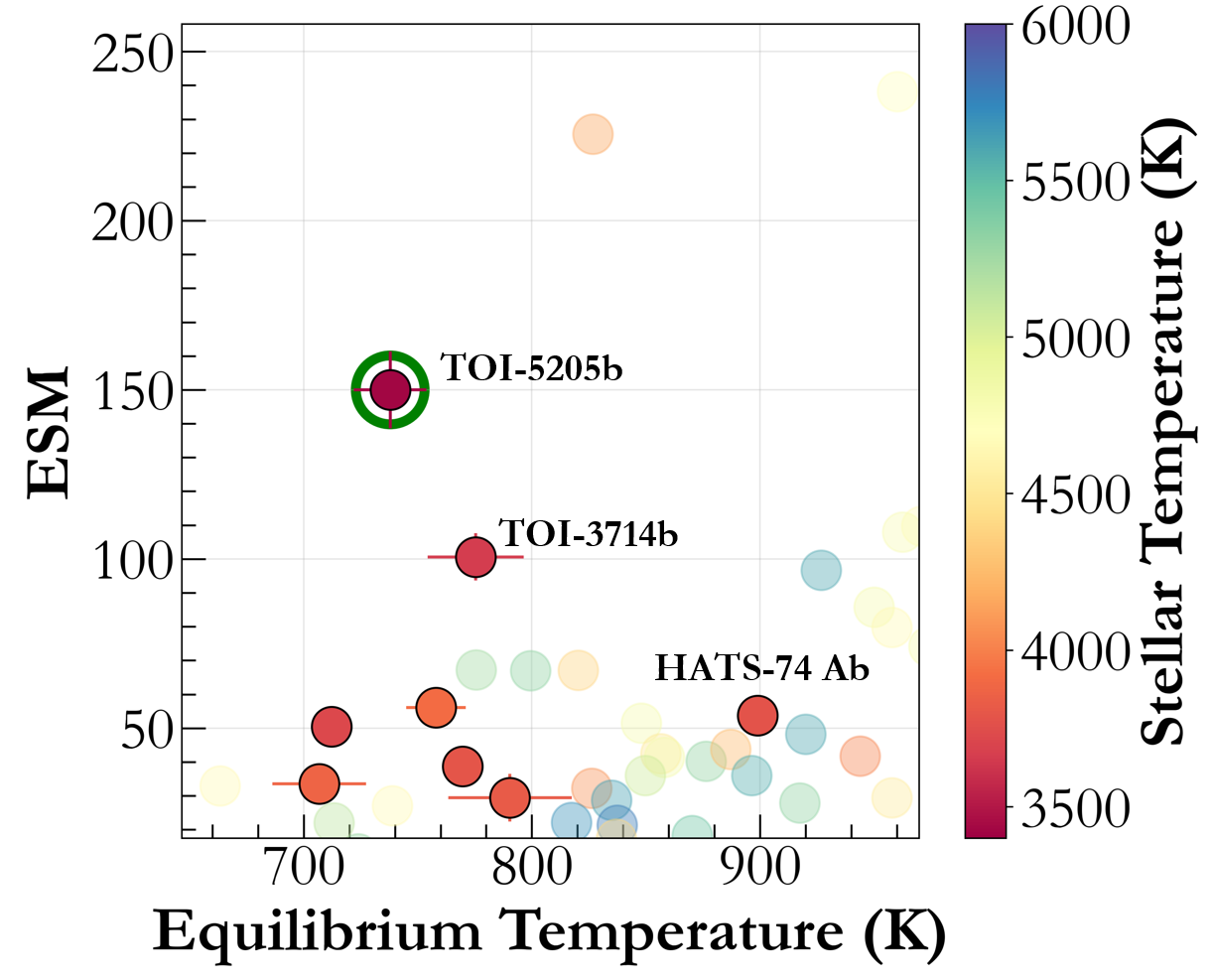}{0.4\textwidth}{ \small b) ESM comparison}} \label{fig:ESM}
\vspace{-0.5cm}          
\gridline{\fig{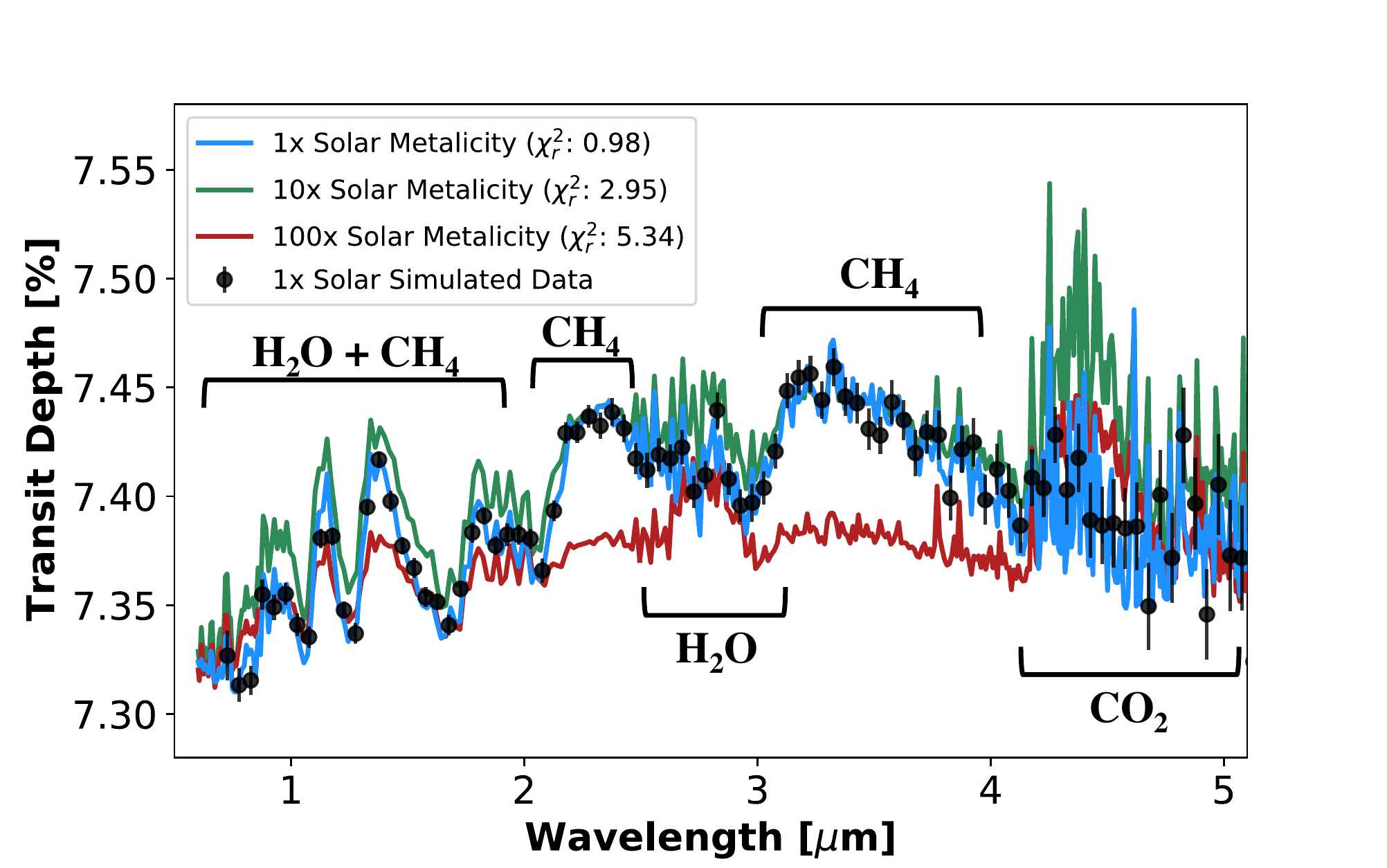}{0.5\textwidth}{{\small c) Simulated transmission spectra}}    \label{fig:transmission}
          \fig{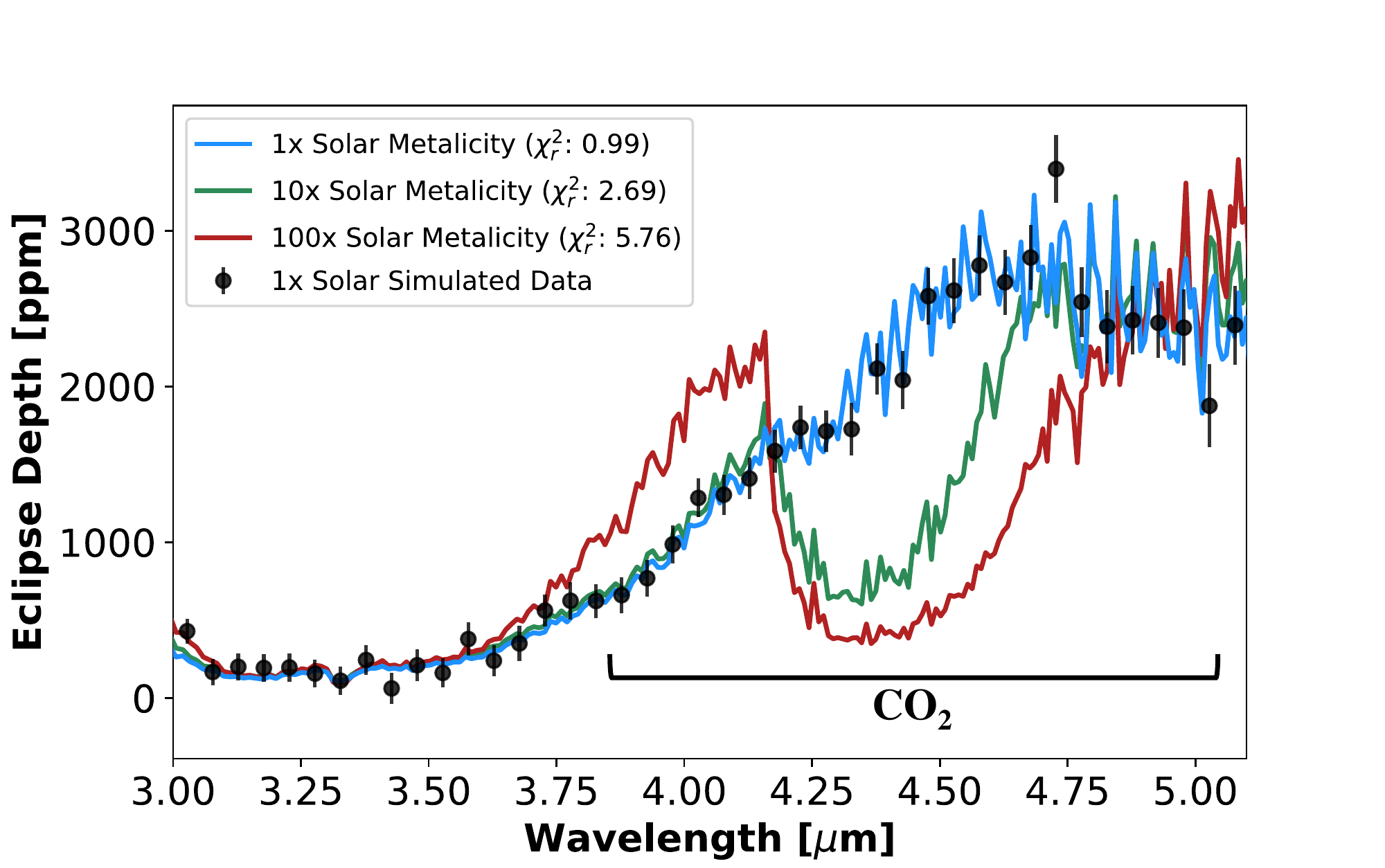}{0.5\textwidth}{ \small d) Simulated emission spectra}} \label{fig:emission}
\caption{\small \textbf{a-b)} We show the TSM and ESM for TOI-5205~b with respect to other M dwarf gas giants, $R_p > 8$ \earthradius{} (solid), while those orbiting FGK stars are in the background. TOI-5205~b (circled in green) has a high TSM ($\sim 100$) and ESM ($\sim 150$) that make it an excellent target for atmospheric characterization to estimate the chemical composition of the planet. \textbf{c-d)} Simulated transmission and thermal emission spectra for three different atmospheric metallicities, along with \texttt{PandExo} predictions for JWST NIRSPEC PRISM spectra for two transits and eclipses, respectively. }\label{fig:Atmosphere}
\end{figure*}

Characterizing the atmosphere of TOI-5205~b may provide clues needed to differentiate between formation mechanisms. Did it form via disk instability or core accretion, furthermore, under core accretion, did it form in-situ or farther out and then migrate inwards through disk or disk-free migration?

% Possible fleshing-out of the above paragraph
Assuming formation via core accretion, TOI-5205~b is expected to have a super-stellar metallicity if it underwent disk migration, or either sub- or super-stellar metallicity if it underwent disk-free migration \citep{madhusudhan_toward_2014}. If TOI-5205~b is metal-enriched, and therefore likely formed via core accretion, the second question surrounds whether TOI-5205~b formed in-situ or further out before migrating inwards. Multiple studies suggest that C/O ratios could provide some indication as to whether a planet formed inside or beyond various disk snowlines \cite[e.g.][]{oberg_effects_2011,madhusudhan_toward_2014}. As molecules ``freeze-out," they remove those elements from the overall gas composition. When water freezes for example, it removes some of the overall oxygen from the gas increasing the C/O ratio beyond the water-ice line  \citep{oberg_effects_2011}. Similarly, \cite{knierim_constraining_2022} show that the ratio of refractory and volatile elements can depend on the migration history of the planet. While \citet{dash_linking_2022} emphasises there are degeneracies and assumptions that must be considered, such as post-formation bombardment, or sublimation of the core, C/O ratios may provide the first insights into where TOI-5205~b originally formed.

Under the disk instability hypothesis, TOI-5205~b would have formed from a collapse of a massive region of the protoplanetary disk prior to migrating inwards. Therefore, from a first approximation, it is assumed that its atmosphere should reflect that of the protoplanetary disk and its host star - i.e. should have the same metallicity and abundances as TOI-5205 \citep[e.g.][]{helled_metallicity_2010, helled_measuring_2014}. However, recent works demonstrate that this initial picture may become complicated both by location of the initial collapse \citep{madhusudhan_toward_2014} or size of particles/objects accreted during this process \citep{helled_giant_2014}. \citet{hobbs_molecular_2022} suggests that comparing abundances of various molecules, notably methane, carbon monoxide/dioxide and hydrogen cyanide, may be a useful method for distinguishing the two formation pathways. Even with these complications, discovering a solar or near-solar metallicity atmosphere (heavy element abundance of $\sim$1\%) for TOI-5205~b would hint at the potential for gravitational instability. In this scenario, the heavy-element mass estimated using the \cite{thorngren_mass-metallicity_2016} sample would be incorrect for TOI-5205~b.

TOI-5205~b is a compelling target scientifically, and with its 7\% transit depth, it is also an object easily accessible with JWST observations. Even though it is a relatively cool (740 K) Jovian world, it still possesses a large Transmission Spectroscopy Metric (TSM) of $\sim 100$ placing it in the second quartile of their giant planet sample (assuming a scale factor of 1.15) from \cite{kempton_framework_2018}. TOI-5205~b also has one of the largest Emission Spectroscopy Metric (ESM) of any planet at $\sim$ 150, in part due to its R$_{p}$/R$_{s}$ and also its bright mid-M dwarf host (\autoref{fig:Atmosphere}). 

We calculate model transmission and thermal emission spectra assuming 1$\times$, 10$\times$, and 100$\times$ solar metallicity. We then simulate JWST NIRSpec PRISM data using \texttt{PandExo} \citep{batalha_pandexo_2017} corresponding to the 1$\times$ solar cases, assuming two transits/secondary eclipses, respectively. The transmission spectra are calculated using \texttt{Exo-Transmit} \citep{kempton_exo-transmit_2017}. We predict that NIRSpec should significantly distinguish between each of the model transmission spectra, as a result of the smaller spectral feature amplitudes for the 100$\times$ Solar metallicity model and the onset of a CO$_{2}$ feature in the 4--5~$\mu$m range between the 1$\times$ and 10$\times$ solar metallicity models. The model thermal emission spectra are generated using the self-consistent atmospheric model \texttt{GENESIS} \citep{gandhi_genesis_2017, gandhi_new_2019, piette_assessing_2020, piette_temperature_2020}. \texttt{GENESIS} calculates full line-by-line radiative transfer under the assumptions of radiative-convective equilibrium, hydrostatic equilibrium, and thermochemical equilibrium. Here, chemical equilibrium abundances are calculated using the analytic prescription of \citet{heng_analytical_2016}.We include opacity due to H$_2$O, CH$_4$, NH$_3$, HCN, CO, CO$_2$, C$_2$H$_2$ and collision-induced absorption (CIA) due to H$_2$-H$_2$ and H$_2$-He. The absorption cross sections for these species are calculated using the methods described in \citet{gandhi_genesis_2017}, using data from ExoMol, HITEMP and HITRAN (H$_2$O, CO and CO$_2$: \citet{Rothman2010}, CH$_4$: \citet{Yurchenko2013,Yurchenko2014a}, NH$_3$: \citet{Yurchenko2011}, HCN: \citet{Harris2006,Barber2014}, C$_2$H$_2$: \citet{rothman_hitran2012_2013}, CIA: \citet{Richard2012}). As shown in \autoref{fig:Atmosphere}, the 1$\times$, 10$\times$, and 100$\times$ Solar metallicity models are easily distinguishable in the $\sim$ 4--5~$\mu$m range due to the onset of an increasingly deep CO$_2$ feature as metallicity increases. Atmospheric characterization of TOI-5205~b with both transmission and thermal emission spectroscopy is therefore a promising avenue to characterise its atmospheric metallicity and C/O ratio, and to place constraints on its formation and evolution.

\section{Summary}\label{sec:conclusion}

We present the discovery of TOI-5205~b, a Jovian exoplanet orbiting a solar metallicity mid-M dwarf. TOI-5205~b was first identified from TESS photometry, and then characterised using a combination of ground-based photometry, radial velocities, spectroscopic observations, and speckle imaging.

The large mass ratio of the planet ($\sim 0.3\%$) necessitates a disk that is $\sim 10\%$ as massive as the host star, thereby stretching our current understanding of protoplanetary disks around M dwarfs. The typical scaling relations used to estimate disk properties are hard-pressed to reproduce the primordial disks that are massive enough to form such a planet. However there is significant scatter in disk dust mass measurements and scaling relations, which could still explain such massive planets around mid-M dwarfs. 

TOI-5205~b has a large transit depth of $7\%$, which makes it an excellent candidate for transmission and emission spectroscopy, both from the ground (high-resolution) and space (JWST). Atmospheric characterization could help constrain the metallicity of the planet and could offer clues about their formation history.

The large sample of M dwarfs being observed by TESS is already improving our understanding of planet formation around M dwarfs. While the first few discoveries were limited to the early M dwarfs, we are now starting to find that it is indeed possible to form these gas giants around mid-M dwarfs. As we go from a sample of these planets around solar-type stars to mid-M dwarfs, there is a unique opportunity to study planet formation at its extremes, spanning more than a 2x range in stellar mass, and 100x in luminosity!

\appendix

\section{Stellar characterization}\label{sec:stellarchar}

\subsection{Spectral classification}

\begin{figure*}[!t] 
\centering
\includegraphics[width=\textwidth]{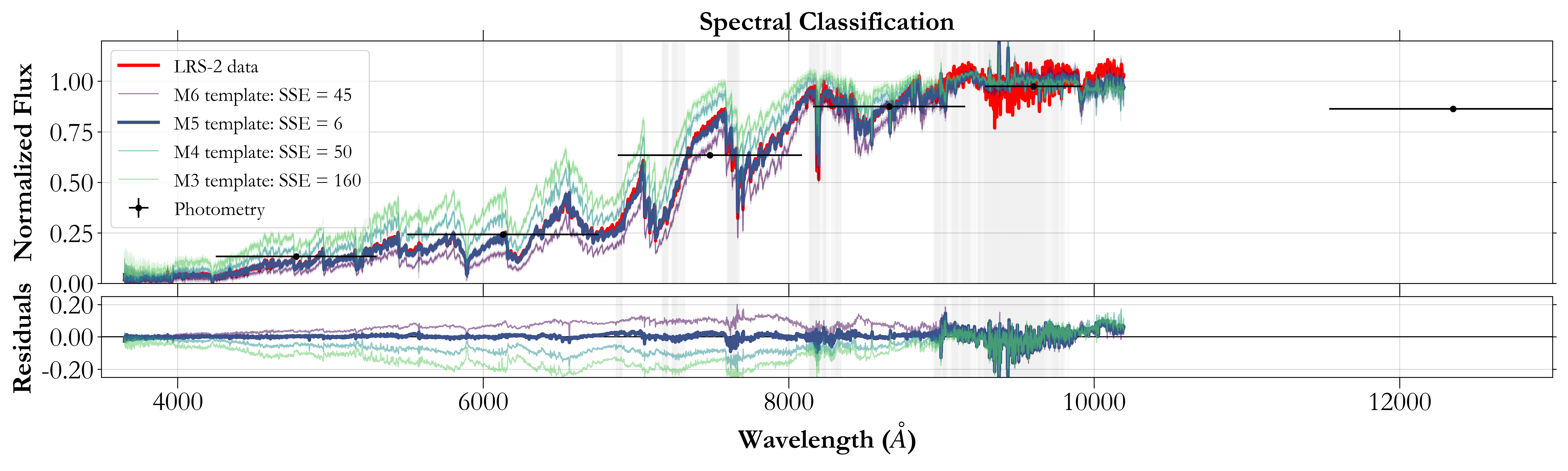}
\caption{Comparing the LRS2 spectra with the empirical templates from \texttt{pyHammer}. In red we show the observed LRS2 spectra after response and telluric correction, while empirical templates from M3 -- M6 are shown in different colours, while the vertical lines in the background denote the regions of significant telluric absorption. The increased noise ($\sim 5\%$) in the measured spectrum around $\sim 9000$ \AA~ can be attributed to the telluric correction. We include the residuals in the lower plot and also the summed square errors (SSE) in the legend showing that the M5 template is the preferred one. We also plot the normalised fluxes based on the photometric magnitudes from PS1 (optical) and the $J$ magnitude from FourStar with the horizontal errorbar depicting the bandpass. We do not include the $H, K$ magnitudes from FourStar in this plot to focus on the optical spectra and template comparison.} \label{fig:pyHammer}
\end{figure*}

\begin{itemize}
    \item Template Matching (\texttt{pyHammer}): We classify the spectral sub-type for TOI-5205 with the LRS2 spectra using \texttt{pyHammer} \citep{roulston_classifying_2020}, which is based on \texttt{The Hammer} \citep{covey_stellar_2007}, and uses an empirical template averaged across many observations. The empirical template is derived from the  MaNGA Stellar Library (MaStar), which consists of well-calibrated optical spectra from SDSS IV \citep{yan_sdss-iv_2019}. The relative calibration for this template is accurate to $< 5\%$. spanning stellar spectral types and metallicity. After applying the response and telluric correction, the combined LRS2 spectra (blue + red) matches a metal-rich M5 spectra the best \citep[based on spectral indices;][]{roulston_classifying_2020}, and also gives the lowest residuals when comparing the entire spectra (\autoref{fig:pyHammer}). 
    \item Spectral ratios: We also use the spectral ratios defined by \cite{kirkpatrick_standard_1991} surrounding CaH, Ti I, Na I and Ca II, to obtain a spectral type of M3 -- M4.5 based on the LRS2 spectra.
    
    \item Photometry relations: We also use relations from \cite{kiman_exploring_2019} to obtain a spectral type using the absolute $G$ magnitude, which suggests an $\sim$ M3.5 spectral type. Using the $G - J$ relation from Figure 13 in \cite{cifuentes_carmenes_2020} corroborates the M4 spectral type estimate for the given $G - J$ colour of $\sim 3$.
    
\end{itemize}

Considering the results from the template matching (M5) and colour relations, we adopt a spectral classification of M4.0 to which we ascribe an error of 1.0 sub-type.

\subsection{Using Photometric Relations}\label{sec:photrelations}

% \subsubsection{Estimating \teff{}}
We obtain $M_G$ of 10.09$^{+0.01}_{-0.03}$ from \cite{anders_photo-astrometric_2022}, which takes into account extinction using estimates from multiple photometric surveys. Using Equation 11 from \cite{rabus_discontinuity_2019},  we estimate a \teff{} from $M_G$ of 3430 K with an error of 54 K, where we propagate the error in $M_G$ to \teff{} and combining in quadrature with the scatter in the the polynomial fit. We do note that our \teff{} estimate is on the hotter end of that expected for an M4 spectral type, however this is not too surprising given the uncertainty of 1 spectral type, and the considerable scatter in theoretical models for mid-type M dwarfs.

% \subsubsection{Estimating R$_*$}
We use the empirically calibrated polynomial relations derived by \cite{mann_how_2015} to estimate the stellar radius. Given the large scatter in the \teff{} vs stellar radius relation \citep[Figure 9 from][]{mann_how_2015}, we use the absolute $K_s$ magnitude -- stellar radius relation instead, and adopt an error of $\sim 3\%$ on the stellar radius based on their cross-validation results.  The full transit obtained for TOI-5205~b from the 3.5 m APO telescope on 2022 July 16 is used to obtain a density constraint on the star of $8.8 \pm 0.4$ \gcmcubed{}, which is also consistent with the $\sim 0.39$ \solradius{} obtained above. We also use $M_G$ along with the bolometric calculator\footnote{\url{https://www.cosmos.esa.int/web/gaia/dr3-bolometric-correction-tool}} for the given \teff{} \citep{creevey_gaia_2022}, to obtain the bolometric magnitude, luminosity, and subsequently a stellar radius of $\sim 0.37 \pm 0.02$ \solradius{}, which is consistent with our radius estimate using the relations from \cite{mann_how_2015}.

Finally, we use the Stefan-Boltzmann law to obtain a stellar luminosity, and the empirically calibrated M-R relationship for main-sequence M dwarfs \citep[Equation 6;][]{schweitzer_carmenes_2019} to obtain a stellar mass (\autoref{tab:stellarparam}). We also verify the stellar mass using photometric relations from \cite{henry_mass-luminosity_1993, delfosse_accurate_2000, benedict_solar_2016, mann_how_2019} and consistently obtain similar results to $\sim$ 1--2 $\sigma$.. Mass-luminosity relations in the optical ($M_V$) from \cite{henry_mass-luminosity_1993} and \cite{benedict_solar_2016} give discrepant results with the $M_K$ mag relations due to the effect of the TiO and VO molecules, especially below 0.4 \solmass{} as discussed by \cite{baraffe_evolutionary_1998}. 

Photometric relations from \cite{bonfils_metallicity_2005}, \cite{schlaufman_physically-motivated_2010} and \cite{neves_metallicity_2012} give a metallicity of 0.02, 0.19 and 0.09 dex respectively, along with a typical uncertainty of 0.2 dex. \cite{maldonado_hades_2020} note that photometric metallicities have systematically lower values than corresponding spectroscopic techniques. However due to the sparse sampling of the \texttt{SpecMatch-Emp} library in \teff{}-[Fe/H] plane for mid-M dwarfs \citep{yee_precision_2017}, and the potential covariance between these two quantities we do not have reliable metallicity estimates from \texttt{SpecMatch-Emp} for this mid-M dwarf. Instead, we adopt a qualitative estimate of solar metallicity for TOI-5205 (\autoref{tab:stellarparam}). See \cite{passegger_metallicities_2022} for a detailed discussion of the complexities in metallicity determination for M dwarfs.

\subsection{Estimating activity level}\label{sec:activity}

\subsubsection{Using H-$\alpha$ from LRS2 spectra}\label{sec:halpha}
Emission in the H$\alpha$ line compared to the overall stellar bolometric luminosity is a powerful stellar activity indicator for M dwarfs \citep{west_activity-rotation_2015}. To estimate $\log(L_{\mathrm{H\alpha}}/L_{\mathrm{bol}}$) for TOI-5205, we measured the pseudo-equivalent width of the H$\alpha$ line from the LRS-2 red channel spectrum. Prior to measuring pEW(H$\alpha$), we shifted the spectra to zero radial velocity, accounting for the barycentric velocity and absolute velocity of the star. We measure the pEW(H$\alpha$) using the following equation:
\begin{equation}
\mathrm{pEW(H\alpha)} = \int_{\lambda_1}^{\lambda_2} \left( 1 - \frac{F(\lambda)}{F_{pc}}\right) d\lambda
\end{equation}
where we integrate over the limit from $\lambda_1 = 6560$\AA\ and $\lambda_2 = 6566$ \AA. $F_{pc}$ is the average of the median flux in the pseudo-continuum in the ranges from $6545-6559$ \AA, and $6567-6580$ \AA\ after removing a linear slope fit to that range seen in the pseudo-continuum surrounding the H$\alpha$ line for late M dwarfs. In doing so, we measure a $\mathrm{pEW(H\alpha)} = -0.81\pm0.01$ \AA, where the error is the statistical uncertainty accounting for the S/N of the observed spectrum. 

To estimate $\log (L_{H\alpha}/L_{bol})$, we use the following equation,
\begin{equation}
\log \left( \frac{L_{H\alpha}}{L_{bol}}\right) = \log \chi + \log(-\mathrm{pEW(H\alpha})),
\end{equation}
where $\log \chi$ is the ratio of the flux in the continuum near H$\alpha$ to the bolometric flux. We use estimate $\chi$ following the methodology in \cite{reiners_chromospheric_2008}, which gives $\chi$ as a function of stellar effective temperature for M dwarfs stars. In doing so, we obtain a $\log(\chi) = -4.3$, and $\log (L_{H\alpha}/L_{\mathrm{bol}})=-4.4$. From the sample of $\log (L_{H\alpha}/L_{\mathrm{bol}})$ values in \cite{west_activity-rotation_2015}, this value for TOI-5205 is suggestive of an M4 star that is not highly active.

\subsubsection{Rotation Period Estimates}\label{sec:rotation}

In Section \ref{sec:joint}, we describe the fitting of the TESS photometry from sectors 15 and 41 with separate stellar rotation kernels that return a rotation period of 3.7$^{+1.3}_{-1.1}$ days from sector 15 and $4.3 \pm 0.6$ days from sector 41.  The kernel consists of two simple harmonic oscillator terms -- one at the rotation period, with the second one at half the period. This observed period is also seen as a peak in a generalised Lomb Scargle (GLS) periodogram \citep{lomb_least-squares_1976,scargle_studies_1982, zechmeister_generalised_2009} on the sector 41 photometry (after masking the transits of TOI-5205~b) using its \texttt{astropy} implementation, and find a significant peak (20\% False Alarm Probability) at $\sim 4.4$ days.

However, similar to TOI-3757 b \citep{kanodia_toi-3757_2022}, we see that this periodic signal is likely an artifact from the photometry reduction of the FFI. The signal is seen a few adjoining pixels in a 8x8 grid centered on the centroid for TOI-5205, which suggests that the signal is not astrophysical in origin. This is further corroborated by the lack of detected rotational broadening in the HPF spectra, with which we can place a limit of  \vsini{} $< 2$ \kms{} on the host star. The corresponding equatorial velocity for a $\sim 4.4$ day rotation period would be $\sim 3.7$ \kms{}. Furthermore, we also check the publicly available data from  ASAS-SN \citep{kochanek_all-sky_2017} in V and $g$, and ATLAS \citep{tonry_atlas_2018} in the cyan (420 -- 650 nm) and orange (560 -- 820 nm) bands using a GLS periodogram, and do not find any significant signals. We did not find any publicly available data from the Zwicky Transient Facility Data Release 12 \citep[ZTF,][]{masci_zwicky_2019}.

Based on the H-$\alpha$ equivalent width estimate and lack of detectable photometric rotation signal for TOI-5205, we classify TOI-5205 as an inactive, old star.

\subsection{Blended sources of contamination}\label{sec:stellar companions}
The stellar density estimated assuming a circular orbit ($8.8 \pm 0.4$ \gcmcubed{}) confirms the mid-M dwarf spectral type for the host (Section \ref{sec:photrelations}). This also rules out the background eclipsing binary scenario around distant giant stars. 

The speckle imaging from NESSI is used to resolve the presence of any objects down to a separation of $0.3 \arcsec$ or about 27 AU (\autoref{fig:FalsePositives}). We then attempt to place constraints on unresolved stellar companions using HPF spectra, \gaia{} astrometry, archival imaging, photometry and the RVs. 

\subsubsection{Background objects}
We  look for background companions by comparing our observations of TOI-5205 from ARCTIC on 2022 April 22 (Section \ref{sec:arctic}) with observations from the Palomar Observatory Sky Survey \citep[POSS-1;][]{harrington_48-inch_1952,minkowski_national_1963} image taken on 1954 June 28. The POSS-1 plate images were taken with Eastman 103a-O spectroscopic plates without a filter and have a limiting magnitude of $\sim 20$. Over this period, TOI-5205 has had a proper motion of $\sim$ 4.5\arcsec, which is comparable to the PSF FWHM for the POSS-1 photographic plate observations. These archival observations rules out background companions that might be blended with TOI-5205 with a contrast of $\Delta V \sim 4$.

The closest companion seen in both images is TIC 1951446034, which is a resolved background star that is $\sim 4\arcsec$ away and not co-moving.

\subsubsection{Co-moving objects}\label{sec:comoving}
We rule out the possibility of a system where TOI-5205~b transits the primary, but is accompanied by a secondary stellar-mass companion orbiting the host star that is redder and fainter than TOI-5205 and would dilute the transit. 

Assuming no unresolved companions, the transit depth ($\delta_0$) for a planet with area $A_p$ crossing a star with area $A_1$ and luminosity $L_1(\lambda)$ is given by

\begin{equation}
    \delta_0 = \left(\frac{A_p  L_1(\lambda)}{A_1}\right) \frac{1}{L_1(\lambda)} = \frac{A_p}{A_1}
\end{equation}

Instead if there was an unresolved companion of later spectral type with luminosity $L_2 (\lambda)$, where $L_2 < L_1$:
    
\begin{align}
    \delta(\lambda) &= \left(\frac{A_p  L_1(\lambda)}{A_1}\right) \frac{1}{L_1(\lambda) + L_2(\lambda)} \\
    \delta(\lambda) &= \delta_0 \left(\frac{L_1(\lambda)}{L_1(\lambda) + L_2(\lambda)} \right) \\
    \delta(\lambda) &\propto \frac{1}{1 + L_2(\lambda)/L_1(\lambda)}
\end{align}

For $\lambda_2 > \lambda_1$,

\begin{align}
    &\frac{L_2}{L_1}\left(\lambda_2\right) > \frac{L_2}{L_1}\left(\lambda_1\right) \\ 
    &\textrm{i.e.,}~ \delta(\lambda_2) <  \delta(\lambda_1)
\end{align}

We obtained precise multi-filter transit photometry from the 3.5 m ARC telescope (\autoref{fig:transits}) in the SDSS \textit{i'} and \textit{g'} filters. If there was a later spectral type unresolved companion (object 2) that was contaminating the photometry of the host star (object 1), it would result in different transit depths across different photometric bands. If we assume that we can compare the two transit depths (in \textit{g'} and \textit{i'}) with a precision of $\epsilon$, where $\epsilon$ is a small number, then - 

\begin{align}
    \frac{\delta(g')}{\delta(i')} = 1 + \epsilon 
\end{align}

Then by this method we can rule out all objects with luminosity lesser than $L_2(\lambda)$,
\begin{align}
    \frac{L_2}{L_1}\bigg(i'\bigg) / \frac{L_2}{L_1}\bigg(g'\bigg) = 1 + \epsilon
\end{align}

\begin{figure}[] 
\centering
\includegraphics[width=0.5\textwidth]{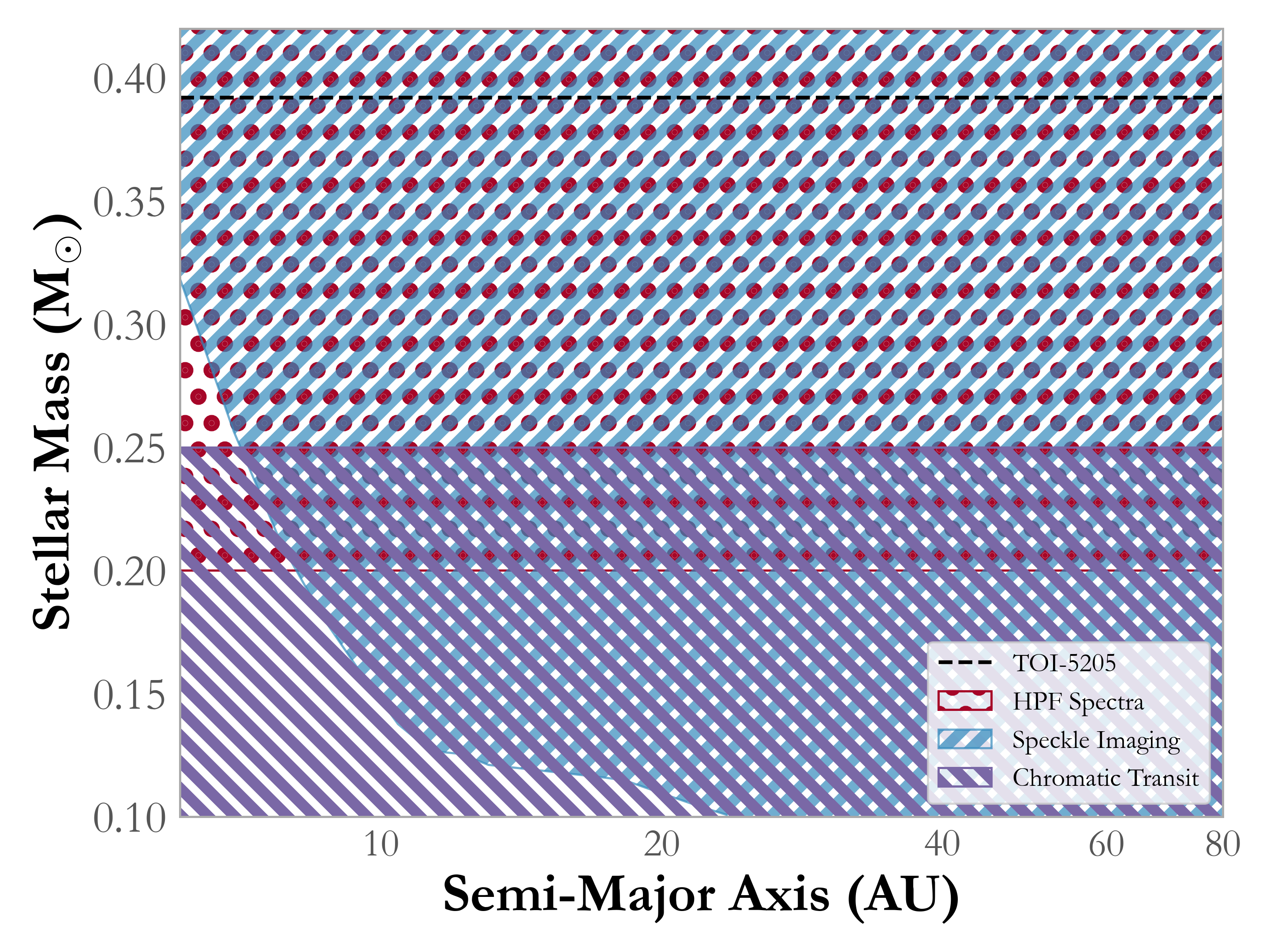}
\caption{We show the limits placed on a blended secondary companion using three different methods. In blue (forward leaning lines) we show the constraints from speckle imaging (Section \ref{sec:nessi}), in purple (backward leaning lines) we show the limits from comparing transit depths in $g'$ and $i'$  (Section \ref{sec:comoving}). Given the short-observing baseline of the HPF RVs, the RV slope can only rule out massive companions with periods $< 40$ days (0.2 AU), and are not shown here. Instead, the flux constraint from HPF spectra (red dots) order 5 ($\sim 8700$ \AA) is used to rule out stars more massive than 0.2 \solmass{} within the HPF aperture (0.85\arcsec~radius $\sim 80$ AU) as long as they have velocity offsets $|\Delta v|$  $> $ 5 \kms{}. The dashed black line is the mass of the primary -- TOI-5205.} \label{fig:FalsePositives}
\end{figure}

Using a separate dilution term for the ARCTIC transit in $g'$, we probe for chromaticity in the transit depth between $g'$ and $i'$, but find the depths to be consistent to $\sim 10\%$, i.e. $\epsilon \sim 0.1$. We then use the SDSS transmission curves for the two filters, and compare the flux within the bandpass using BT-Settl CIFIST theoretical stellar spectra for a range of stellar masses \citep{allard_model_2011, allard_models_2012}. We conclude that this method would be sensitive to transit depth variations for a unresolved companion cooler than $\sim$ 3100 K or roughly 0.25 \solmass{} (\autoref{fig:FalsePositives}).

If there was a secondary stellar-mass object present in the system, i.e. a hierarchical system, it would be a source of dilution that would suggest a radius larger than the $\sim$ 1~ $R_J$ estimated here. Due to the electron degeneracy pressure, objects around this size can range from Jovian planets to very low mass stars \citep[M7-M8;][]{zapolsky_mass-radius_1969, burrows_theory_2001}. Therefore, if TOI-5205~b had a larger radius (due to unaccounted dilution), it would have to be a late-type M dwarf or larger, which would make it at least 100x more massive than the $\sim 1$~$M_J$ we measure (\autoref{tab:planetprop}).

Finally, we put additional constraints on the possibility of a bound stellar companion, as follows:

\begin{itemize}
    \item Constraints from HPF spectra: We follow the procedure outlined in \cite{kanodia_toi-1728b_2020} to place limits on any spatially unresolved stellar companion to TOI-5205 using the HPF spectra to quantify the lack of flux from a secondary object. We combine the spectra from a single epoch to obtain a higher S/N template for comparison, and then model the test spectra (TOI-3757) as a linear combination of a primary M dwarf (GJ 273) and a secondary companions (GJ 9066, GJ 1072, GJ 1111 and LSPM J0510+2714). The flux ratio between the secondary and primary star, $F$, is calculated as:
        
    \begin{eqnarray}
    S_{\mathrm{obs}} &=& A \left( (1-x)S_{\mathrm{primary}} + (x)S_{\mathrm{secondary}} \right) \label{eq:spectra} \\
    F &=& \frac{x}{1-x} \label{eq:fluxratio}
    \end{eqnarray}
    \noindent where $S_{\mathrm{obs}}$ is the observed spectrum,  $S_{\mathrm{primary}}$ is the primary spectrum, $S_{\mathrm{secondary}}$ represents the secondary spectrum, and $A$ is the normalization constant. For a given primary and secondary template, we (i) perform a $\chi^2$ minimization to shift the secondary spectrum in velocity space, (ii) add this shifted secondary spectrum to the primary, and (iii) fit for the value of $x$ (and $A$) that best fits the observed spectrum. We perform this for a range of spectral types for the secondary from M4.5 to M7 spanning velocity offsets of $\pm 150$ \kms{}. We place a conservative upper limit for a secondary companion of flux ratio $<$ 0.2 or $\Delta \rm{mag} \simeq 1.8$ for $|\Delta v|$  $> $ 5 \kms{}, using HPF order index 5 spanning $8650 - 8770$ \AA. The lower limit coincides with HPF's spectral resolution ($R \sim 55,000 \approx 5.5$~\kms{}). At lower velocity offsets, the degeneracy between the primary and secondary spectra prevents any meaningful flux ratio constraints.
    
    \item Constraints from \gaia{} astrometry: Gaia DR3 \citep{vallenari_gaia_2022} provides an additional astrometric constraint on the presence of unresolved bound companions using the re-normalised unit weight error (RUWE) metric. RUWE is sensitive to the change in the position of the primary target due to reflex motion caused by unresolved bound companions. For the single-star astrometric solution in use for Gaia DR3, this astrometric motion of the primary star around the center of mass would manifest as noise \citep{kervella_stellar_2019}, especially for orbital periods much shorter than the observing baseline for Gaia DR3 ($\sim 34$ months). The commonly accepted threshold in literature for this is RUWE $\gtrsim 1.4$, which correlates with the presence of a bound stellar companion in recent studies of stellar binaries \citep{penoyre_binary_2020, belokurov_unresolved_2020, gandhi_astrometric_2021}. For TOI-5205, Gaia DR3 reports a RUWE of $\sim 1.03$, which is in agreement with a single-star astrometric solution.
    
    % \iffalse
    \item Constraints from RVs: A joint fit of the photometry and RVs is used to estimate the planetary and system properties (Section \ref{sec:joint}). We also include a linear RV trend in the orbital solution while fitting the RVs. We estimate this to be consistent with 0, with an the estimated RV trend $\sim$ 0.05$^{+4.92}_{-5.08}$ \ms{} yr$^{-1}$. Assuming a circular orbit for a unresolved companion star, the maximum is at phase 0 (conjunction) and 180$^{\circ}$, where the amplitude would be $2\pi K/P$, where $K$ is the RV-semi amplitude on the primary star due to a hypothetical secondary, and $P$ is its orbital period. However, given our short observing period ($\sim 30$ days), we use this to only constrain companions with a maximum orbital period of $\sim 60$ days, or a semi-major axis of 0.2 AU.
    % \fi

\end{itemize}

\subsection{Galactic kinematics}

Using the systemic velocity from HPF and proper motion from Gaia DR3, we calculate the \textit{UVW} velocities in the barycentric frame using \texttt{GALPY} \citep{bovy_galpy_2015}\footnote{With \textit{U} towards the Galactic center, \textit{V} towards the direction of Galactic spin, and \textit{W} towards the North Galactic Pole  \citep{johnson_calculating_1987}.}. We provide these velocities in \autoref{tab:stellarparam}, including those in the local standard of rest using the offsets from \cite{schonrich_local_2010}. Using the BANYAN tool \citep{gagne_banyan_2018}, we classify TOI-5205 as a field star in the thin disk with very high probability \citep[$> 99 \%$;][]{bensby_exploring_2014}.

% Probability of Thin Disk to Thick Disk = 16.615
% Probability of Thin Disk to Halo= 83566.453

\section{Acknowledgements}
% \begin{acknowledgments}

We thank the anonymous referee for the valuable feedback which has improved the quality of this manuscript.

SK thanks Rocio Kiman for help with spectral typing using Gaia colours; Alycia J. Weinberger for illuminating discussions and references regarding protoplanetary disks around M dwarfs; Peter Gao for discussions regarding the atmospheric characterization of these planets and for proof-reading sections of this manuscript. SK acknowledges research support from Carnegie Institution of Science through the Carnegie Fellowship.

% PSU Land acknowledgement
The Pennsylvania State University campuses are located on the original homelands of the Erie, Haudenosaunee (Seneca, Cayuga, Onondaga, Oneida, Mohawk, and Tuscarora), Lenape (Delaware Nation, Delaware Tribe, Stockbridge-Munsee), Shawnee (Absentee, Eastern, and Oklahoma), Susquehannock, and Wahzhazhe (Osage) Nations.  As a land grant institution, we acknowledge and honor the traditional caretakers of these lands and strive to understand and model their responsible stewardship. We also acknowledge the longer history of these lands and our place in that history.

% HET/HPF, also needs footnote
These results are based on observations obtained with the Habitable-zone Planet Finder Spectrograph on the HET. We acknowledge support from NSF grants AST-1006676, AST-1126413, AST-1310885, AST-1310875,  ATI 2009889, ATI-2009982, AST-2108512, AST-2108801 and the NASA Astrobiology Institute (NNA09DA76A) in the pursuit of precision radial velocities in the NIR. The HPF team also acknowledges support from the Heising-Simons Foundation via grant 2017-0494. The Low Resolution Spectrograph 2 (LRS2) was developed and funded by the University of Texas at Austin McDonald Observatory and Department of Astronomy and by Pennsylvania State University. We thank the Leibniz-Institut für Astrophysik Potsdam (AIP) and the Institut für Astrophysik Göttingen (IAG) for their contributions to the construction of the integral field units.  The Hobby-Eberly Telescope is a joint project of the University of Texas at Austin, the Pennsylvania State University, Ludwig-Maximilians-Universität München, and Georg-August Universität Gottingen. The HET is named in honor of its principal benefactors, William P. Hobby and Robert E. Eberly. The HET collaboration acknowledges the support and resources from the Texas Advanced Computing Center. We thank the Resident astronomers and Telescope Operators at the HET for the skillful execution of our observations with HPF. We would like to acknowledge that the HET is built on Indigenous land. Moreover, we would like to acknowledge and pay our respects to the Carrizo \& Comecrudo, Coahuiltecan, Caddo, Tonkawa, Comanche, Lipan Apache, Alabama-Coushatta, Kickapoo, Tigua Pueblo, and all the American Indian and Indigenous Peoples and communities who have been or have become a part of these lands and territories in Texas, here on Turtle Island.

% Diffuser grant
We acknowledge support from NSF grants AST-1910954, AST-1907622, AST-1909506, AST-1909682 for the ultra-precise photometry effort.

% From Bill Cochran - AST-2108801

% LRS2
The Low Resolution Spectrograph 2 (LRS2) was developed and funded by the University of Texas at Austin McDonald Observatory and Department of Astronomy and by Pennsylvania State University. We thank the Leibniz-Institut für Astrophysik Potsdam (AIP) and the Institut für Astrophysik Göttingen (IAG) for their contributions to the construction of the integral field units.

%WIYN acknowledgment
WIYN is a joint facility of the University of Wisconsin-Madison, Indiana University, NSF's NOIRLab, the Pennsylvania State University, Purdue University, University of California-Irvine, and the University of Missouri. 
%Land acknowledgment
The authors are honored to be permitted to conduct astronomical research on Iolkam Du'ag (Kitt Peak), a mountain with particular significance to the Tohono O'odham. Data presented herein were obtained at the WIYN Observatory from telescope time allocated to NN-EXPLORE through the scientific partnership of NASA, the NSF, and NOIRLab.

Deepest gratitude to Zade Arnold, Joe Davis, Michelle Edwards, John Ehret, Tina Juan, Brian Pisarek, Aaron Rowe, Fred Wortman, the Eastern Area Incident Management Team, and all of the firefighters and air support crew who fought the recent Contreras fire. Against great odds, you saved Kitt Peak National Observatory.

% NESSI
Some of the observations in this paper made use of the NN-EXPLORE Exoplanet and Stellar Speckle Imager (NESSI). NESSI was funded by the NASA Exoplanet Exploration Program and the NASA Ames Research Center. NESSI was built at the Ames Research Center by Steve B. Howell, Nic Scott, Elliott P. Horch, and Emmett Quigley.

% Gaia
This work has made use of data from the European Space Agency (ESA) mission Gaia (\url{https://www.cosmos.esa.int/gaia}), processed by the Gaia Data Processing and Analysis Consortium (DPAC, \url{https://www.cosmos.esa.int/web/gaia/dpac/consortium}). Funding for the DPAC has been provided by national institutions, in particular the institutions participating in the Gaia Multilateral Agreement.

% ZTF
% Some of the observations in this paper were obtained with the Samuel Oschin Telescope 48-inch and the 60-inch Telescope at the Palomar Observatory as part of the ZTF project. ZTF is supported by the NSF under Grant No. AST-2034437 and a collaboration including Caltech, IPAC, the Weizmann Institute for Science, the Oskar Klein Center at Stockholm University, the University of Maryland, Deutsches Elektronen-Synchrotron and Humboldt University, the TANGO Consortium of Taiwan, the University of Wisconsin at Milwaukee, Trinity College Dublin, Lawrence Livermore National Laboratories, and IN2P3, France. Operations are conducted by COO, IPAC, and UW.

% ACI
Computations for this research were performed on the Pennsylvania State University’s Institute for Computational and Data Sciences Advanced CyberInfrastructure (ICDS-ACI).  This content is solely the responsibility of the authors and does not necessarily represent the views of the Institute for Computational and Data Sciences.

% John Wisneiswki
% We acknowledge support from NSF grant AST-1907622 in the pursuit of precise photometric observations from the ground.

% CEHW 
The Center for Exoplanets and Habitable Worlds is supported by the Pennsylvania State University, the Eberly College of Science, and the Pennsylvania Space Grant Consortium. 

% MAST
Some of the data presented in this paper were obtained from MAST at STScI. Support for MAST for non-HST data is provided by the NASA Office of Space Science via grant NNX09AF08G and by other grants and contracts.
% Kepler/TESS
This work includes data collected by the TESS mission, which are publicly available from MAST. Funding for the TESS mission is provided by the NASA Science Mission directorate. 
% NASA Exoplanet Archive, ADS, 2MASS
This research made use of the (i) NASA Exoplanet Archive, which is operated by Caltech, under contract with NASA under the Exoplanet Exploration Program, (ii) SIMBAD database, operated at CDS, Strasbourg, France, (iii) NASA's Astrophysics Data System Bibliographic Services, and (iv) data from 2MASS, a joint project of the University of Massachusetts and IPAC at Caltech, funded by NASA and the NSF.

%ADS
This research has made use of the SIMBAD database, operated at CDS, Strasbourg, France, 
and NASA's Astrophysics Data System Bibliographic Services.

% Exofop 
This research has made use of the Exoplanet Follow-up Observation Program website, which is operated by the California Institute of Technology, under contract with the National Aeronautics and Space Administration under the Exoplanet Exploration Program

% Sam, JPL
The research was carried out (in part) at the Jet Propulsion Laboratory, California Institute of Technology, under a contract with the National Aeronautics and Space Administration. 

% Caleb
CIC acknowledges support by an appointment to the NASA Postdoctoral Program at the Goddard Space Flight Center, administered by USRA through a contract with NASA.

SK would like to acknowledge Theodora and Rafa for help with this project.

This research made use of \textsf{exoplanet} \citep{foreman-mackey_exoplanet-devexoplanet_2021, foreman-mackey_exoplanet_2021} and its
dependencies \citep{agol_analytic_2020, foreman-mackey_fast_2017, foreman-mackey_scalable_2018, kumar_arviz_2019, robitaille_astropy_2013, astropy_collaboration_astropy_2018, kipping_efficient_2013, luger_starry_2019, the_theano_development_team_theano_2016, salvatier_probabilistic_2016}

% \end{acknowledgments}

\facilities{\gaia{}, HET (HPF), APO (ARCTIC), WIYN 3.5 m (NESSI), Magellan (FourStar) RBO \tess{}, Exoplanet Archive}
\software{
\texttt{ArviZ} \citep{kumar_arviz_2019}, 
AstroImageJ \citep{collins_astroimagej_2017}, 
\texttt{astroquery} \citep{ginsburg_astroquery_2019}, 
\texttt{astropy} \citep{robitaille_astropy_2013, astropy_collaboration_astropy_2018},
\texttt{barycorrpy} \citep{kanodia_python_2018}, 
\texttt{celerite2} \citep{foreman-mackey_fast_2017, foreman-mackey_scalable_2018},
\texttt{daophot} \citep{stetson_daophot_1987, stetson_ccd_1988},
\texttt{eleanor} \citep{feinstein_eleanor_2019},
\texttt{exoplanet} \citep{foreman-mackey_exoplanet-devexoplanet_2021, foreman-mackey_exoplanet_2021},
\texttt{Exo-Transmit} \citet{kempton_exo-transmit_2017},
\texttt{GENESIS} \citep{gandhi_genesis_2017},
\texttt{HxRGproc} \citep{ninan_habitable-zone_2018},
\texttt{ipython} \citep{perez_ipython_2007},
\texttt{lightkurve} \citep{lightkurve_collaboration_lightkurve_2018},
\texttt{matplotlib} \citep{hunter_matplotlib_2007},
\texttt{numpy} \citep{oliphant_numpy_2006},
\texttt{Panacea}\footnote{https://github.com/grzeimann/Panacea},
\texttt{pandas} \citep{mckinney_data_2010},
\texttt{PandExo} \citep{batalha_pandexo_2017},
\texttt{PyMC3} \citep{salvatier_probabilistic_2016},
\texttt{pyHammer}\citep{roulston_classifying_2020},
\texttt{scipy} \citep{oliphant_python_2007, virtanen_scipy_2020},
\texttt{SERVAL} \citep{zechmeister_spectrum_2018},
\texttt{starry} \citep{luger_starry_2019, agol_analytic_2020},
\texttt{Theano} \citep{the_theano_development_team_theano_2016}.
}

\bibliography{references, ManualReferences}

\begin{thebibliography}{}
\expandafter\ifx\csname natexlab\endcsname\relax\def\natexlab#1{#1}\fi
\providecommand{\url}[1]{\href{#1}{#1}}

\bibitem[{Agol {et~al.}(2020)Agol, Luger, \&
  Foreman-Mackey}]{agol_analytic_2020}
Agol, E., Luger, R., \& Foreman-Mackey, D. 2020, The Astronomical Journal, 159,
  123.
\newblock \url{https://ui.adsabs.harvard.edu/abs/2020AJ....159..123A}

\bibitem[{Allard {et~al.}(2011)Allard, Homeier, \& Freytag}]{allard_model_2011}
Allard, F., Homeier, D., \& Freytag, B. 2011, 448, 91, conference Name: 16th
  Cambridge Workshop on Cool Stars, Stellar Systems, and the Sun Place: eprint:
  arXiv:1011.5405 ADS Bibcode: 2011ASPC..448...91A.
\newblock \url{https://ui.adsabs.harvard.edu/abs/2011ASPC..448...91A}

\bibitem[{Allard {et~al.}(2012)Allard, Homeier, \&
  Freytag}]{allard_models_2012}
---. 2012, Philosophical Transactions of the Royal Society of London Series A,
  370, 2765.
\newblock \url{http://adsabs.harvard.edu/abs/2012RSPTA.370.2765A}

\bibitem[{Anders {et~al.}(2022)Anders, Khalatyan, Queiroz, Chiappini, Ardèvol,
  Casamiquela, Figueras, Jiménez-Arranz, Jordi, Monguió, Romero-Gómez,
  Altamirano, Antoja, Assaad, Cantat-Gaudin, Castro-Ginard, Enke, Girardi,
  Guiglion, Khan, Luri, Miglio, Minchev, Ramos, Santiago, \&
  Steinmetz}]{anders_photo-astrometric_2022}
Anders, F., Khalatyan, A., Queiroz, A. B.~A., {et~al.} 2022, Astronomy \&amp;
  Astrophysics, Volume 658, id.A91,
  {\textless}NUMPAGES{\textgreater}27{\textless}/NUMPAGES{\textgreater} pp.,
  658, A91.
\newblock
  \url{https://ui.adsabs.harvard.edu/abs/2022A%26A...658A..91A/abstract}

\bibitem[{Anderson {et~al.}(2022)Anderson, Cleeves, Blake, Bergin, Zhang,
  Carpenter, \& Schwarz}]{anderson_new_2022}
Anderson, D.~E., Cleeves, L.~I., Blake, G.~A., {et~al.} 2022, The Astrophysical
  Journal, 927, 229, aDS Bibcode: 2022ApJ...927..229A.
\newblock \url{https://ui.adsabs.harvard.edu/abs/2022ApJ...927..229A}

\bibitem[{Andrews {et~al.}(2013)Andrews, Rosenfeld, Kraus, \&
  Wilner}]{andrews_mass_2013}
Andrews, S.~M., Rosenfeld, K.~A., Kraus, A.~L., \& Wilner, D.~J. 2013, The
  Astrophysical Journal, 771, 129, aDS Bibcode: 2013ApJ...771..129A.
\newblock \url{https://ui.adsabs.harvard.edu/abs/2013ApJ...771..129A}

\bibitem[{Andrews \& Williams(2005)}]{andrews_circumstellar_2005}
Andrews, S.~M., \& Williams, J.~P. 2005, The Astrophysical Journal, 631, 1134,
  aDS Bibcode: 2005ApJ...631.1134A.
\newblock \url{https://ui.adsabs.harvard.edu/abs/2005ApJ...631.1134A}

\bibitem[{Anglada-Escudé \& Butler(2012)}]{anglada-escude_harps-terra_2012}
Anglada-Escudé, G., \& Butler, R.~P. 2012, The Astrophysical Journal
  Supplement Series, 200, 15.
\newblock \url{https://doi.org/10.1088%2F0067-0049%2F200%2F2%2F15}

\bibitem[{Ansdell {et~al.}(2017)Ansdell, Williams, Manara, Miotello, Facchini,
  van~der Marel, Testi, \& van Dishoeck}]{ansdell_alma_2017}
Ansdell, M., Williams, J.~P., Manara, C.~F., {et~al.} 2017, The Astronomical
  Journal, 153, 240, aDS Bibcode: 2017AJ....153..240A.
\newblock \url{https://ui.adsabs.harvard.edu/abs/2017AJ....153..240A}

\bibitem[{Ansdell {et~al.}(2016)Ansdell, Williams, van~der Marel, Carpenter,
  Guidi, Hogerheijde, Mathews, Manara, Miotello, Natta, Oliveira, Tazzari,
  Testi, van Dishoeck, \& van Terwisga}]{ansdell_alma_2016}
Ansdell, M., Williams, J.~P., van~der Marel, N., {et~al.} 2016, The
  Astrophysical Journal, 828, 46, aDS Bibcode: 2016ApJ...828...46A.
\newblock \url{https://ui.adsabs.harvard.edu/abs/2016ApJ...828...46A}

\bibitem[{{Astropy Collaboration} {et~al.}(2018){Astropy Collaboration},
  Price-Whelan, Sipőcz, Günther, Lim, Crawford, Conseil, Shupe, Craig,
  Dencheva, Ginsburg, VanderPlas, Bradley, Pérez-Suárez, de~Val-Borro,
  Aldcroft, Cruz, Robitaille, Tollerud, Ardelean, Babej, Bach, Bachetti,
  Bakanov, Bamford, Barentsen, Barmby, Baumbach, Berry, Biscani, Boquien,
  Bostroem, Bouma, Brammer, Bray, Breytenbach, Buddelmeijer, Burke, Calderone,
  Cano~Rodríguez, Cara, Cardoso, Cheedella, Copin, Corrales, Crichton,
  D'Avella, Deil, Depagne, Dietrich, Donath, Droettboom, Earl, Erben, Fabbro,
  Ferreira, Finethy, Fox, Garrison, Gibbons, Goldstein, Gommers, Greco,
  Greenfield, Groener, Grollier, Hagen, Hirst, Homeier, Horton, Hosseinzadeh,
  Hu, Hunkeler, Ivezić, Jain, Jenness, Kanarek, Kendrew, Kern, Kerzendorf,
  Khvalko, King, Kirkby, Kulkarni, Kumar, Lee, Lenz, Littlefair, Ma, Macleod,
  Mastropietro, McCully, Montagnac, Morris, Mueller, Mumford, Muna, Murphy,
  Nelson, Nguyen, Ninan, Nöthe, Ogaz, Oh, Parejko, Parley, Pascual, Patil,
  Patil, Plunkett, Prochaska, Rastogi, Reddy~Janga, Sabater, Sakurikar,
  Seifert, Sherbert, Sherwood-Taylor, Shih, Sick, Silbiger, Singanamalla,
  Singer, Sladen, Sooley, Sornarajah, Streicher, Teuben, Thomas, Tremblay,
  Turner, Terrón, van Kerkwijk, de~la Vega, Watkins, Weaver, Whitmore,
  Woillez, Zabalza, \& {Astropy
  Contributors}}]{astropy_collaboration_astropy_2018}
{Astropy Collaboration}, Price-Whelan, A.~M., Sipőcz, B.~M., {et~al.} 2018,
  The Astronomical Journal, 156, 123.
\newblock \url{https://ui.adsabs.harvard.edu/abs/2018AJ....156..123A}

\bibitem[{Astudillo-Defru {et~al.}(2017)Astudillo-Defru, Forveille, Bonfils,
  Ségransan, Bouchy, Delfosse, Lovis, Mayor, Murgas, Pepe, Santos, Udry, \&
  Wünsche}]{astudillo-defru_harps_2017}
Astudillo-Defru, N., Forveille, T., Bonfils, X., {et~al.} 2017, Astronomy and
  Astrophysics, 602, A88.
\newblock \url{https://ui.adsabs.harvard.edu/abs/2017A&A...602A..88A/abstract}

\bibitem[{Baraffe \& Chabrier(2018)}]{baraffe_closer_2018}
Baraffe, I., \& Chabrier, G. 2018, Astronomy and Astrophysics, 619, A177.
\newblock \url{https://ui.adsabs.harvard.edu/abs/2018A&A...619A.177B/abstract}

\bibitem[{Baraffe {et~al.}(1998)Baraffe, Chabrier, Allard, \&
  Hauschildt}]{baraffe_evolutionary_1998}
Baraffe, I., Chabrier, G., Allard, F., \& Hauschildt, P.~H. 1998, Astronomy and
  Astrophysics, 337, 403.
\newblock \url{https://ui.adsabs.harvard.edu/abs/1998A&A...337..403B/abstract}

\bibitem[{{Barber} {et~al.}(2014){Barber}, {Strange}, {Hill}, {Polyansky},
  {Mellau}, {Yurchenko}, \& {Tennyson}}]{Barber2014}
{Barber}, R.~J., {Strange}, J.~K., {Hill}, C., {et~al.} 2014, \mnras, 437, 1828

\bibitem[{Barenfeld {et~al.}(2016)Barenfeld, Carpenter, Ricci, \&
  Isella}]{barenfeld_alma_2016}
Barenfeld, S.~A., Carpenter, J.~M., Ricci, L., \& Isella, A. 2016, The
  Astrophysical Journal, 827, 142, aDS Bibcode: 2016ApJ...827..142B.
\newblock \url{https://ui.adsabs.harvard.edu/abs/2016ApJ...827..142B}

\bibitem[{Batalha {et~al.}(2017)Batalha, Mandell, Pontoppidan, Stevenson,
  Lewis, Kalirai, Earl, Greene, Albert, \& Nielsen}]{batalha_pandexo_2017}
Batalha, N.~E., Mandell, A., Pontoppidan, K., {et~al.} 2017, Publications of
  the Astronomical Society of the Pacific, 129, 064501, aDS Bibcode:
  2017PASP..129f4501B.
\newblock \url{https://ui.adsabs.harvard.edu/abs/2017PASP..129f4501B}

\bibitem[{Bayliss {et~al.}(2018)Bayliss, Gillen, Eigmüller, McCormac,
  Alexander, Armstrong, Booth, Bouchy, Burleigh, Cabrera, Casewell, Chaushev,
  Chazelas, Csizmadia, Erikson, Faedi, Foxell, Gänsicke, Goad, Grange,
  Günther, Hodgkin, Jackman, Jenkins, Lambert, Louden, Metrailler, Moyano,
  Pollacco, Poppenhaeger, Queloz, Raddi, Rauer, Raynard, Smith, Soto, Thompson,
  Titz-Weider, Udry, Walker, Watson, West, \& Wheatley}]{bayliss_ngts-1b_2018}
Bayliss, D., Gillen, E., Eigmüller, P., {et~al.} 2018, Monthly Notices of the
  Royal Astronomical Society, 475, 4467, aDS Bibcode: 2018MNRAS.475.4467B.
\newblock \url{https://ui.adsabs.harvard.edu/abs/2018MNRAS.475.4467B}

\bibitem[{Beaugé \& Nesvorný(2012)}]{beauge_multiple-planet_2012}
Beaugé, C., \& Nesvorný, D. 2012, The Astrophysical Journal, 751, 119, aDS
  Bibcode: 2012ApJ...751..119B.
\newblock \url{https://ui.adsabs.harvard.edu/abs/2012ApJ...751..119B}

\bibitem[{Belokurov {et~al.}(2020)Belokurov, Penoyre, Oh, Iorio, Hodgkin,
  Evans, Everall, Koposov, Tout, Izzard, Clarke, \&
  Brown}]{belokurov_unresolved_2020}
Belokurov, V., Penoyre, Z., Oh, S., {et~al.} 2020, Monthly Notices of the Royal
  Astronomical Society, 496, 1922, aDS Bibcode: 2020MNRAS.496.1922B.
\newblock \url{https://ui.adsabs.harvard.edu/abs/2020MNRAS.496.1922B}

\bibitem[{Benedict {et~al.}(2016)Benedict, Henry, Franz, McArthur, Wasserman,
  Jao, Cargile, Dieterich, Bradley, Nelan, \& Whipple}]{benedict_solar_2016}
Benedict, G.~F., Henry, T.~J., Franz, O.~G., {et~al.} 2016, The Astronomical
  Journal, 152, 141.
\newblock \url{http://adsabs.harvard.edu/abs/2016AJ....152..141B}

\bibitem[{Bensby {et~al.}(2014)Bensby, Feltzing, \&
  Oey}]{bensby_exploring_2014}
Bensby, T., Feltzing, S., \& Oey, M.~S. 2014, Astronomy and Astrophysics, 562,
  A71.
\newblock \url{http://adsabs.harvard.edu/abs/2014A%26A...562A..71B}

\bibitem[{Bohlin {et~al.}(1978)Bohlin, Savage, \& Drake}]{bohlin_survey_1978}
Bohlin, R.~C., Savage, B.~D., \& Drake, J.~F. 1978, The Astrophysical Journal,
  224, 132, aDS Bibcode: 1978ApJ...224..132B.
\newblock \url{https://ui.adsabs.harvard.edu/abs/1978ApJ...224..132B}

\bibitem[{Bonfils {et~al.}(2005)Bonfils, Delfosse, Udry, Santos, Forveille, \&
  Ségransan}]{bonfils_metallicity_2005}
Bonfils, X., Delfosse, X., Udry, S., {et~al.} 2005, Astronomy and Astrophysics,
  Volume 442, Issue 2, November I 2005, pp.635-642, 442, 635.
\newblock
  \url{https://ui.adsabs.harvard.edu/abs/2005A%26A...442..635B/abstract}

\bibitem[{Boss(2006)}]{boss_rapid_2006}
Boss, A.~P. 2006, The Astrophysical Journal, 643, 501.
\newblock \url{http://adsabs.harvard.edu/abs/2006ApJ...643..501B}

\bibitem[{Boss(2011)}]{boss_formation_2011}
---. 2011, The Astrophysical Journal, 731, 74, aDS Bibcode:
  2011ApJ...731...74B.
\newblock \url{https://ui.adsabs.harvard.edu/abs/2011ApJ...731...74B}

\bibitem[{Bovy(2015)}]{bovy_galpy_2015}
Bovy, J. 2015, The Astrophysical Journal Supplement Series, 216, 29.
\newblock \url{http://adsabs.harvard.edu/abs/2015ApJS..216...29B}

\bibitem[{Broyden(1970)}]{broyden_convergence_1970}
Broyden, C.~G. 1970, IMA Journal of Applied Mathematics, 6, 76.
\newblock \url{https://doi.org/10.1093/imamat/6.1.76}

\bibitem[{Burn {et~al.}(2021)Burn, Schlecker, Mordasini, Emsenhuber, Alibert,
  Henning, Klahr, \& Benz}]{burn_new_2021}
Burn, R., Schlecker, M., Mordasini, C., {et~al.} 2021, Astronomy and
  Astrophysics, 656, A72.
\newblock \url{https://ui.adsabs.harvard.edu/abs/2021A&A...656A..72B/abstract}

\bibitem[{Burrows {et~al.}(2001)Burrows, Hubbard, Lunine, \&
  Liebert}]{burrows_theory_2001}
Burrows, A., Hubbard, W.~B., Lunine, J.~I., \& Liebert, J. 2001, Reviews of
  Modern Physics, 73, 719, aDS Bibcode: 2001RvMP...73..719B.
\newblock \url{https://ui.adsabs.harvard.edu/abs/2001RvMP...73..719B}

\bibitem[{Burt {et~al.}(2020)Burt, Nielsen, Quinn, Mamajek, Matthews, Zhou,
  Seidel, Huang, Lopez, Soto, Otegi, Stassun, Kreidberg, Collins, Eastman,
  Rodriguez, Vanderburg, Halverson, Teske, Wang, Butler, Bouchy, Dumusque,
  Segransen, Shectman, Crane, Feng, Montet, Feinstein, Beletski, Flowers,
  Günther, Daylan, Collins, Conti, Gan, Jensen, Kielkopf, Tan, Helled, Dorn,
  Haldemann, Lissauer, Ricker, Vanderspek, Latham, Seager, Winn, Jenkins,
  Twicken, Smith, Tenenbaum, Cartwright, Barclay, Pepper, Esquerdo, \&
  Fong}]{burt_toi-824_2020}
Burt, J.~A., Nielsen, L.~D., Quinn, S.~N., {et~al.} 2020, The Astronomical
  Journal, 160, 153.
\newblock \url{https://ui.adsabs.harvard.edu/abs/2020AJ....160..153B}

\bibitem[{Carpenter {et~al.}(2006)Carpenter, Mamajek, Hillenbrand, \&
  Meyer}]{carpenter_evidence_2006}
Carpenter, J.~M., Mamajek, E.~E., Hillenbrand, L.~A., \& Meyer, M.~R. 2006, The
  Astrophysical Journal, 651, L49, aDS Bibcode: 2006ApJ...651L..49C.
\newblock \url{https://ui.adsabs.harvard.edu/abs/2006ApJ...651L..49C}

\bibitem[{Cañas {et~al.}(2020)Cañas, Stefansson, Kanodia, Mahadevan, Cochran,
  Endl, Robertson, Bender, Ninan, Beard, Lubin, Gupta, Everett, Monson, Wilson,
  Lewis, Brewer, Majewski, Hebb, Dawson, Diddams, Ford, Fredrick, Halverson,
  Hearty, Lin, Metcalf, Rajagopal, Ramsey, Roy, Schwab, Terrien, \&
  Wright}]{canas_warm_2020}
Cañas, C.~I., Stefansson, G., Kanodia, S., {et~al.} 2020, The Astronomical
  Journal, 160, 147.
\newblock \url{http://adsabs.harvard.edu/abs/2020AJ....160..147C}

\bibitem[{Cañas {et~al.}(2022)Cañas, Kanodia, Bender, Mahadevan, Stefánsson,
  Cochran, Lin, Hwang, Powers, Monson, Green, Parker, Swaby, Kobulnicky,
  Wisniewski, Gupta, Everett, Jones, Anjakos, Beard, Blake, Diddams, Dong,
  Fredrick, Hakemiamjad, Hebb, Libby-Roberts, Logsdon, McElwain, Metcalf,
  Ninan, Rajagopal, Ramsey, Robertson, Roy, Ruhle, Schwab, Terrien, \&
  Wright}]{canas_toi-3714_2022}
Cañas, C.~I., Kanodia, S., Bender, C.~F., {et~al.} 2022, The Astronomical
  Journal, 164, 50, aDS Bibcode: 2022AJ....164...50C.
\newblock \url{https://ui.adsabs.harvard.edu/abs/2022AJ....164...50C}

\bibitem[{Chambers {et~al.}(2016)Chambers, Magnier, Metcalfe, Flewelling,
  Huber, Waters, Denneau, Draper, Farrow, Finkbeiner, Holmberg, Koppenhoefer,
  Price, Rest, Saglia, Schlafly, Smartt, Sweeney, Wainscoat, Burgett, Chastel,
  Grav, Heasley, Hodapp, Jedicke, Kaiser, Kudritzki, Luppino, Lupton, Monet,
  Morgan, Onaka, Shiao, Stubbs, Tonry, White, Bañados, Bell, Bender, Bernard,
  Boegner, Boffi, Botticella, Calamida, Casertano, Chen, Chen, Cole, Deacon,
  Frenk, Fitzsimmons, Gezari, Gibbs, Goessl, Goggia, Gourgue, Goldman, Grant,
  Grebel, Hambly, Hasinger, Heavens, Heckman, Henderson, Henning, Holman, Hopp,
  Ip, Isani, Jackson, Keyes, Koekemoer, Kotak, Le, Liska, Long, Lucey, Liu,
  Martin, Masci, McLean, Mindel, Misra, Morganson, Murphy, Obaika, Narayan,
  Nieto-Santisteban, Norberg, Peacock, Pier, Postman, Primak, Rae, Rai, Riess,
  Riffeser, Rix, Röser, Russel, Rutz, Schilbach, Schultz, Scolnic, Strolger,
  Szalay, Seitz, Small, Smith, Soderblom, Taylor, Thomson, Taylor, Thakar,
  Thiel, Thilker, Unger, Urata, Valenti, Wagner, Walder, Walter, Watters,
  Werner, Wood-Vasey, \& Wyse}]{chambers_pan-starrs1_2016}
Chambers, K.~C., Magnier, E.~A., Metcalfe, N., {et~al.} 2016, The
  {Pan}-{STARRS1} {Surveys}, Tech. rep., publication Title: arXiv e-prints ADS
  Bibcode: 2016arXiv161205560C Type: article.
\newblock \url{https://ui.adsabs.harvard.edu/abs/2016arXiv161205560C}

\bibitem[{Chonis {et~al.}(2016)Chonis, Hill, Lee, Tuttle, Vattiat, Drory,
  Indahl, Peterson, \& Ramsey}]{chonis_lrs2_2016}
Chonis, T.~S., Hill, G.~J., Lee, H., {et~al.} 2016, 9908, 99084C, conference
  Name: Ground-based and Airborne Instrumentation for Astronomy VI ADS Bibcode:
  2016SPIE.9908E..4CC.
\newblock \url{https://ui.adsabs.harvard.edu/abs/2016SPIE.9908E..4CC}

\bibitem[{Cifuentes {et~al.}(2020)Cifuentes, Caballero, Cortés-Contreras,
  Montes, Abellán, Dorda, Holgado, Zapatero~Osorio, Morales, Amado, Passegger,
  Quirrenbach, Reiners, Ribas, Sanz-Forcada, Schweitzer, Seifert, \&
  Solano}]{cifuentes_carmenes_2020}
Cifuentes, C., Caballero, J.~A., Cortés-Contreras, M., {et~al.} 2020,
  Astronomy and Astrophysics, 642, A115.
\newblock \url{https://ui.adsabs.harvard.edu/abs/2020A&A...642A.115C/abstract}

\bibitem[{Clough {et~al.}(2005)Clough, Shephard, Mlawer, Delamere, Iacono,
  Cady-Pereira, Boukabara, \& Brown}]{clough_atmospheric_2005}
Clough, S.~A., Shephard, M.~W., Mlawer, E.~J., {et~al.} 2005, Journal of
  Quantitative Spectroscopy and Radiative Transfer, 91, 233.
\newblock \url{http://adsabs.harvard.edu/abs/2005JQSRT..91..233C}

\bibitem[{Coleman \& Haworth(2020)}]{coleman_peter_2020}
Coleman, G. A.~L., \& Haworth, T.~J. 2020, Monthly Notices of the Royal
  Astronomical Society, 496, L111, aDS Bibcode: 2020MNRAS.496L.111C.
\newblock \url{https://ui.adsabs.harvard.edu/abs/2020MNRAS.496L.111C}

\bibitem[{Collins {et~al.}(2017)Collins, Kielkopf, Stassun, \&
  Hessman}]{collins_astroimagej_2017}
Collins, K.~A., Kielkopf, J.~F., Stassun, K.~G., \& Hessman, F.~V. 2017, The
  Astronomical Journal, 153, 77.
\newblock \url{http://adsabs.harvard.edu/abs/2017AJ....153...77C}

\bibitem[{Covey {et~al.}(2007)Covey, Ivezic, Schlegel, Finkbeiner, Padmanabhan,
  Lupton, Agüeros, Bochanski, Hawley, West, Seth, Kimball, Gogarten, Claire,
  Haggard, Kaib, Schneider, \& Sesar}]{covey_stellar_2007}
Covey, K.~R., Ivezic, Z., Schlegel, D., {et~al.} 2007, The Astronomical
  Journal, 134, 2398, aDS Bibcode: 2007AJ....134.2398C.
\newblock \url{https://ui.adsabs.harvard.edu/abs/2007AJ....134.2398C}

\bibitem[{Creevey {et~al.}(2022)Creevey, Sordo, Pailler, Frémat, Heiter,
  Thévenin, Andrae, Fouesneau, Lobel, Bailer-Jones, Garabato, Bellas-Velidis,
  Brugaletta, Lorca, Ordenovic, Palicio, Sarro, Delchambre, Drimmel, Rybizki,
  Torralba~Elipe, Korn, Recio-Blanco, Schultheis, De~Angeli, Montegriffo,
  Abreu~Aramburu, Accart, Álvarez, Bakker, Brouillet, Burlacu, Carballo,
  Casamiquela, Chiavassa, Contursi, Cooper, Dafonte, Dapergolas, de~Laverny,
  Dharmawardena, Edvardsson, Le~Fustec, García-Lario, García-Torres, Gomez,
  González-Santamaría, Hatzidimitriou, Jean-Antoine~Piccolo, Kontizas,
  Kordopatis, Lanzafame, Lebreton, Licata, Lindstrøm, Livanou,
  Magdaleno~Romeo, Manteiga, Marocco, Marshall, Mary, Nicolas, Pallas-Quintela,
  Panem, Pichon, Poggio, Riclet, Robin, Santoveña, Silvelo, Slezak, Smart,
  Soubiran, Süveges, Ulla, Utrilla, Vallenari, Zhao, Zorec, Barrado, Bijaoui,
  Bouret, Blomme, Brott, Cassisi, Kochukhov, Martayan, Shulyak, \&
  Silvester}]{creevey_gaia_2022}
Creevey, O.~L., Sordo, R., Pailler, F., {et~al.} 2022, Gaia {Data} {Release} 3:
  {Astrophysical} parameters inference system ({Apsis}) {I} -- methods and
  content overview, Tech. rep., publication Title: arXiv e-prints ADS Bibcode:
  2022arXiv220605864C Type: article.
\newblock \url{https://ui.adsabs.harvard.edu/abs/2022arXiv220605864C}

\bibitem[{Dash {et~al.}(2022)Dash, Majumdar, Willacy, Tsai, Turner, Rimmer,
  Gudipati, Lyra, \& Bhardwaj}]{dash_linking_2022}
Dash, S., Majumdar, L., Willacy, K., {et~al.} 2022, Linking atmospheric
  chemistry of the hot {Jupiter} {HD} 209458b to its formation location through
  infrared transmission and emission spectra, Tech. rep., publication Title:
  arXiv e-prints ADS Bibcode: 2022arXiv220404103D Type: article.
\newblock \url{https://ui.adsabs.harvard.edu/abs/2022arXiv220404103D}

\bibitem[{Dawson \& Johnson(2012)}]{dawson_photoeccentric_2012}
Dawson, R.~I., \& Johnson, J.~A. 2012, The Astrophysical Journal, 756, 122, aDS
  Bibcode: 2012ApJ...756..122D.
\newblock \url{https://ui.adsabs.harvard.edu/abs/2012ApJ...756..122D}

\bibitem[{Dawson \& Johnson(2018)}]{dawson_origins_2018}
---. 2018, Annual Review of Astronomy and Astrophysics, 56, 175.
\newblock \url{http://adsabs.harvard.edu/abs/2018ARA%26A..56..175D}

\bibitem[{Delfosse {et~al.}(2000)Delfosse, Forveille, Ségransan, Beuzit, Udry,
  Perrier, \& Mayor}]{delfosse_accurate_2000}
Delfosse, X., Forveille, T., Ségransan, D., {et~al.} 2000, Astronomy and
  Astrophysics, 364, 217.
\newblock \url{https://ui.adsabs.harvard.edu/abs/2000A&A...364..217D/abstract}

\bibitem[{Eisner {et~al.}(2018)Eisner, Arce, Ballering, Bally, Andrews, Boyden,
  Di~Francesco, Fang, Johnstone, Kim, Mann, Matthews, Pascucci, Ricci, Sheehan,
  \& Williams}]{eisner_protoplanetary_2018}
Eisner, J.~A., Arce, H.~G., Ballering, N.~P., {et~al.} 2018, The Astrophysical
  Journal, 860, 77, aDS Bibcode: 2018ApJ...860...77E.
\newblock \url{https://ui.adsabs.harvard.edu/abs/2018ApJ...860...77E}

\bibitem[{Feiden {et~al.}(2021)Feiden, Skidmore, \& Jao}]{feiden_gaia_2021}
Feiden, G.~A., Skidmore, K., \& Jao, W.-C. 2021, The Astrophysical Journal,
  907, 53, aDS Bibcode: 2021ApJ...907...53F.
\newblock \url{https://ui.adsabs.harvard.edu/abs/2021ApJ...907...53F}

\bibitem[{Feinstein {et~al.}(2019)Feinstein, Montet, Foreman-Mackey, Bedell,
  Saunders, Bean, Christiansen, Hedges, Luger, Scolnic, \&
  Cardoso}]{feinstein_eleanor_2019}
Feinstein, A.~D., Montet, B.~T., Foreman-Mackey, D., {et~al.} 2019,
  Publications of the Astronomical Society of the Pacific, 131, 094502.
\newblock \url{https://ui.adsabs.harvard.edu/abs/2019PASP..131i4502F}

\bibitem[{Feng {et~al.}(2020)Feng, Shectman, Clement, Vogt, Tuomi, Teske, Burt,
  Crane, Holden, Wang, Thompson, Diaz, \& Butler}]{feng_search_2020}
Feng, F., Shectman, S.~A., Clement, M.~S., {et~al.} 2020, arXiv:2008.07998
  [astro-ph], arXiv: 2008.07998.
\newblock \url{http://arxiv.org/abs/2008.07998}

\bibitem[{Fischer \& Valenti(2005)}]{fischer_planet-metallicity_2005}
Fischer, D.~A., \& Valenti, J. 2005, The Astrophysical Journal, 622, 1102, aDS
  Bibcode: 2005ApJ...622.1102F.
\newblock \url{https://ui.adsabs.harvard.edu/abs/2005ApJ...622.1102F}

\bibitem[{Fletcher(1970)}]{fletcher_new_1970}
Fletcher, R. 1970, The Computer Journal, 13, 317.
\newblock \url{https://doi.org/10.1093/comjnl/13.3.317}

\bibitem[{Ford(2006)}]{ford_improving_2006}
Ford, E.~B. 2006, The Astrophysical Journal, 642, 505.
\newblock \url{http://adsabs.harvard.edu/abs/2006ApJ...642..505F}

\bibitem[{Foreman-Mackey(2018)}]{foreman-mackey_scalable_2018}
Foreman-Mackey, D. 2018, Research Notes of the American Astronomical Society,
  2, 31

\bibitem[{Foreman-Mackey {et~al.}(2017)Foreman-Mackey, Agol, Ambikasaran, \&
  Angus}]{foreman-mackey_fast_2017}
Foreman-Mackey, D., Agol, E., Ambikasaran, S., \& Angus, R. 2017, The
  Astronomical Journal, 154, 220, arXiv: 1703.09710.
\newblock \url{http://arxiv.org/abs/1703.09710}

\bibitem[{Foreman-Mackey {et~al.}(2021{\natexlab{a}})Foreman-Mackey, Savel,
  Luger, Czekala, Agol, Price-Whelan, Gilbert, Brandt, Barclay, \&
  Bouma}]{foreman-mackey_exoplanet-devexoplanet_2021}
Foreman-Mackey, D., Savel, A., Luger, R., {et~al.} 2021{\natexlab{a}},
  exoplanet-dev/exoplanet v0.4.4, , , doi:10.5281/zenodo.1998447.
\newblock \url{https://doi.org/10.5281/zenodo.1998447}

\bibitem[{Foreman-Mackey {et~al.}(2021{\natexlab{b}})Foreman-Mackey, Luger,
  Agol, Barclay, Bouma, Brandt, Czekala, David, Dong, Gilbert, Gordon, Hedges,
  Hey, Morris, Price-Whelan, \& Savel}]{foreman-mackey_exoplanet_2021}
Foreman-Mackey, D., Luger, R., Agol, E., {et~al.} 2021{\natexlab{b}}, The
  Journal of Open Source Software, 6, 3285.
\newblock \url{https://ui.adsabs.harvard.edu/abs/2021JOSS....6.3285F}

\bibitem[{Gagné {et~al.}(2018)Gagné, Mamajek, Malo, Riedel, Rodriguez,
  Lafrenière, Faherty, Roy-Loubier, Pueyo, Robin, \&
  Doyon}]{gagne_banyan_2018}
Gagné, J., Mamajek, E.~E., Malo, L., {et~al.} 2018, The Astrophysical Journal,
  856, 23.
\newblock \url{http://adsabs.harvard.edu/abs/2018ApJ...856...23G}

\bibitem[{Gaidos {et~al.}(2022)Gaidos, Mann, Rojas-Ayala, Feiden, Wood,
  Narayanan, Ansdell, Jacobs, \& LaCourse}]{gaidos_planetesimals_2022}
Gaidos, E., Mann, A.~W., Rojas-Ayala, B., {et~al.} 2022, Monthly Notices of the
  Royal Astronomical Society, 514, 1386, aDS Bibcode: 2022MNRAS.514.1386G.
\newblock \url{https://ui.adsabs.harvard.edu/abs/2022MNRAS.514.1386G}

\bibitem[{Gan {et~al.}(2022)Gan, Lin, Wang, Mao, Fouqué, Fan, Bedell, Stassun,
  Giacalone, Fukui, Murgas, Ciardi, Howell, Collins, Shporer, Arnold, Barclay,
  Charbonneau, Christiansen, Crossfield, Dressing, Elliott, Esparza-Borges,
  Evans, Gnilka, Gonzales, Howard, Isogai, Kawauchi, Kurita, Liu, Livingston,
  Matson, Narita, Palle, Parviainen, Rackham, Rodriguez, Rose, Rudat,
  Schlieder, Scott, Vezie, Ricker, Vanderspek, Latham, Seager, Winn, \&
  Jenkins}]{gan_toi-530b_2022}
Gan, T., Lin, Z., Wang, S.~X., {et~al.} 2022, Monthly Notices of the Royal
  Astronomical Society, 511, 83, aDS Bibcode: 2022MNRAS.511...83G.
\newblock \url{https://ui.adsabs.harvard.edu/abs/2022MNRAS.511...83G}

\bibitem[{Gandhi {et~al.}(2021)Gandhi, Buckley, Charles, Hodgkin, Scaringi,
  Knigge, Rao, Paice, \& Zhao}]{gandhi_astrometric_2021}
Gandhi, P., Buckley, D. A.~H., Charles, P.~A., {et~al.} 2021, Monthly Notices
  of the Royal Astronomical Society, doi:10.1093/mnras/stab3771, aDS Bibcode:
  2021MNRAS.tmp.3450G.
\newblock \url{https://ui.adsabs.harvard.edu/abs/2021MNRAS.tmp.3450G}

\bibitem[{Gandhi \& Madhusudhan(2017)}]{gandhi_genesis_2017}
Gandhi, S., \& Madhusudhan, N. 2017, Monthly Notices of the Royal Astronomical
  Society, 472, 2334, aDS Bibcode: 2017MNRAS.472.2334G.
\newblock \url{https://ui.adsabs.harvard.edu/abs/2017MNRAS.472.2334G}

\bibitem[{Gandhi \& Madhusudhan(2019)}]{gandhi_new_2019}
---. 2019, Monthly Notices of the Royal Astronomical Society, 485, 5817, aDS
  Bibcode: 2019MNRAS.485.5817G.
\newblock \url{https://ui.adsabs.harvard.edu/abs/2019MNRAS.485.5817G}

\bibitem[{Ghezzi {et~al.}(2010)Ghezzi, Cunha, Smith, de~Araújo, Schuler, \&
  de~la Reza}]{ghezzi_stellar_2010}
Ghezzi, L., Cunha, K., Smith, V.~V., {et~al.} 2010, The Astrophysical Journal,
  720, 1290.
\newblock \url{http://adsabs.harvard.edu/abs/2010ApJ...720.1290G}

\bibitem[{Ghezzi {et~al.}(2018)Ghezzi, Montet, \&
  Johnson}]{ghezzi_retired_2018}
Ghezzi, L., Montet, B.~T., \& Johnson, J.~A. 2018, The Astrophysical Journal,
  860, 109, aDS Bibcode: 2018ApJ...860..109G.
\newblock \url{https://ui.adsabs.harvard.edu/abs/2018ApJ...860..109G}

\bibitem[{Ginsburg {et~al.}(2019)Ginsburg, Sipőcz, Brasseur, Cowperthwaite,
  Craig, Deil, Guillochon, Guzman, Liedtke, Lian~Lim, Lockhart, Mommert,
  Morris, Norman, Parikh, Persson, Robitaille, Segovia, Singer, Tollerud,
  de~Val-Borro, Valtchanov, Woillez, {Astroquery Collaboration}, \& {a subset
  of astropy Collaboration}}]{ginsburg_astroquery_2019}
Ginsburg, A., Sipőcz, B.~M., Brasseur, C.~E., {et~al.} 2019, The Astronomical
  Journal, 157, 98.
\newblock \url{http://adsabs.harvard.edu/abs/2019AJ....157...98G}

\bibitem[{Goldfarb(1970)}]{goldfarb_family_1970}
Goldfarb, D. 1970, Mathematics of Computation, 24, 23.
\newblock \url{https://www.ams.org/mcom/1970-24-109/S0025-5718-1970-0258249-6/}

\bibitem[{Gonzalez(1997)}]{gonzalez_stellar_1997}
Gonzalez, G. 1997, Monthly Notices of the Royal Astronomical Society, 285, 403.
\newblock \url{http://adsabs.harvard.edu/abs/1997MNRAS.285..403G}

\bibitem[{Greaves \& Rice(2010)}]{greaves_have_2010}
Greaves, J.~S., \& Rice, W. K.~M. 2010, Monthly Notices of the Royal
  Astronomical Society, 407, 1981, aDS Bibcode: 2010MNRAS.407.1981G.
\newblock \url{https://ui.adsabs.harvard.edu/abs/2010MNRAS.407.1981G}

\bibitem[{Gullikson {et~al.}(2014)Gullikson, Dodson-Robinson, \&
  Kraus}]{gullikson_correcting_2014}
Gullikson, K., Dodson-Robinson, S., \& Kraus, A. 2014, The Astronomical
  Journal, 148, 53.
\newblock \url{http://adsabs.harvard.edu/abs/2014AJ....148...53G}

\bibitem[{Harrington(1952)}]{harrington_48-inch_1952}
Harrington, R.~G. 1952, Publications of the Astronomical Society of the
  Pacific, 64, 275.
\newblock \url{http://adsabs.harvard.edu/abs/1952PASP...64..275H}

\bibitem[{{Harris} {et~al.}(2006){Harris}, {Tennyson}, {Kaminsky}, {Pavlenko},
  \& {Jones}}]{Harris2006}
{Harris}, G.~J., {Tennyson}, J., {Kaminsky}, B.~M., {Pavlenko}, Y.~V., \&
  {Jones}, H.~R.~A. 2006, \mnras, 367, 400

\bibitem[{Hartman {et~al.}(2015)Hartman, Bayliss, Brahm, Bakos, Mancini,
  Jordán, Penev, Rabus, Zhou, Butler, Espinoza, de~Val-Borro, Bhatti, Csubry,
  Ciceri, Henning, Schmidt, Arriagada, Shectman, Crane, Thompson, Suc, Csák,
  Tan, Noyes, Lázár, Papp, \& Sári}]{hartman_hats-6b_2015}
Hartman, J.~D., Bayliss, D., Brahm, R., {et~al.} 2015, The Astronomical
  Journal, 149, 166, aDS Bibcode: 2015AJ....149..166H.
\newblock \url{https://ui.adsabs.harvard.edu/abs/2015AJ....149..166H}

\bibitem[{Helled \& Bodenheimer(2010)}]{helled_metallicity_2010}
Helled, R., \& Bodenheimer, P. 2010, Icarus, 207, 503, aDS Bibcode:
  2010Icar..207..503H.
\newblock \url{https://ui.adsabs.harvard.edu/abs/2010Icar..207..503H}

\bibitem[{Helled \& Lunine(2014)}]{helled_measuring_2014}
Helled, R., \& Lunine, J. 2014, Monthly Notices of the Royal Astronomical
  Society, 441, 2273, aDS Bibcode: 2014MNRAS.441.2273H.
\newblock \url{https://ui.adsabs.harvard.edu/abs/2014MNRAS.441.2273H}

\bibitem[{Helled \& Morbidelli(2021)}]{helled_planet_2021}
Helled, R., \& Morbidelli, A. 2021, Planet {Formation}, Tech. rep., publication
  Title: arXiv e-prints ADS Bibcode: 2021arXiv210907790H Type: article.
\newblock \url{https://ui.adsabs.harvard.edu/abs/2021arXiv210907790H}

\bibitem[{Helled {et~al.}(2014)Helled, Bodenheimer, Podolak, Boley, Meru,
  Nayakshin, Fortney, Mayer, Alibert, \& Boss}]{helled_giant_2014}
Helled, R., Bodenheimer, P., Podolak, M., {et~al.} 2014, Giant planet
  formation, evolution, and internal structure, ed. H.~Beuther, R.~S. Klessen,
  C.~P. Dullemond, \& T.~Henning, arXiv: 1311.1142 [astro-ph.EP] tex.adsnote:
  Provided by the SAO/NASA Astrophysics Data System tex.adsurl:
  https://ui.adsabs.harvard.edu/abs/2014prpl.conf..643H

\bibitem[{Heng \& Tsai(2016)}]{heng_analytical_2016}
Heng, K., \& Tsai, S.-M. 2016, The Astrophysical Journal, 829, 104, aDS
  Bibcode: 2016ApJ...829..104H.
\newblock \url{https://ui.adsabs.harvard.edu/abs/2016ApJ...829..104H}

\bibitem[{Henry {et~al.}(2006)Henry, Jao, Subasavage, Beaulieu, Ianna, Costa,
  \& Méndez}]{henry_solar_2006}
Henry, T.~J., Jao, W.-C., Subasavage, J.~P., {et~al.} 2006, The Astronomical
  Journal, 132, 2360.
\newblock \url{http://adsabs.harvard.edu/abs/2006AJ....132.2360H}

\bibitem[{Henry \& McCarthy(1993)}]{henry_mass-luminosity_1993}
Henry, T.~J., \& McCarthy, Jr., D.~W. 1993, The Astronomical Journal, 106, 773,
  aDS Bibcode: 1993AJ....106..773H.
\newblock \url{https://ui.adsabs.harvard.edu/abs/1993AJ....106..773H}

\bibitem[{Hobbs {et~al.}(2022)Hobbs, Shorttle, {Madhusudhan}, \&
  {Nikku}}]{hobbs_molecular_2022}
Hobbs, R., Shorttle, O., {Madhusudhan}, \& {Nikku}. 2022, Monthly Notices of
  the Royal Astronomical Society, doi:10.1093/mnras/stac2106, aDS Bibcode:
  2022MNRAS.tmp.2036H.
\newblock \url{https://ui.adsabs.harvard.edu/abs/2022MNRAS.tmp.2036H}

\bibitem[{Hoffman \& Gelman(2014)}]{hoffman_no-u-turn_2014}
Hoffman, M.~D., \& Gelman, A. 2014, Journal of Machine Learning Research, 15,
  1593.
\newblock \url{http://jmlr.org/papers/v15/hoffman14a.html}

\bibitem[{Howell {et~al.}(2011)Howell, Everett, Sherry, Horch, \&
  Ciardi}]{howell_speckle_2011}
Howell, S.~B., Everett, M.~E., Sherry, W., Horch, E., \& Ciardi, D.~R. 2011,
  The Astronomical Journal, 142, 19, aDS Bibcode: 2011AJ....142...19H.
\newblock \url{https://ui.adsabs.harvard.edu/abs/2011AJ....142...19H}

\bibitem[{Huang {et~al.}(2020)Huang, Vanderburg, Pál, Sha, Yu, Fong,
  Fausnaugh, Shporer, Guerrero, Vanderspek, \& Ricker}]{huang_photometry_2020}
Huang, C.~X., Vanderburg, A., Pál, A., {et~al.} 2020, Research Notes of the
  AAS, 4, 204, publisher: American Astronomical Society.
\newblock \url{https://doi.org/10.3847/2515-5172/abca2e}

\bibitem[{Huehnerhoff {et~al.}(2016)Huehnerhoff, Ketzeback, Bradley, Dembicky,
  Doughty, Hawley, Johnson, Klaene, Leon, McMillan, Owen, Sayres, Sheen, \&
  Shugart}]{huehnerhoff_astrophysical_2016}
Huehnerhoff, J., Ketzeback, W., Bradley, A., {et~al.} 2016, 9908, 99085H,
  conference Name: Ground-based and Airborne Instrumentation for Astronomy VI.
\newblock \url{http://adsabs.harvard.edu/abs/2016SPIE.9908E..5HH}

\bibitem[{Hunter(2007)}]{hunter_matplotlib_2007}
Hunter, J.~D. 2007, Computing in Science Engineering, 9, 90

\bibitem[{Ida \& Lin(2005)}]{ida_toward_2005}
Ida, S., \& Lin, D. N.~C. 2005, The Astrophysical Journal, 626, 1045.
\newblock \url{http://adsabs.harvard.edu/abs/2005ApJ...626.1045I}

\bibitem[{Jao \& Feiden(2020)}]{jao_fine_2020}
Jao, W.-C., \& Feiden, G.~A. 2020, The Astronomical Journal, 160, 102, aDS
  Bibcode: 2020AJ....160..102J.
\newblock \url{https://ui.adsabs.harvard.edu/abs/2020AJ....160..102J}

\bibitem[{Jao {et~al.}(2018)Jao, Henry, Gies, \& Hambly}]{jao_gap_2018}
Jao, W.-C., Henry, T.~J., Gies, D.~R., \& Hambly, N.~C. 2018, The Astrophysical
  Journal, 861, L11, aDS Bibcode: 2018ApJ...861L..11J.
\newblock \url{https://ui.adsabs.harvard.edu/abs/2018ApJ...861L..11J}

\bibitem[{Jenkins {et~al.}(2010)Jenkins, Caldwell, Chandrasekaran, Twicken,
  Bryson, Quintana, Clarke, Li, Allen, Tenenbaum, Wu, Klaus, Middour, Cote,
  McCauliff, Girouard, Gunter, Wohler, Sommers, Hall, Uddin, Wu, Bhavsar,
  Van~Cleve, Pletcher, Dotson, Haas, Gilliland, Koch, \&
  Borucki}]{jenkins_overview_2010}
Jenkins, J.~M., Caldwell, D.~A., Chandrasekaran, H., {et~al.} 2010, The
  Astrophysical Journal, 713, L87, aDS Bibcode: 2010ApJ...713L..87J.
\newblock \url{https://ui.adsabs.harvard.edu/abs/2010ApJ...713L..87J}

\bibitem[{Johnson \& Soderblom(1987)}]{johnson_calculating_1987}
Johnson, D. R.~H., \& Soderblom, D.~R. 1987, The Astronomical Journal, 93, 864.
\newblock \url{https://ui.adsabs.harvard.edu/abs/1987AJ.....93..864J/abstract}

\bibitem[{Johnson {et~al.}(2010)Johnson, Howard, Marcy, Bowler, Henry, Fischer,
  Apps, Isaacson, \& Wright}]{johnson_california_2010}
Johnson, J.~A., Howard, A.~W., Marcy, G.~W., {et~al.} 2010, Publications of the
  Astronomical Society of the Pacific, 122, 149, aDS Bibcode:
  2010PASP..122..149J.
\newblock \url{https://ui.adsabs.harvard.edu/abs/2010PASP..122..149J}

\bibitem[{Johnson {et~al.}(2012)Johnson, Gazak, Apps, Muirhead, Crepp,
  Crossfield, Boyajian, von Braun, Rojas-Ayala, Howard, Covey, Schlawin,
  Hamren, Morton, Marcy, \& Lloyd}]{johnson_characterizing_2012}
Johnson, J.~A., Gazak, J.~Z., Apps, K., {et~al.} 2012, The Astronomical
  Journal, 143, 111, aDS Bibcode: 2012AJ....143..111J.
\newblock \url{https://ui.adsabs.harvard.edu/abs/2012AJ....143..111J}

\bibitem[{Jordán {et~al.}(2022)Jordán, Hartman, Bayliss, Bakos, Brahm,
  Bryant, Csubry, Henning, Hobson, Mancini, Penev, Rabus, Suc, de~Val-Borro,
  Wallace, Barkaoui, Ciardi, Collins, Esparza-Borges, Furlan, Gan, Benkhaldoun,
  Ghachoui, Gillon, Howell, Jehin, Fukui, Kawauchi, Livingston, Luque, Matson,
  Matthews, Osborn, Murgas, Narita, Palle, Parvianen, \&
  Waalkes}]{jordan_hats-74ab_2022}
Jordán, A., Hartman, J.~D., Bayliss, D., {et~al.} 2022, The Astronomical
  Journal, 163, 125, aDS Bibcode: 2022AJ....163..125J.
\newblock \url{https://ui.adsabs.harvard.edu/abs/2022AJ....163..125J}

\bibitem[{Kanodia \& Wright(2018)}]{kanodia_python_2018}
Kanodia, S., \& Wright, J. 2018, Research Notes of the AAS, 2, 4, publisher:
  American Astronomical Society.
\newblock \url{https://doi.org/10.3847/2515-5172/aaa4b7}

\bibitem[{Kanodia {et~al.}(2018)Kanodia, Mahadevan, Ramsey, Stefansson, Monson,
  Hearty, Blakeslee, Lubar, Bender, Ninan, Sterner, Roy, Halverson, \&
  Robertson}]{kanodia_overview_2018}
Kanodia, S., Mahadevan, S., Ramsey, L.~W., {et~al.} 2018, SPIE Proceedings,
  0702, 107026Q, conference Name: Ground-based and Airborne Instrumentation for
  Astronomy VII ISBN: 9781510619579 Place: eprint: arXiv:1808.00557.
\newblock \url{http://adsabs.harvard.edu/abs/2018SPIE10702E..6QK}

\bibitem[{Kanodia {et~al.}(2020)Kanodia, Cañas, Stefansson, Ninan, Hebb, Lin,
  Baran, Maney, Terrien, Mahadevan, Cochran, Endl, Dong, Bender, Diddams, Ford,
  Fredrick, Halverson, Hearty, Metcalf, Monson, Ramsey, Robertson, Roy, Schwab,
  \& Wright}]{kanodia_toi-1728b_2020}
Kanodia, S., Cañas, C.~I., Stefansson, G., {et~al.} 2020, The Astrophysical
  Journal, 899, 29.
\newblock \url{http://adsabs.harvard.edu/abs/2020ApJ...899...29K}

\bibitem[{Kanodia {et~al.}(2022)Kanodia, Libby-Roberts, Cañas, Ninan,
  Mahadevan, Stefansson, Lin, Jones, Monson, Parker, Kobulnicky, Swaby, Powers,
  Beard, Bender, Blake, Cochran, Dong, Diddams, Fredrick, Gupta, Halverson,
  Hearty, Logsdon, Metcalf, McElwain, Morley, Rajagopal, Ramsey, Robertson,
  Roy, Schwab, Terrien, Wisniewski, \& Wright}]{kanodia_toi-3757_2022}
Kanodia, S., Libby-Roberts, J., Cañas, C.~I., {et~al.} 2022, The Astronomical
  Journal, 164, 81, aDS Bibcode: 2022AJ....164...81K.
\newblock \url{https://ui.adsabs.harvard.edu/abs/2022AJ....164...81K}

\bibitem[{Kasper {et~al.}(2016)Kasper, Ellis, Yeigh, Kobulnicky, Jang-Condell,
  Kelley, Bucher, \& Weger}]{kasper_remote_2016}
Kasper, D.~H., Ellis, T.~G., Yeigh, R.~R., {et~al.} 2016, Publications of the
  Astronomical Society of the Pacific, 128, 105005.
\newblock \url{https://ui.adsabs.harvard.edu/abs/2016PASP..128j5005K}

\bibitem[{Kempton {et~al.}(2017)Kempton, Lupu, Owusu-Asare, Slough, \&
  Cale}]{kempton_exo-transmit_2017}
Kempton, E. M.~R., Lupu, R., Owusu-Asare, A., Slough, P., \& Cale, B. 2017,
  Publications of the Astronomical Society of the Pacific, 129, 044402, aDS
  Bibcode: 2017PASP..129d4402K.
\newblock \url{https://ui.adsabs.harvard.edu/abs/2017PASP..129d4402K}

\bibitem[{Kempton {et~al.}(2018)Kempton, Bean, Louie, Deming, Koll, Mansfield,
  Christiansen, López-Morales, Swain, Zellem, Ballard, Barclay, Barstow,
  Batalha, Beatty, Berta-Thompson, Birkby, Buchhave, Charbonneau, Cowan,
  Crossfield, de~Val-Borro, Doyon, Dragomir, Gaidos, Heng, Hu, Kane, Kreidberg,
  Mallonn, Morley, Narita, Nascimbeni, Pallé, Quintana, Rauscher, Seager,
  Shkolnik, Sing, Sozzetti, Stassun, Valenti, \& von
  Essen}]{kempton_framework_2018}
Kempton, E. M.~R., Bean, J.~L., Louie, D.~R., {et~al.} 2018, Publications of
  the Astronomical Society of the Pacific, 130, 114401, aDS Bibcode:
  2018PASP..130k4401K.
\newblock \url{https://ui.adsabs.harvard.edu/abs/2018PASP..130k4401K}

\bibitem[{Kervella {et~al.}(2019)Kervella, Arenou, Mignard, \&
  Thévenin}]{kervella_stellar_2019}
Kervella, P., Arenou, F., Mignard, F., \& Thévenin, F. 2019, Astronomy and
  Astrophysics, 623, A72.
\newblock \url{http://adsabs.harvard.edu/abs/2019A%26A...623A..72K}

\bibitem[{Kiman {et~al.}(2019)Kiman, Schmidt, Angus, Cruz, Faherty, \&
  Rice}]{kiman_exploring_2019}
Kiman, R., Schmidt, S.~J., Angus, R., {et~al.} 2019, The Astronomical Journal,
  157, 231, aDS Bibcode: 2019AJ....157..231K.
\newblock \url{https://ui.adsabs.harvard.edu/abs/2019AJ....157..231K}

\bibitem[{Kipping(2013)}]{kipping_efficient_2013}
Kipping, D.~M. 2013, {\textbackslash}mnras, 435, 2152

\bibitem[{Kirkpatrick {et~al.}(1991)Kirkpatrick, Henry, \&
  McCarthy}]{kirkpatrick_standard_1991}
Kirkpatrick, J.~D., Henry, T.~J., \& McCarthy, Jr., D.~W. 1991, The
  Astrophysical Journal Supplement Series, 77, 417, aDS Bibcode:
  1991ApJS...77..417K.
\newblock \url{https://ui.adsabs.harvard.edu/abs/1991ApJS...77..417K}

\bibitem[{Kley \& Nelson(2012)}]{kley_planet-disk_2012}
Kley, W., \& Nelson, R.~P. 2012, Annual Review of Astronomy and Astrophysics,
  vol. 50, p.211-249, 50, 211.
\newblock
  \url{https://ui.adsabs.harvard.edu/abs/2012ARA%26A..50..211K/abstract}

\bibitem[{Knierim {et~al.}(2022)Knierim, Shibata, \&
  Helled}]{knierim_constraining_2022}
Knierim, H., Shibata, S., \& Helled, R. 2022, Constraining the {Origin} of
  {Giant} {Exoplanets} via {Elemental} {Abundance} {Measurements},  arXiv,
  arXiv:2209.01240 [astro-ph], doi:10.48550/arXiv.2209.01240.
\newblock \url{http://arxiv.org/abs/2209.01240}

\bibitem[{Kochanek {et~al.}(2017)Kochanek, Shappee, Stanek, Holoien, Thompson,
  Prieto, Dong, Shields, Will, Britt, Perzanowski, \&
  Pojmański}]{kochanek_all-sky_2017}
Kochanek, C.~S., Shappee, B.~J., Stanek, K.~Z., {et~al.} 2017, Publications of
  the Astronomical Society of the Pacific, 129, 104502, aDS Bibcode:
  2017PASP..129j4502K.
\newblock \url{https://ui.adsabs.harvard.edu/abs/2017PASP..129j4502K}

\bibitem[{Kroupa \& Tout(1997)}]{kroupa_theoretical_1997}
Kroupa, P., \& Tout, C.~A. 1997, Monthly Notices of the Royal Astronomical
  Society, 287, 402, aDS Bibcode: 1997MNRAS.287..402K.
\newblock \url{https://ui.adsabs.harvard.edu/abs/1997MNRAS.287..402K}

\bibitem[{Kroupa {et~al.}(1990)Kroupa, Tout, \&
  Gilmore}]{kroupa_low-luminosity_1990}
Kroupa, P., Tout, C.~A., \& Gilmore, G. 1990, Monthly Notices of the Royal
  Astronomical Society, 244, 76, aDS Bibcode: 1990MNRAS.244...76K.
\newblock \url{https://ui.adsabs.harvard.edu/abs/1990MNRAS.244...76K}

\bibitem[{Kumar {et~al.}(2019)Kumar, Carroll, Hartikainen, \&
  Martin}]{kumar_arviz_2019}
Kumar, R., Carroll, C., Hartikainen, A., \& Martin, O.~A. 2019, The Journal of
  Open Source Software, doi:10.21105/joss.01143.
\newblock \url{http://joss.theoj.org/papers/10.21105/joss.01143}

\bibitem[{Kunimoto {et~al.}(2022)Kunimoto, Daylan, Guerrero, Fong, Bryson,
  Ricker, Fausnaugh, Huang, Sha, Shporer, Vanderburg, Vanderspek, \&
  Yu}]{kunimoto_tess_2022}
Kunimoto, M., Daylan, T., Guerrero, N., {et~al.} 2022, The Astrophysical
  Journal Supplement Series, 259, 33, aDS Bibcode: 2022ApJS..259...33K.
\newblock \url{https://ui.adsabs.harvard.edu/abs/2022ApJS..259...33K}

\bibitem[{Laughlin {et~al.}(2004)Laughlin, Bodenheimer, \&
  Adams}]{laughlin_core_2004}
Laughlin, G., Bodenheimer, P., \& Adams, F.~C. 2004, The Astrophysical Journal
  Letters, 612, L73.
\newblock \url{http://adsabs.harvard.edu/abs/2004ApJ...612L..73L}

\bibitem[{Lee {et~al.}(2010)Lee, Chonis, Hill, DePoy, Marshall, \&
  Vattiat}]{lee_lrs2_2010}
Lee, H., Chonis, T.~S., Hill, G.~J., {et~al.} 2010, in Ground-based and
  {Airborne} {Instrumentation} for {Astronomy} {III}, Vol. 7735 (International
  Society for Optics and Photonics), 77357H.
\newblock
  \url{https://www.spiedigitallibrary.org/conference-proceedings-of-spie/7735/77357H/LRS2--a-new-low-resolution-spectrograph-for-the-Hobby/10.1117/12.857201.short}

\bibitem[{Lee {et~al.}(2020)Lee, Song, \& Murphy}]{lee_2mass_2020}
Lee, J., Song, I., \& Murphy, S. 2020, Monthly Notices of the Royal
  Astronomical Society, 494, 62, aDS Bibcode: 2020MNRAS.494...62L.
\newblock \url{https://ui.adsabs.harvard.edu/abs/2020MNRAS.494...62L}

\bibitem[{{Lightkurve Collaboration} {et~al.}(2018){Lightkurve Collaboration},
  Cardoso, Hedges, Gully-Santiago, Saunders, Cody, Barclay, Hall, Sagear,
  Turtelboom, Zhang, Tzanidakis, Mighell, Coughlin, Bell, Berta-Thompson,
  Williams, Dotson, \& Barentsen}]{lightkurve_collaboration_lightkurve_2018}
{Lightkurve Collaboration}, Cardoso, J. V. d.~M., Hedges, C., {et~al.} 2018,
  Astrophysics Source Code Library, ascl:1812.013.
\newblock \url{https://ui.adsabs.harvard.edu/abs/2018ascl.soft12013L}

\bibitem[{Limber(1958)}]{limber_structure_1958}
Limber, D.~N. 1958, The Astrophysical Journal, 127, 363, aDS Bibcode:
  1958ApJ...127..363L.
\newblock \url{https://ui.adsabs.harvard.edu/abs/1958ApJ...127..363L}

\bibitem[{Limber(1960)}]{limber_dwarf_1960}
---. 1960, Leaflet of the Astronomical Society of the Pacific, 8, 127.
\newblock \url{http://adsabs.harvard.edu/abs/1960ASPL....8..127L}

\bibitem[{Lin {et~al.}(2018)Lin, Lee, \& Chiang}]{lin_balanced_2018}
Lin, J.~W., Lee, E.~J., \& Chiang, E. 2018, Monthly Notices of the Royal
  Astronomical Society, 480, 4338, aDS Bibcode: 2018MNRAS.480.4338L.
\newblock \url{https://ui.adsabs.harvard.edu/abs/2018MNRAS.480.4338L}

\bibitem[{Lomb(1976)}]{lomb_least-squares_1976}
Lomb, N.~R. 1976, Astrophysics and Space Science, 39, 447.
\newblock \url{http://adsabs.harvard.edu/abs/1976Ap%26SS..39..447L}

\bibitem[{Luger {et~al.}(2019)Luger, Agol, Foreman-Mackey, Fleming,
  Lustig-Yaeger, \& Deitrick}]{luger_starry_2019}
Luger, R., Agol, E., Foreman-Mackey, D., {et~al.} 2019, The Astronomical
  Journal, 157, 64.
\newblock \url{https://ui.adsabs.harvard.edu/abs/2019AJ....157...64L}

\bibitem[{MacDonald \& Gizis(2018)}]{macdonald_explanation_2018}
MacDonald, J., \& Gizis, J. 2018, Monthly Notices of the Royal Astronomical
  Society, 480, 1711, aDS Bibcode: 2018MNRAS.480.1711M.
\newblock \url{https://ui.adsabs.harvard.edu/abs/2018MNRAS.480.1711M}

\bibitem[{Madhusudhan {et~al.}(2014)Madhusudhan, Amin, \&
  Kennedy}]{madhusudhan_toward_2014}
Madhusudhan, N., Amin, M.~A., \& Kennedy, G.~M. 2014, The Astrophysical
  Journal, 794, L12, aDS Bibcode: 2014ApJ...794L..12M.
\newblock \url{https://ui.adsabs.harvard.edu/abs/2014ApJ...794L..12M}

\bibitem[{Mahadevan {et~al.}(2012)Mahadevan, Ramsey, Bender, Terrien, Wright,
  Halverson, Hearty, Nelson, Burton, Redman, Osterman, Diddams, Kasting, Endl,
  \& Deshpande}]{mahadevan_habitable-zone_2012}
Mahadevan, S., Ramsey, L., Bender, C., {et~al.} 2012, SPIE, 8446, 84461S,
  conference Name: Ground-based and Airborne Instrumentation for Astronomy IV
  Place: eprint: arXiv:1209.1686.
\newblock \url{https://ui.adsabs.harvard.edu/abs/2012SPIE.8446E..1SM}

\bibitem[{Mahadevan {et~al.}(2014)Mahadevan, Ramsey, Terrien, Halverson, Roy,
  Hearty, Levi, Stefansson, Robertson, Bender, Schwab, \&
  Nelson}]{mahadevan_habitable-zone_2014}
Mahadevan, S., Ramsey, L.~W., Terrien, R., {et~al.} 2014, SPIE, 9147, 91471G.
\newblock \url{http://adsabs.harvard.edu/abs/2014SPIE.9147E..1GM}

\bibitem[{Maldonado {et~al.}(2019)Maldonado, Villaver, Eiroa, \&
  Micela}]{maldonado_connecting_2019}
Maldonado, J., Villaver, E., Eiroa, C., \& Micela, G. 2019, Astronomy \&amp;
  Astrophysics, Volume 624, id.A94,
  {\textless}NUMPAGES{\textgreater}7{\textless}/NUMPAGES{\textgreater} pp.,
  624, A94.
\newblock
  \url{https://ui.adsabs.harvard.edu/abs/2019A%26A...624A..94M/abstract}

\bibitem[{Maldonado {et~al.}(2020)Maldonado, Micela, Baratella, D'Orazi, Affer,
  Biazzo, Lanza, Maggio, González~Hernández, Perger, Pinamonti, Scandariato,
  Sozzetti, Locci, Di~Maio, Bignamini, Claudi, Molinari, Rebolo, Ribas,
  Toledo-Padrón, Covino, Desidera, Herrero, Morales, Suárez-Mascareño,
  Pagano, Petralia, Piotto, \& Poretti}]{maldonado_hades_2020}
Maldonado, J., Micela, G., Baratella, M., {et~al.} 2020, Astronomy and
  Astrophysics, 644, A68.
\newblock \url{https://ui.adsabs.harvard.edu/abs/2020A&A...644A..68M/abstract}

\bibitem[{Mandel \& Agol(2002)}]{mandel_analytic_2002}
Mandel, K., \& Agol, E. 2002, The Astrophysical Journal Letters, 580, L171.
\newblock \url{http://adsabs.harvard.edu/abs/2002ApJ...580L.171M}

\bibitem[{Mann {et~al.}(2015)Mann, Feiden, Gaidos, Boyajian, \& von
  Braun}]{mann_how_2015}
Mann, A.~W., Feiden, G.~A., Gaidos, E., Boyajian, T., \& von Braun, K. 2015,
  The Astrophysical Journal, 804, 64.
\newblock \url{http://adsabs.harvard.edu/abs/2015ApJ...804...64M}

\bibitem[{Mann {et~al.}(2016)Mann, Feiden, Gaidos, Boyajian, \& von
  Braun}]{mann_erratum_2016}
---. 2016, The Astrophysical Journal, 819, 87, aDS Bibcode:
  2016ApJ...819...87M.
\newblock \url{https://ui.adsabs.harvard.edu/abs/2016ApJ...819...87M}

\bibitem[{Mann {et~al.}(2019)Mann, Dupuy, Kraus, Gaidos, Ansdell, Ireland,
  Rizzuto, Hung, Dittmann, Factor, Feiden, Martinez, Ruíz-Rodríguez, \&
  Chia~Thao}]{mann_how_2019}
Mann, A.~W., Dupuy, T., Kraus, A.~L., {et~al.} 2019, The Astrophysical Journal,
  871, 63.
\newblock \url{http://adsabs.harvard.edu/abs/2019ApJ...871...63M}

\bibitem[{Mann {et~al.}(2022)Mann, Wood, Schmidt, Barber, Owen, Tofflemire,
  Newton, Mamajek, Bush, Mace, Kraus, Thao, Vanderburg, Llama, Johns-Krull,
  Prato, Stahl, Tang, Fields, Collins, Collins, Gan, Jensen, Kamler, Schwarz,
  Furlan, Gnilka, Howell, Lester, Owens, Suarez, Mekarnia, Guillot, Abe,
  Triaud, Johnson, Milburn, Rizzuto, Quinn, Kerr, Ricker, Vanderspek, Latham,
  Seager, Winn, Jenkins, Guerrero, Shporer, Schlieder, McLean, \&
  Wohler}]{mann_tess_2022}
Mann, A.~W., Wood, M.~L., Schmidt, S.~P., {et~al.} 2022, The Astronomical
  Journal, 163, 156, publisher: American Astronomical Society.
\newblock \url{https://doi.org/10.3847/1538-3881/ac511d}

\bibitem[{Masci {et~al.}(2019)Masci, Laher, Rusholme, Shupe, Groom, Surace,
  Jackson, Monkewitz, Beck, Flynn, Terek, Landry, Hacopians, Desai, Howell,
  Brooke, Imel, Wachter, Ye, Lin, Cenko, Cunningham, Rebbapragada, Bue, Miller,
  Mahabal, Bellm, Patterson, Jurić, Golkhou, Ofek, Walters, Graham, Kasliwal,
  Dekany, Kupfer, Burdge, Cannella, Barlow, Van~Sistine, Giomi, Fremling,
  Blagorodnova, Levitan, Riddle, Smith, Helou, Prince, \&
  Kulkarni}]{masci_zwicky_2019}
Masci, F.~J., Laher, R.~R., Rusholme, B., {et~al.} 2019, Publications of the
  Astronomical Society of the Pacific, 131, 018003.
\newblock \url{http://adsabs.harvard.edu/abs/2019PASP..131a8003M}

\bibitem[{McKinney(2010)}]{mckinney_data_2010}
McKinney, W. 2010, in Proceedings of the 9th {Python} in {Science}
  {Conference}, ed. S.~v.~d. Walt \& J.~Millman, 56 -- 61

\bibitem[{Metcalf {et~al.}(2019)Metcalf, Anderson, Bender, Blakeslee, Brand,
  Carlson, Cochran, Diddams, Endl, Fredrick, Halverson, Hickstein, Hearty,
  Jennings, Kanodia, Kaplan, Levi, Lubar, Mahadevan, Monson, Ninan, Nitroy,
  Osterman, Papp, Quinlan, Ramsey, Robertson, Roy, Schwab, Sigurdsson,
  Srinivasan, Stefansson, Sterner, Terrien, Wolszczan, Wright, \&
  Ycas}]{metcalf_stellar_2019}
Metcalf, A.~J., Anderson, T., Bender, C.~F., {et~al.} 2019, Optica, 6, 233.
\newblock \url{https://ui.adsabs.harvard.edu/abs/2019Optic...6..233M}

\bibitem[{Minkowski \& Abell(1963)}]{minkowski_national_1963}
Minkowski, R.~L., \& Abell, G.~O. 1963, Basic Astronomical Data: Stars and
  Stellar Systems, 481.
\newblock \url{http://adsabs.harvard.edu/abs/1963bad..book..481M}

\bibitem[{Miotello {et~al.}(2017)Miotello, van Dishoeck, Williams, Ansdell,
  Guidi, Hogerheijde, Manara, Tazzari, Testi, van~der Marel, \& van
  Terwisga}]{miotello_lupus_2017}
Miotello, A., van Dishoeck, E.~F., Williams, J.~P., {et~al.} 2017, Astronomy
  \&amp; Astrophysics, Volume 599, id.A113,
  {\textless}NUMPAGES{\textgreater}10{\textless}/NUMPAGES{\textgreater} pp.,
  599, A113.
\newblock
  \url{https://ui.adsabs.harvard.edu/abs/2017A%26A...599A.113M/abstract}

\bibitem[{Monson {et~al.}(2017)Monson, Beaton, Scowcroft, Freedman, Madore,
  Rich, Seibert, Kollmeier, \& Clementini}]{monson_standard_2017}
Monson, A.~J., Beaton, R.~L., Scowcroft, V., {et~al.} 2017, The Astronomical
  Journal, 153, 96.
\newblock \url{http://adsabs.harvard.edu/abs/2017AJ....153...96M}

\bibitem[{Morales {et~al.}(2019)Morales, Mustill, Ribas, Davies, Reiners,
  Bauer, Kossakowski, Herrero, Rodríguez, López-González, Rodríguez-López,
  Béjar, González-Cuesta, Luque, Pallé, Perger, Baroch, Johansen, Klahr,
  Mordasini, Anglada-Escudé, Caballero, Cortés-Contreras, Dreizler, Lafarga,
  Nagel, Passegger, Reffert, Rosich, Schweitzer, Tal-Or, Trifonov, Zechmeister,
  Quirrenbach, Amado, Guenther, Hagen, Henning, Jeffers, Kaminski, Kürster,
  Montes, Seifert, Abellán, Abril, Aceituno, Aceituno, Alonso-Floriano,
  Ammler-von Eiff, Antona, Arroyo-Torres, Azzaro, Barrado, Becerril-Jarque,
  Benítez, Berdiñas, Bergond, Brinkmöller, del Burgo, Burn, Calvo-Ortega,
  Cano, Cárdenas, Cardona~Guillén, Carro, Casal, Casanova, Casasayas-Barris,
  Chaturvedi, Cifuentes, Claret, Colomé, Czesla, Díez-Alonso, Dorda,
  Emsenhuber, Fernández, Fernández-Martín, Ferro, Fuhrmeister,
  Galadí-Enríquez, Gallardo~Cava, García~Vargas, Garcia-Piquer, Gesa,
  González-Álvarez, González~Hernández, González-Peinado, Guàrdia,
  Guijarro, de~Guindos, Hatzes, Hauschildt, Hedrosa, Hermelo, Hernández~Arabi,
  Hernández~Otero, Hintz, Holgado, Huber, Huke, Johnson, de~Juan, Kehr,
  Kemmer, Kim, Klüter, Klutsch, Labarga, Labiche, Lalitha, Lampón, Lara,
  Launhardt, Lázaro, Lizon, Llamas, Lodieu, López~del Fresno, López~Salas,
  López-Santiago, Magán~Madinabeitia, Mall, Mancini, Mandel, Marfil,
  Marín~Molina, Martín, Martín-Fernández, Martín-Ruiz,
  Martínez-Rodríguez, Marvin, Mirabet, Moya, Naranjo, Nelson, Nortmann,
  Nowak, Ofir, Pascual, Pavlov, Pedraz, Pérez~Medialdea, Pérez-Calpena,
  Perryman, Rabaza, Ramón~Ballesta, Rebolo, Redondo, Rix, Rodler,
  Rodríguez~Trinidad, Sabotta, Sadegi, Salz, Sánchez-Blanco,
  Sánchez~Carrasco, Sánchez-López, Sanz-Forcada, Sarkis, Sarmiento,
  Schäfer, Schlecker, Schmitt, Schöfer, Solano, Sota, Stahl, Stock, Stuber,
  Stürmer, Suárez, Tabernero, Tulloch, Veredas, Vico-Linares, Vilardell,
  Wagner, Winkler, Wolthoff, Yan, \& Zapatero~Osorio}]{morales_giant_2019}
Morales, J.~C., Mustill, A.~J., Ribas, I., {et~al.} 2019, Science, 365, 1441.
\newblock \url{http://adsabs.harvard.edu/abs/2019Sci...365.1441M}

\bibitem[{Mulders {et~al.}(2015)Mulders, Pascucci, \&
  Apai}]{mulders_increase_2015}
Mulders, G.~D., Pascucci, I., \& Apai, D. 2015, The Astrophysical Journal, 814,
  130.
\newblock \url{http://adsabs.harvard.edu/abs/2015ApJ...814..130M}

\bibitem[{Najita \& Kenyon(2014)}]{najita_mass_2014}
Najita, J.~R., \& Kenyon, S.~J. 2014, Monthly Notices of the Royal Astronomical
  Society, 445, 3315, aDS Bibcode: 2014MNRAS.445.3315N.
\newblock \url{https://ui.adsabs.harvard.edu/abs/2014MNRAS.445.3315N}

\bibitem[{{NASA Exoplanet Archive}(2022)}]{PSCompPars}
{NASA Exoplanet Archive}. 2022, Planetary Systems Composite Parameters,
  vVersion: 2022-08-20 00:00,  NExScI-Caltech/IPAC, doi:10.26133/NEA13.
\newblock \url{https://catcopy.ipac.caltech.edu/dois/doi.php?id=10.26133/NEA13}

\bibitem[{Neves {et~al.}(2012)Neves, Bonfils, Santos, Delfosse, Forveille,
  Allard, Natário, Fernandes, \& Udry}]{neves_metallicity_2012}
Neves, V., Bonfils, X., Santos, N.~C., {et~al.} 2012, Astronomy \&amp;
  Astrophysics, Volume 538, id.A25,
  {\textless}NUMPAGES{\textgreater}10{\textless}/NUMPAGES{\textgreater} pp.,
  538, A25.
\newblock
  \url{https://ui.adsabs.harvard.edu/abs/2012A%26A...538A..25N/abstract}

\bibitem[{Ninan {et~al.}(2018)Ninan, Bender, Mahadevan, Ford, Monson, Kaplan,
  Terrien, Roy, Robertson, Kanodia, \& Stefansson}]{ninan_habitable-zone_2018}
Ninan, J.~P., Bender, C.~F., Mahadevan, S., {et~al.} 2018, Proceedings of the
  SPIE, 0709, 107092U.
\newblock \url{http://adsabs.harvard.edu/abs/2018SPIE10709E..2UN}

\bibitem[{Oliphant(2006)}]{oliphant_numpy_2006}
Oliphant, T. 2006, {NumPy}: {A} guide to {NumPy}, published: USA: Trelgol
  Publishing.
\newblock \url{http://www.numpy.org/}

\bibitem[{Oliphant(2007)}]{oliphant_python_2007}
Oliphant, T.~E. 2007, Computing in Science Engineering, 9, 10

\bibitem[{Parviainen {et~al.}(2021)Parviainen, Palle, Zapatero-Osorio, Nowak,
  Fukui, Murgas, Narita, Stassun, Livingston, Collins, Hidalgo~Soto, Béjar,
  Korth, Monelli, Montanes~Rodriguez, Casasayas-Barris, Chen, Crouzet, de~Leon,
  Hernandez, Kawauchi, Klagyivik, Kusakabe, Luque, Mori, Nishiumi,
  Prieto-Arranz, Tamura, Watanabe, Gan, Collins, Jensen, Barclay, Doty,
  Jenkins, Latham, Paegert, Ricker, Rodriguez, Seager, Shporer, Vanderspek,
  Villaseñor, Winn, Wohler, \& Wong}]{parviainen_toi-519_2021}
Parviainen, H., Palle, E., Zapatero-Osorio, M.~R., {et~al.} 2021, Astronomy and
  Astrophysics, 645, A16.
\newblock \url{https://ui.adsabs.harvard.edu/abs/2021A&A...645A..16P/abstract}

\bibitem[{Pascucci {et~al.}(2016)Pascucci, Testi, Herczeg, Long, Manara,
  Hendler, Mulders, Krijt, Ciesla, Henning, Mohanty, Drabek-Maunder, Apai,
  Szűcs, Sacco, \& Olofsson}]{pascucci_steeper_2016}
Pascucci, I., Testi, L., Herczeg, G.~J., {et~al.} 2016, The Astrophysical
  Journal, 831, 125, aDS Bibcode: 2016ApJ...831..125P.
\newblock \url{https://ui.adsabs.harvard.edu/abs/2016ApJ...831..125P}

\bibitem[{Passegger {et~al.}(2022)Passegger, Bello-García, Ordieres-Meré,
  Antoniadis-Karnavas, Marfil, Duque-Arribas, Amado, Delgado-Mena, Montes,
  Rojas-Ayala, Schweitzer, Tabernero, Béjar, Caballero, Hatzes, Henning,
  Pedraz, Quirrenbach, Reiners, \& Ribas}]{passegger_metallicities_2022}
Passegger, V.~M., Bello-García, A., Ordieres-Meré, J., {et~al.} 2022,
  Astronomy and Astrophysics, 658, A194.
\newblock \url{https://ui.adsabs.harvard.edu/abs/2022A&A...658A.194P/abstract}

\bibitem[{Pecaut \& Mamajek(2016)}]{pecaut_star_2016}
Pecaut, M.~J., \& Mamajek, E.~E. 2016, Monthly Notices of the Royal
  Astronomical Society, 461, 794, aDS Bibcode: 2016MNRAS.461..794P.
\newblock \url{https://ui.adsabs.harvard.edu/abs/2016MNRAS.461..794P}

\bibitem[{Penoyre {et~al.}(2020)Penoyre, Belokurov, Wyn Evans, Everall, \&
  Koposov}]{penoyre_binary_2020}
Penoyre, Z., Belokurov, V., Wyn Evans, N., Everall, A., \& Koposov, S.~E.
  2020, Monthly Notices of the Royal Astronomical Society, 495, 321.
\newblock \url{https://doi.org/10.1093/mnras/staa1148}

\bibitem[{Persson {et~al.}(2013)Persson, Murphy, Smee, Birk, Monson, Uomoto,
  Koch, Shectman, Barkhouser, Orndorff, Hammond, Harding, Scharfstein, Kelson,
  Marshall, \& McCarthy}]{persson_fourstar_2013}
Persson, S.~E., Murphy, D.~C., Smee, S., {et~al.} 2013, Publications of the
  Astronomical Society of the Pacific, 125, 654, aDS Bibcode:
  2013PASP..125..654P.
\newblock \url{https://ui.adsabs.harvard.edu/abs/2013PASP..125..654P}

\bibitem[{Piette \& Madhusudhan(2020)}]{piette_temperature_2020}
Piette, A. A.~A., \& Madhusudhan, N. 2020, arXiv:2009.11290 [astro-ph], arXiv:
  2009.11290.
\newblock \url{http://arxiv.org/abs/2009.11290}

\bibitem[{Piette {et~al.}(2020)Piette, Madhusudhan, McKemmish, Gandhi,
  Masseron, \& Welbanks}]{piette_assessing_2020}
Piette, A. A.~A., Madhusudhan, N., McKemmish, L.~K., {et~al.} 2020, Monthly
  Notices of the Royal Astronomical Society, 496, 3870, aDS Bibcode:
  2020MNRAS.496.3870P.
\newblock \url{https://ui.adsabs.harvard.edu/abs/2020MNRAS.496.3870P}

\bibitem[{Pollack {et~al.}(1996)Pollack, Hubickyj, Bodenheimer, Lissauer,
  Podolak, \& Greenzweig}]{pollack_formation_1996}
Pollack, J.~B., Hubickyj, O., Bodenheimer, P., {et~al.} 1996, Icarus, 124, 62.
\newblock \url{http://adsabs.harvard.edu/abs/1996Icar..124...62P}

\bibitem[{Pérez \& Granger(2007)}]{perez_ipython_2007}
Pérez, F., \& Granger, B.~E. 2007, Computing in Science and Engineering, 9,
  21.
\newblock \url{https://ipython.org}

\bibitem[{Quirrenbach {et~al.}(2022)Quirrenbach, Passegger, Trifonov, Amado,
  Caballero, Reiners, Ribas, Aceituno, Bejar, Chaturvedi, Gonzalez-Cuesta,
  Henning, Herrero, Kaminski, Kuerster, Lalitha, Lodieu, Lopez-Gonzalez,
  Montes, Palle, Perger, Pollacco, Reffert, Rodriguez, Rodriguez~Lopez, Shan,
  Tal-Or, Zapatero~Osorio, \& Zechmeister}]{quirrenbach_carmenes_2022}
Quirrenbach, A., Passegger, V.~M., Trifonov, T., {et~al.} 2022, The {CARMENES}
  search for exoplanets around {M} dwarfs: {Two} {Saturn}-mass planets orbiting
  active stars, Tech. rep., publication Title: arXiv e-prints ADS Bibcode:
  2022arXiv220316504Q Type: article.
\newblock \url{https://ui.adsabs.harvard.edu/abs/2022arXiv220316504Q}

\bibitem[{Rabus {et~al.}(2019)Rabus, Lachaume, Jordán, Brahm, Boyajian, von
  Braun, Espinoza, Berger, Le~Bouquin, \& Absil}]{rabus_discontinuity_2019}
Rabus, M., Lachaume, R., Jordán, A., {et~al.} 2019, Monthly Notices of the
  Royal Astronomical Society, 484, 2674, aDS Bibcode: 2019MNRAS.484.2674R.
\newblock \url{https://ui.adsabs.harvard.edu/abs/2019MNRAS.484.2674R}

\bibitem[{Ramsey {et~al.}(1998)Ramsey, Adams, Barnes, Booth, Cornell, Fowler,
  Gaffney, Glaspey, Good, Hill, Kelton, Krabbendam, Long, MacQueen, Ray,
  Ricklefs, Sage, Sebring, Spiesman, \& Steiner}]{ramsey_early_1998}
Ramsey, L.~W., Adams, M.~T., Barnes, T.~G., {et~al.} 1998, 3352, 34, conference
  Name: Advanced Technology Optical/IR Telescopes VI.
\newblock \url{http://adsabs.harvard.edu/abs/1998SPIE.3352...34R}

\bibitem[{Reiners \& Basri(2008)}]{reiners_chromospheric_2008}
Reiners, A., \& Basri, G. 2008, The Astrophysical Journal, 684, 1390, aDS
  Bibcode: 2008ApJ...684.1390R.
\newblock \url{https://ui.adsabs.harvard.edu/abs/2008ApJ...684.1390R}

\bibitem[{Reylé {et~al.}(2021)Reylé, Jardine, Fouqué, Caballero, Smart, \&
  Sozzetti}]{reyle_10_2021}
Reylé, C., Jardine, K., Fouqué, P., {et~al.} 2021, Astronomy and
  Astrophysics, 650, A201.
\newblock \url{https://ui.adsabs.harvard.edu/abs/2021A&A...650A.201R/abstract}

\bibitem[{Ribas {et~al.}(2014)Ribas, Merin, Bouy, \& Maud}]{ribas_disk_2014}
Ribas, A., Merin, B., Bouy, H., \& Maud, L.~T. 2014, Astronomy and
  Astrophysics, 561, A54.
\newblock \url{https://ui.adsabs.harvard.edu/abs/2014A&A...561A..54R/abstract}

\bibitem[{Richard {et~al.}(2012)Richard, Gordon, Rothman, Abel, Frommhold,
  Gustafsson, Hartmann, Hermans, Lafferty, Orton, Smith, \& Tran}]{Richard2012}
Richard, C., Gordon, I., Rothman, L., {et~al.} 2012, Journal of Quantitative
  Spectroscopy and Radiative Transfer, 113, 1276 , three Leaders in
  Spectroscopy.
\newblock
  \url{http://www.sciencedirect.com/science/article/pii/S0022407311003773}

\bibitem[{Ricker {et~al.}(2014)Ricker, Winn, Vanderspek, Latham, Bakos, Bean,
  Berta-Thompson, Brown, Buchhave, Butler, Butler, Chaplin, Charbonneau,
  Christensen-Dalsgaard, Clampin, Deming, Doty, Lee, Dressing, Dunham, Endl,
  Fressin, Ge, Henning, Holman, Howard, Ida, Jenkins, Jernigan, Johnson,
  Kaltenegger, Kawai, Kjeldsen, Laughlin, Levine, Lin, Lissauer, MacQueen,
  Marcy, McCullough, Morton, Narita, Paegert, Palle, Pepe, Pepper, Quirrenbach,
  Rinehart, Sasselov, Sato, Seager, Sozzetti, Stassun, Sullivan, Szentgyorgyi,
  Torres, Udry, \& Villasenor}]{ricker_transiting_2014}
Ricker, G.~R., Winn, J.~N., Vanderspek, R., {et~al.} 2014, Journal of
  Astronomical Telescopes, Instruments, and Systems, 1, 014003.
\newblock
  \url{https://www.spiedigitallibrary.org/journals/Journal-of-Astronomical-Telescopes-Instruments-and-Systems/volume-1/issue-1/014003/Transiting-Exoplanet-Survey-Satellite/10.1117/1.JATIS.1.1.014003.short}

\bibitem[{Robitaille {et~al.}(2013)Robitaille, Tollerud, Greenfield,
  Droettboom, Bray, Aldcroft, Davis, Ginsburg, Price-Whelan, Kerzendorf,
  Conley, Crighton, Barbary, Muna, Ferguson, Grollier, Parikh, Nair, Günther,
  Deil, Woillez, Conseil, Kramer, Turner, Singer, Fox, Weaver, Zabalza,
  Edwards, Bostroem, Burke, Casey, Crawford, Dencheva, Ely, Jenness, Labrie,
  Lim, Pierfederici, Pontzen, Ptak, Refsdal, Servillat, \&
  Streicher}]{robitaille_astropy_2013}
Robitaille, T.~P., Tollerud, E.~J., Greenfield, P., {et~al.} 2013, Astronomy \&
  Astrophysics, 558, A33.
\newblock
  \url{https://www.aanda.org/articles/aa/abs/2013/10/aa22068-13/aa22068-13.html}

\bibitem[{Rosotti {et~al.}(2017)Rosotti, Clarke, Manara, \&
  Facchini}]{rosotti_constraining_2017}
Rosotti, G.~P., Clarke, C.~J., Manara, C.~F., \& Facchini, S. 2017, Monthly
  Notices of the Royal Astronomical Society, 468, 1631, aDS Bibcode:
  2017MNRAS.468.1631R.
\newblock \url{https://ui.adsabs.harvard.edu/abs/2017MNRAS.468.1631R}

\bibitem[{{Rothman} {et~al.}(2010){Rothman}, {Gordon}, {Barber}, {Dothe},
  {Gamache}, {Goldman}, {Perevalov}, {Tashkun}, \& {Tennyson}}]{Rothman2010}
{Rothman}, L.~S., {Gordon}, I.~E., {Barber}, R.~J., {et~al.} 2010, \jqsrt, 111,
  2139

\bibitem[{Rothman {et~al.}(2013)Rothman, Gordon, Babikov, Barbe, Chris~Benner,
  Bernath, Birk, Bizzocchi, Boudon, Brown, Campargue, Chance, Cohen, Coudert,
  Devi, Drouin, Fayt, Flaud, Gamache, Harrison, Hartmann, Hill, Hodges,
  Jacquemart, Jolly, Lamouroux, Le~Roy, Li, Long, Lyulin, Mackie, Massie,
  Mikhailenko, Müller, Naumenko, Nikitin, Orphal, Perevalov, Perrin,
  Polovtseva, Richard, Smith, Starikova, Sung, Tashkun, Tennyson, Toon,
  Tyuterev, \& Wagner}]{rothman_hitran2012_2013}
Rothman, L.~S., Gordon, I.~E., Babikov, Y., {et~al.} 2013, Journal of
  Quantitative Spectroscopy and Radiative Transfer, 130, 4.
\newblock \url{https://ui.adsabs.harvard.edu/abs/2013JQSRT.130....4R/abstract}

\bibitem[{Roulston {et~al.}(2020)Roulston, Green, \&
  Kesseli}]{roulston_classifying_2020}
Roulston, B.~R., Green, P.~J., \& Kesseli, A.~Y. 2020, The Astrophysical
  Journal Supplement Series, 249, 34, aDS Bibcode: 2020ApJS..249...34R.
\newblock \url{https://ui.adsabs.harvard.edu/abs/2020ApJS..249...34R}

\bibitem[{Salvatier {et~al.}(2016)Salvatier, Wiecki, \&
  Fonnesbeck}]{salvatier_probabilistic_2016}
Salvatier, J., Wiecki, T.~V., \& Fonnesbeck, C. 2016, PeerJ Computer Science,
  2, e55, publisher: PeerJ Inc.

\bibitem[{Santos {et~al.}(2001)Santos, Israelian, \&
  Mayor}]{santos_metal-rich_2001}
Santos, N.~C., Israelian, G., \& Mayor, M. 2001, Astronomy and Astrophysics,
  373, 1019.
\newblock \url{http://adsabs.harvard.edu/abs/2001A%26A...373.1019S}

\bibitem[{Santos {et~al.}(2017)Santos, Adibekyan, Figueira, Andreasen, Barros,
  Delgado-Mena, Demangeon, Faria, Oshagh, Sousa, Viana, \&
  Ferreira}]{santos_observational_2017}
Santos, N.~C., Adibekyan, V., Figueira, P., {et~al.} 2017, Astronomy and
  Astrophysics, 603, A30.
\newblock \url{http://adsabs.harvard.edu/abs/2017A%26A...603A..30S}

\bibitem[{Scargle(1982)}]{scargle_studies_1982}
Scargle, J.~D. 1982, The Astrophysical Journal, 263, 835.
\newblock \url{http://adsabs.harvard.edu/abs/1982ApJ...263..835S}

\bibitem[{Schlaufman(2018)}]{schlaufman_evidence_2018}
Schlaufman, K.~C. 2018, The Astrophysical Journal, 853, 37.
\newblock \url{https://ui.adsabs.harvard.edu/abs/2018ApJ...853...37S}

\bibitem[{Schlaufman \& Laughlin(2010)}]{schlaufman_physically-motivated_2010}
Schlaufman, K.~C., \& Laughlin, G. 2010, Astronomy and Astrophysics, 519, A105.
\newblock \url{http://adsabs.harvard.edu/abs/2010A%26A...519A.105S}

\bibitem[{Schlecker {et~al.}(2022)Schlecker, Burn, Sabotta, Seifert, Henning,
  Emsenhuber, Mordasini, Reffert, Shan, \& Klahr}]{schlecker_rv-detected_2022}
Schlecker, M., Burn, R., Sabotta, S., {et~al.} 2022, {RV}-detected planets
  around {M} dwarfs: {Challenges} for core accretion models, Tech. rep.,
  publication Title: arXiv e-prints ADS Bibcode: 2022arXiv220512971S Type:
  article.
\newblock \url{https://ui.adsabs.harvard.edu/abs/2022arXiv220512971S}

\bibitem[{Schweitzer {et~al.}(2019)Schweitzer, Passegger, Cifuentes, Béjar,
  Cortés-Contreras, Caballero, del Burgo, Czesla, Kürster, Montes,
  Zapatero~Osorio, Ribas, Reiners, Quirrenbach, Amado, Aceituno,
  Anglada-Escudé, Bauer, Dreizler, Jeffers, Guenther, Henning, Kaminski,
  Lafarga, Marfil, Morales, Schmitt, Seifert, Solano, Tabernero, \&
  Zechmeister}]{schweitzer_carmenes_2019}
Schweitzer, A., Passegger, V.~M., Cifuentes, C., {et~al.} 2019, Astronomy and
  Astrophysics, 625, A68.
\newblock \url{http://adsabs.harvard.edu/abs/2019A%26A...625A..68S}

\bibitem[{Schönrich {et~al.}(2010)Schönrich, Binney, \&
  Dehnen}]{schonrich_local_2010}
Schönrich, R., Binney, J., \& Dehnen, W. 2010, Monthly Notices of the Royal
  Astronomical Society, 403, 1829.
\newblock \url{https://academic.oup.com/mnras/article/403/4/1829/1054839}

\bibitem[{Scott {et~al.}(2018)Scott, Howell, Horch, \&
  Everett}]{scott_nn-explore_2018}
Scott, N.~J., Howell, S.~B., Horch, E.~P., \& Everett, M.~E. 2018, Publications
  of the Astronomical Society of the Pacific, 130, 054502, aDS Bibcode:
  2018PASP..130e4502S.
\newblock \url{https://ui.adsabs.harvard.edu/abs/2018PASP..130e4502S}

\bibitem[{Shanno(1970)}]{shanno_conditioning_1970}
Shanno, D.~F. 1970, Mathematics of Computation, 24, 647.
\newblock \url{https://www.ams.org/mcom/1970-24-111/S0025-5718-1970-0274029-X/}

\bibitem[{Shetrone {et~al.}(2007)Shetrone, Cornell, Fowler, Gaffney, Laws,
  Mader, Mason, Odewahn, Roman, Rostopchin, Schneider, Umbarger, \&
  Westfall}]{shetrone_ten_2007}
Shetrone, M., Cornell, M.~E., Fowler, J.~R., {et~al.} 2007, Publications of the
  Astronomical Society of the Pacific, 119, 556.
\newblock \url{http://adsabs.harvard.edu/abs/2007PASP..119..556S}

\bibitem[{Silverberg {et~al.}(2020)Silverberg, Wisniewski, Kuchner, Lawson,
  Bans, Debes, Biggs, Bosch, Doll, Luca, Enachioaie, Hamilton, Holden, \&
  Hyogo}]{silverberg_peter_2020}
Silverberg, S.~M., Wisniewski, J.~P., Kuchner, M.~J., {et~al.} 2020, The
  Astrophysical Journal, 890, 106, aDS Bibcode: 2020ApJ...890..106S.
\newblock \url{https://ui.adsabs.harvard.edu/abs/2020ApJ...890..106S}

\bibitem[{Silverstein {et~al.}(2022)Silverstein, Schlieder, Barclay, Hord, Jao,
  Vrijmoet, Henry, Cloutier, Kostov, Kruse, Winters, Irwin, Kane, Stassun,
  Huang, Kunimoto, Tey, Vanderburg, Astudillo-Defru, Bonfils, Brasseur,
  Charbonneau, Ciardi, Collins, Collins, Conti, Crossfield, Daylan, Doty,
  Dressing, Gilbert, Horne, Jenkins, Latham, Mann, Matthews, Paredes, Quinn,
  Ricker, Schwarz, Seager, Sefako, Shporer, Smith, Stockdale, Tan, Torres,
  Twicken, Vanderspek, Wang, \& Winn}]{silverstein_lhs_2022}
Silverstein, M.~L., Schlieder, J.~E., Barclay, T., {et~al.} 2022, The
  Astronomical Journal, 163, 151, aDS Bibcode: 2022AJ....163..151S.
\newblock \url{https://ui.adsabs.harvard.edu/abs/2022AJ....163..151S}

\bibitem[{Sousa {et~al.}(2011)Sousa, Santos, Israelian, Mayor, \&
  Udry}]{sousa_spectroscopic_2011}
Sousa, S.~G., Santos, N.~C., Israelian, G., Mayor, M., \& Udry, S. 2011,
  Astronomy and Astrophysics, 533, A141.
\newblock \url{http://adsabs.harvard.edu/abs/2011A%26A...533A.141S}

\bibitem[{Stassun {et~al.}(2018)Stassun, Oelkers, Pepper, Paegert, De~Lee,
  Torres, Latham, Charpinet, Dressing, Huber, Kane, Lépine, Mann, Muirhead,
  Rojas-Ayala, Silvotti, Fleming, Levine, \& Plavchan}]{stassun_tess_2018}
Stassun, K.~G., Oelkers, R.~J., Pepper, J., {et~al.} 2018, The Astronomical
  Journal, 156, 102.
\newblock \url{http://adsabs.harvard.edu/abs/2018AJ....156..102S}

\bibitem[{Stassun {et~al.}(2019)Stassun, Oelkers, Paegert, Torres, Pepper,
  De~Lee, Collins, Latham, Muirhead, Chittidi, Rojas-Ayala, Fleming, Rose,
  Tenenbaum, Ting, Kane, Barclay, Bean, Brassuer, Charbonneau, Ge, Lissauer,
  Mann, McLean, Mullally, Narita, Plavchan, Ricker, Sasselov, Seager, Sharma,
  Shiao, Sozzetti, Stello, Vanderspek, Wallace, \& Winn}]{stassun_revised_2019}
Stassun, K.~G., Oelkers, R.~J., Paegert, M., {et~al.} 2019, The Astronomical
  Journal, 158, 138, aDS Bibcode: 2019AJ....158..138S.
\newblock \url{https://ui.adsabs.harvard.edu/abs/2019AJ....158..138S}

\bibitem[{Stefansson {et~al.}(2016)Stefansson, Hearty, Robertson, Mahadevan,
  Anderson, Levi, Bender, Nelson, Monson, Blank, Halverson, Henderson, Ramsey,
  Roy, Schwab, \& Terrien}]{stefansson_versatile_2016}
Stefansson, G., Hearty, F., Robertson, P., {et~al.} 2016, The Astrophysical
  Journal, 833, 175.
\newblock \url{http://adsabs.harvard.edu/abs/2016ApJ...833..175S}

\bibitem[{Stefansson {et~al.}(2017)Stefansson, Mahadevan, Hebb, Wisniewski,
  Huehnerhoff, Morris, Halverson, Zhao, Wright, O'rourke, Knutson, Hawley,
  Kanodia, Li, Hagen, Liu, Beatty, Bender, Robertson, Dembicky, Gray,
  Ketzeback, McMillan, \& Rudyk}]{stefansson_toward_2017}
Stefansson, G., Mahadevan, S., Hebb, L., {et~al.} 2017, The Astrophysical
  Journal, 848, 9.
\newblock \url{http://adsabs.harvard.edu/abs/2017ApJ...848....9S}

\bibitem[{Stefansson {et~al.}(2020)Stefansson, Cañas, Wisniewski, Robertson,
  Mahadevan, Maney, Kanodia, Beard, Bender, Brunt, Clemens, Cochran, Diddams,
  Endl, Ford, Fredrick, Halverson, Hearty, Hebb, Huehnerhoff, Jennings, Kaplan,
  Levi, Lubar, Metcalf, Monson, Morris, Ninan, Nitroy, Ramsey, Roy, Schwab,
  Sigurdsson, Terrien, \& Wright}]{stefansson_sub-neptune-sized_2020}
Stefansson, G., Cañas, C., Wisniewski, J., {et~al.} 2020, The Astronomical
  Journal, 159, 100.
\newblock \url{http://adsabs.harvard.edu/abs/2020AJ....159..100S}

\bibitem[{Stetson(1987)}]{stetson_daophot_1987}
Stetson, P.~B. 1987, Publications of the Astronomical Society of the Pacific,
  99, 191, aDS Bibcode: 1987PASP...99..191S.
\newblock \url{https://ui.adsabs.harvard.edu/abs/1987PASP...99..191S}

\bibitem[{Stetson \& Harris(1988)}]{stetson_ccd_1988}
Stetson, P.~B., \& Harris, W.~E. 1988, The Astronomical Journal, 96, 909, aDS
  Bibcode: 1988AJ.....96..909S.
\newblock \url{https://ui.adsabs.harvard.edu/abs/1988AJ.....96..909S}

\bibitem[{{The Theano Development Team} {et~al.}(2016){The Theano Development
  Team}, Al-Rfou, Alain, Almahairi, Angermueller, Bahdanau, Ballas, Bastien,
  Bayer, Belikov, Belopolsky, Bengio, Bergeron, Bergstra, Bisson,
  Bleecher~Snyder, Bouchard, Boulanger-Lewandowski, Bouthillier, de~Brébisson,
  Breuleux, Carrier, Cho, Chorowski, Christiano, Cooijmans, Côté, Côté,
  Courville, Dauphin, Delalleau, Demouth, Desjardins, Dieleman, Dinh, Ducoffe,
  Dumoulin, Ebrahimi~Kahou, Erhan, Fan, Firat, Germain, Glorot, Goodfellow,
  Graham, Gulcehre, Hamel, Harlouchet, Heng, Hidasi, Honari, Jain, Jean, Jia,
  Korobov, Kulkarni, Lamb, Lamblin, Larsen, Laurent, Lee, Lefrancois, Lemieux,
  Léonard, Lin, Livezey, Lorenz, Lowin, Ma, Manzagol, Mastropietro, McGibbon,
  Memisevic, van Merriënboer, Michalski, Mirza, Orlandi, Pal, Pascanu,
  Pezeshki, Raffel, Renshaw, Rocklin, Romero, Roth, Sadowski, Salvatier,
  Savard, Schlüter, Schulman, Schwartz, Vlad~Serban, Serdyuk, Shabanian,
  Simon, Spieckermann, Ramana~Subramanyam, Sygnowski, Tanguay, van Tulder,
  Turian, Urban, Vincent, Visin, de~Vries, Warde-Farley, Webb, Willson, Xu,
  Xue, Yao, Zhang, \& Zhang}]{the_theano_development_team_theano_2016}
{The Theano Development Team}, Al-Rfou, R., Alain, G., {et~al.} 2016, arXiv
  e-prints, arXiv:1605.02688.
\newblock \url{https://ui.adsabs.harvard.edu/abs/2016arXiv160502688T}

\bibitem[{Thorngren {et~al.}(2016)Thorngren, Fortney, Murray-Clay, \&
  Lopez}]{thorngren_mass-metallicity_2016}
Thorngren, D.~P., Fortney, J.~J., Murray-Clay, R.~A., \& Lopez, E.~D. 2016, The
  Astrophysical Journal, 831, 64, aDS Bibcode: 2016ApJ...831...64T.
\newblock \url{https://ui.adsabs.harvard.edu/abs/2016ApJ...831...64T}

\bibitem[{Tonry {et~al.}(2018)Tonry, Denneau, Heinze, Stalder, Smith, Smartt,
  Stubbs, Weiland, \& Rest}]{tonry_atlas_2018}
Tonry, J.~L., Denneau, L., Heinze, A.~N., {et~al.} 2018, Publications of the
  Astronomical Society of the Pacific, 130, 064505, aDS Bibcode:
  2018PASP..130f4505T.
\newblock \url{https://ui.adsabs.harvard.edu/abs/2018PASP..130f4505T}

\bibitem[{Trifonov {et~al.}(2018)Trifonov, Kürster, Zechmeister, Tal-Or,
  Caballero, Quirrenbach, Amado, Ribas, Reiners, Reffert, Dreizler, Hatzes,
  Kaminski, Launhardt, Henning, Montes, Béjar, Mundt, Pavlov, Schmitt,
  Seifert, Morales, Nowak, Jeffers, Rodríguez-López, del Burgo,
  Anglada-Escudé, López-Santiago, Mathar, Ammler-von Eiff, Guenther, Barrado,
  González~Hernández, Mancini, Stürmer, Abril, Aceituno, Alonso-Floriano,
  Antona, Anwand-Heerwart, Arroyo-Torres, Azzaro, Baroch, Bauer, Becerril,
  Benítez, Berdiñas, Bergond, Blümcke, Brinkmöller, Cano,
  Cárdenas~Vázquez, Casal, Cifuentes, Claret, Colomé, Cortés-Contreras,
  Czesla, Díez-Alonso, Feiz, Fernández, Ferro, Fuhrmeister,
  Galadí-Enríquez, Garcia-Piquer, García~Vargas, Gesa, Gómez~Galera,
  González-Peinado, Grözinger, Grohnert, Guàrdia, Guijarro, de~Guindos,
  Gutiérrez-Soto, Hagen, Hauschildt, Hedrosa, Helmling, Hermelo,
  Hernández~Arabí, Hernández~Castaño, Hernández~Hernando, Herrero, Huber,
  Huke, Johnson, de~Juan, Kim, Klein, Klüter, Klutsch, Lafarga, Lampón, Lara,
  Laun, Lemke, Lenzen, López~del Fresno, López-González, López-Puertas,
  López~Salas, Luque, Magán~Madinabeitia, Mall, Mandel, Marfil,
  Marín~Molina, Maroto~Fernández, Martín, Martín-Ruiz, Marvin, Mirabet,
  Moya, Moreno-Raya, Nagel, Naranjo, Nortmann, Ofir, Oreiro, Pallé, Panduro,
  Pascual, Passegger, Pedraz, Pérez-Calpena, Pérez~Medialdea, Perger,
  Perryman, Pluto, Rabaza, Ramón, Rebolo, Redondo, Reinhardt, Rhode, Rix,
  Rodler, Rodríguez, Rodríguez~Trinidad, Rohloff, Rosich, Sadegi,
  Sánchez-Blanco, Sánchez~Carrasco, Sánchez-López, Sanz-Forcada, Sarkis,
  Sarmiento, Schäfer, Schiller, Schöfer, Schweitzer, Solano, Stahl, Strachan,
  Suárez, Tabernero, Tala, Tulloch, Veredas, Vico~Linares, Vilardell, Wagner,
  Winkler, Wolthoff, Xu, Yan, \& Zapatero~Osorio}]{trifonov_carmenes_2018}
Trifonov, T., Kürster, M., Zechmeister, M., {et~al.} 2018, Astronomy and
  Astrophysics, 609, A117.
\newblock \url{https://ui.adsabs.harvard.edu/abs/2018A&A...609A.117T/abstract}

\bibitem[{Vallenari {et~al.}(2022)Vallenari, Brown, \&
  Prusti}]{vallenari_gaia_2022}
Vallenari, A., Brown, A. G.~A., \& Prusti, T. 2022, Astronomy \& Astrophysics,
  doi:10.1051/0004-6361/202243940, publisher: EDP Sciences.
\newblock
  \url{https://www.aanda.org/articles/aa/abs/forth/aa43940-22/aa43940-22.html}

\bibitem[{van Saders \& Pinsonneault(2012)}]{van_saders_3he-driven_2012}
van Saders, J.~L., \& Pinsonneault, M.~H. 2012, The Astrophysical Journal, 751,
  98, aDS Bibcode: 2012ApJ...751...98V.
\newblock \url{https://ui.adsabs.harvard.edu/abs/2012ApJ...751...98V}

\bibitem[{VanderPlas(2018)}]{vanderplas_understanding_2018}
VanderPlas, J.~T. 2018, The Astrophysical Journal Supplement Series, 236, 16,
  arXiv: 1703.09824.
\newblock \url{http://arxiv.org/abs/1703.09824}

\bibitem[{Virtanen {et~al.}(2020)Virtanen, Gommers, Oliphant, Haberland, Reddy,
  Cournapeau, Burovski, Peterson, Weckesser, Bright, van~der Walt, Brett,
  Wilson, Jarrod~Millman, Mayorov, Nelson, Jones, Kern, Larson, Carey, Polat,
  Feng, Moore, Vand~erPlas, Laxalde, Perktold, Cimrman, Henriksen, Quintero,
  Harris, Archibald, Ribeiro, Pedregosa, van Mulbregt, \&
  Contributors}]{virtanen_scipy_2020}
Virtanen, P., Gommers, R., Oliphant, T.~E., {et~al.} 2020, Nature Methods, 17,
  261

\bibitem[{Vorobyov(2011)}]{vorobyov_embedded_2011}
Vorobyov, E.~I. 2011, The Astrophysical Journal, 729, 146, aDS Bibcode:
  2011ApJ...729..146V.
\newblock \url{https://ui.adsabs.harvard.edu/abs/2011ApJ...729..146V}

\bibitem[{West {et~al.}(2015)West, Weisenburger, Irwin, Berta-Thompson,
  Charbonneau, Dittmann, \& Pineda}]{west_activity-rotation_2015}
West, A.~A., Weisenburger, K.~L., Irwin, J., {et~al.} 2015, The Astrophysical
  Journal, 812, 3, aDS Bibcode: 2015ApJ...812....3W.
\newblock \url{https://ui.adsabs.harvard.edu/abs/2015ApJ...812....3W}

\bibitem[{Williams \& Best(2014)}]{williams_parametric_2014}
Williams, J.~P., \& Best, W. M.~J. 2014, The Astrophysical Journal, 788, 59,
  aDS Bibcode: 2014ApJ...788...59W.
\newblock \url{https://ui.adsabs.harvard.edu/abs/2014ApJ...788...59W}

\bibitem[{Wittenmyer {et~al.}(2014)Wittenmyer, Tuomi, Butler, Jones,
  Anglada-Escudé, Horner, Tinney, Marshall, Carter, Bailey, Salter, O'Toole,
  Wright, Crane, Schectman, Arriagada, Thompson, Minniti, Jenkins, \&
  Diaz}]{wittenmyer_gj_2014}
Wittenmyer, R.~A., Tuomi, M., Butler, R.~P., {et~al.} 2014, The Astrophysical
  Journal, 791, 114, aDS Bibcode: 2014ApJ...791..114W.
\newblock \url{https://ui.adsabs.harvard.edu/abs/2014ApJ...791..114W}

\bibitem[{Wright \& Eastman(2014)}]{wright_barycentric_2014}
Wright, J.~T., \& Eastman, J.~D. 2014, Publications of the Astronomical Society
  of the Pacific, 126, 838.
\newblock \url{https://ui.adsabs.harvard.edu/abs/2014PASP..126..838W}

\bibitem[{Yan {et~al.}(2019)Yan, Chen, Lazarz, Bizyaev, Maraston, Stringfellow,
  McCarthy, Meneses-Goytia, Law, Thomas, Falcon~Barroso, Sánchez-Gallego,
  Schlafly, Zheng, Argudo-Fernández, Beaton, Beers, Bershady, Blanton,
  Brownstein, Bundy, Chambers, Cherinka, De~Lee, Drory, Galbany, Holtzman,
  Imig, Kaiser, Kinemuchi, Liu, Luo, Magnier, Majewski, Nair, Oravetz, Oravetz,
  Pan, Sobeck, Stassun, Talbot, Tremonti, Waters, Weijmans, Wilhelm, Zasowski,
  Zhao, \& Zhao}]{yan_sdss-iv_2019}
Yan, R., Chen, Y., Lazarz, D., {et~al.} 2019, The Astrophysical Journal, 883,
  175, aDS Bibcode: 2019ApJ...883..175Y.
\newblock \url{https://ui.adsabs.harvard.edu/abs/2019ApJ...883..175Y}

\bibitem[{Yee {et~al.}(2017)Yee, Petigura, \& Braun}]{yee_precision_2017}
Yee, S.~W., Petigura, E.~A., \& Braun, K.~v. 2017, The Astrophysical Journal,
  836, 77.
\newblock \url{https://doi.org/10.3847%2F1538-4357%2F836%2F1%2F77}

\bibitem[{{Yurchenko} {et~al.}(2011){Yurchenko}, {Barber}, \&
  {Tennyson}}]{Yurchenko2011}
{Yurchenko}, S.~N., {Barber}, R.~J., \& {Tennyson}, J. 2011, \mnras, 413, 1828

\bibitem[{{Yurchenko} \& {Tennyson}(2014)}]{Yurchenko2014a}
{Yurchenko}, S.~N., \& {Tennyson}, J. 2014, \mnras, 440, 1649

\bibitem[{{Yurchenko} {et~al.}(2013){Yurchenko}, {Tennyson}, {Barber}, \&
  {Thiel}}]{Yurchenko2013}
{Yurchenko}, S.~N., {Tennyson}, J., {Barber}, R.~J., \& {Thiel}, W. 2013,
  Journal of Molecular Spectroscopy, 291, 69

\bibitem[{Zapolsky \& Salpeter(1969)}]{zapolsky_mass-radius_1969}
Zapolsky, H.~S., \& Salpeter, E.~E. 1969, The Astrophysical Journal, 158, 809,
  aDS Bibcode: 1969ApJ...158..809Z.
\newblock \url{https://ui.adsabs.harvard.edu/abs/1969ApJ...158..809Z}

\bibitem[{Zechmeister \& Kürster(2009)}]{zechmeister_generalised_2009}
Zechmeister, M., \& Kürster, M. 2009, Astronomy and Astrophysics, 496, 577.
\newblock \url{http://adsabs.harvard.edu/abs/2009A%26A...496..577Z}

\bibitem[{Zechmeister {et~al.}(2018)Zechmeister, Reiners, Amado, Azzaro, Bauer,
  Béjar, Caballero, Guenther, Hagen, Jeffers, Kaminski, Kürster, Launhardt,
  Montes, Morales, Quirrenbach, Reffert, Ribas, Seifert, Tal-Or, \&
  Wolthoff}]{zechmeister_spectrum_2018}
Zechmeister, M., Reiners, A., Amado, P.~J., {et~al.} 2018, Astronomy and
  Astrophysics, 609, A12.
\newblock \url{http://adsabs.harvard.edu/abs/2018A%26A...609A..12Z}

\bibitem[{Öberg {et~al.}(2011)Öberg, Murray-Clay, \&
  Bergin}]{oberg_effects_2011}
Öberg, K.~I., Murray-Clay, R., \& Bergin, E.~A. 2011, The Astrophysical
  Journal, 743, L16, aDS Bibcode: 2011ApJ...743L..16O.
\newblock \url{https://ui.adsabs.harvard.edu/abs/2011ApJ...743L..16O}

\end{thebibliography}

\listofchanges
\end{document}